\documentclass[twocolumn]{aastex63}
\usepackage{amsmath}
\usepackage[caption=false]{subfig}
\usepackage{placeins}
\usepackage{natbib}
\usepackage{hyperref}

\newcommand{\Rfiveoo}{\ensuremath{R_{500}}}
\newcommand{\Mfiveoo}{\ensuremath{M_{500}}}

\received{-}
\revised{-}
\accepted{-}
\submitjournal{ApJ}

\shorttitle{SPT - XMM}
\shortauthors{Ghirardini et al.}

\newcommand{\xmm}{\emph{XMM-Newton}}
\newcommand{\chandra}{\emph{Chandra}}

\begin{document}
\title{\bf Evolution of the Thermodynamic Properties of Clusters of Galaxies out to Redshift of 1.8}

\correspondingauthor{Vittorio Ghirardini}
\email{vittorio.ghirardini@cfa.harvard.edu}

\author[0000-0002-3736-8058]{Vittorio Ghirardini}
\affiliation{Center for Astrophysics $|$ Harvard \& Smithsonian, 60 Garden Street, MA 02138, USA }

\author{Esra Bulbul}
\affiliation{Center for Astrophysics $|$ Harvard \& Smithsonian, 60 Garden Street, MA 02138, USA }

\author{Ralph Kraft}
\affiliation{Center for Astrophysics $|$ Harvard \& Smithsonian, 60 Garden Street, MA 02138, USA }

\author{Matt Bayliss}
\affiliation{Department of Physics, University of Cincinnati, Cincinnati, OH 45221, USA}

\author{Bradford Benson}
\affiliation{Fermi National Accelerator Laboratory, P. O. Box 500, Batavia, IL 60510, USA}
\affiliation{Kavli Institute for Cosmological Physics, University of Chicago, 5640 South Ellis Avenue, Chicago, IL 60637, USA}
\affiliation{Department of Astronomy and Astrophysics, University of Chicago, 5640 South Ellis Avenue, Chicago, IL 60637, USA}

\author{Lindsey Bleem}
\affiliation{High Energy Physics Division, Argonne National Laboratory, 9700 South Cass Avenue, Lemont, IL 60439, USA}
\affiliation{Kavli Institute for Cosmological Physics, University of Chicago, 5640 South Ellis Avenue, Chicago, IL 60637, USA}

\author{Sebastian Bocquet}
\affiliation{Faculty of Physics, Ludwig-Maximilians-Universit¨at, Scheinerstr. 1, 81679 Munich, Germany}
\affiliation{HEP Division, Argonne National Laboratory, Argonne, IL 60439, USA}
\affiliation{Kavli Institute for Cosmological Physics, University of Chicago, 5640 South Ellis Avenue, Chicago, IL 60637, USA}

\author{Micheal Calzadilla}
\affiliation{Kavli Institute for Astrophysics and Space Research, Massachusetts Institute of Technology, 77 Massachusetts Avenue, Cambridge, MA 02139, USA }

\author{Dominique Eckert}
\affiliation{Department of Astronomy, University of Geneva, ch. d’Ecogia 16, 1290 Versoix, Switzerland}

\author{William Forman}
\affiliation{Center for Astrophysics $|$ Harvard \& Smithsonian, 60 Garden Street, MA 02138, USA }

\author{Juan David Remolina González}
\affiliation{Department of Astronomy, University of Michigan, 1085 S. University Ave, Ann Arbor, MI 48109, USA}

\author{Gourav Khullar}
\affiliation{Kavli Institute for Cosmological Physics, University of Chicago, 5640 South Ellis Avenue, Chicago, IL 60637, USA}
\affiliation{Department of Astronomy and Astrophysics, University of Chicago, Chicago, IL 60637, USA}

\author{Guillaume Mahler}
\affiliation{Department of Astronomy, University of Michigan, 1085 S. University Ave, Ann Arbor, MI 48109, USA}

\author{Michael McDonald}
\affiliation{Kavli Institute for Astrophysics and Space Research, Massachusetts Institute of Technology, 77 Massachusetts Avenue, Cambridge, MA 02139, USA }



\begin{abstract}

The thermodynamic properties of the hot plasma in galaxy clusters retains information on the processes leading to the formation and evolution of the gas in their deep, dark matter potential wells. These processes are dictated not only by gravity but also by gas physics, e.g. AGN feedback and turbulence.
In this work, we study the thermodynamic properties, e.g. density, temperature, pressure, and entropy, of the most massive and the most distant ($z > 1.2$) SPT-selected clusters, and compare them with those of the nearby clusters ($z<0.1$) to constrain  their evolution as a function of time and radius. We find that thermodynamic properties in the outskirts of high redshift clusters are remarkably similar to the low redshift clusters, and their evolution follows the prediction of the self-similar model. Their intrinsic scatter is larger, indicating that the physical properties that lead to the formation and virialization of cluster outskirts show evolving variance. On the other hand, thermodynamic properties in the cluster cores deviates significantly from self-similarity indicating that the processes that regulate the core are already in place in these very high redshift clusters. This result is supported by the unevolving physical scatter of all thermodynamic quantities in cluster cores.

\end{abstract}


%
\section{Introduction} 
\label{sec:intro}
Clusters of galaxies are the largest gravitationally-bound objects in the Universe and are ideal laboratories to study how cosmic structures form and evolve in time. While the majority of their mass is in the form of dark matter,  the hot fully ionized plasma, i.e. the intra-cluster medium (ICM), retains most of the baryonic component,  with only a small contribution from stars and cold gas \citep[3-5\%;][]{gonzalez+13}. 
The ICM is observable in the  X-ray band mainly through its emission via thermal Bremsstrahlung and radiative recombination processes. X-ray observations of clusters of galaxies provide in depth information about the ICM's thermodynamic properties. The thermal Sunyaev-Zeldovich (SZ) effect, a spectral distortion of the cosmic microwave background caused by the ICM, provides a complementary tool for finding clusters at all redshifts and examining their properties.

X-ray studies of clusters of galaxies provided constraints on thermodynamic properties of the ICM in nearby clusters with redshifts of $<$0.3 \citep[e.g.][]{croston+06,degrandi2002,cavagnolo+09,vikhlini+06,arnaud+10,pratt+10, bulbul12}. X-ray observations have also provided the serendipitous detection of single high redshift clusters \citep[$z > 1$;][]{fabian+03,tozzi+15,brodwin+16}, however these studies are prone to X-ray selection biases \citep[e.g. the cool-core bias, ][]{Eckert+11}. The majority of theoretical studies in the literature also focus on predicting thermodynamic properties of the ICM in nearby clusters \citep{kravtsov12}. 
In recent years, owing to the wide area sky surveys performed with the current SZ telescopes, e.g. the South Pole Telescope \citep[SPT; ][]{Carlstrom11}, the Atacama Cosmology Telescope \citep{Fowler+07}, and the Planck mission \citep{planck+16}, it has become possible to detect clusters out to much higher redshifts (z$\sim$1.8) with a simpler selection function, i.e., the SZ signal tightly correlates with mass \citep{planck_14_scaling,Bocquet+19}. Therefore, X-ray follow-up observations of the SZ selected clusters provide a unique opportunity to study the evolution of ICM properties in a uniform way. 

Integrated X-ray properties of the SPT-selected clusters spanning a large redshift range have been studied in the literature \citep{mcdonald+14, Sanders+18, bulbul19}. \cite{bartalucci+17_general, bartalucci17} examined the electron number density distribution of the ICM by combining the \chandra\ and \xmm\ follow-up observations of a handful of high redshift clusters (z~$\sim\,$~1) detected by SPT and ACT. Studies of the evolution of the ICM properties in large SZ-selected cluster samples have become possible with large targeted X-ray follow-up programs, e.g. \chandra\ Large Program (LP). \cite{mcdonald+13,mcdonald+14} have reported that the evolution in the electron number density is consistent with the self-similar expectation, where only gravitational forces dominate the formation and evolution of the ICM in the intermediate regions ($0.15R_{500}\,-\,R_{500}$)\footnote{$
\Rfiveoo = \left( \frac{3 \Mfiveoo}{4\pi \times 500 \rho_{\mathrm{crit}}(z)}\right)^{1/3},
$ is the overdensity radius within which the mean density is 500 times the critical density of the Universe} of the SPT-selected clusters of galaxies in the redshift range of $0.2<z<1.2$. The authors also found a clear deviation from self-similarity in the evolution of the core density of these clusters. Deeper \chandra\ observations of 8 high redshift SPT-selected clusters beyond redshift of 1.2 confirm earlier results of no evolution in the cluster cores indicating that Active Galactic Nuclei (AGN) feedback is tightly regulated since this early epoch and self-similar evolution is followed in intermediate regions \citep[][hereafter MD17]{mcdonald17}. Recently, \cite{Sanders+18} reported a self-similar evolution of the thermodynamic properties at all radii for the same large sample but using a different center and a slightly different analysis scheme out to $R_{500}$.

In this work, we combine deep \chandra\ and \xmm\ observations of a sample of the highest redshift and most massive 7 SPT-selected galaxy clusters beyond redshift of 1.2 to study the thermodynamic properties of the ICM and their evolution. We take advantage of the sharp point spread function (PSF) of \chandra\ to study the small scales (at this redshift,beyond 1.2, \chandra\ resolution of 0.5 arcsec corresponds to about 5 kpc), while the large effective area of \xmm\ provides the required photon statistics to measure densities and temperatures out to large scales. 
Thus, the combination of \chandra\ and \xmm\ allows us to obtain precise and extended density profiles, and sufficient photon statistics to measure temperature profiles required to probe the evolution of the ICM properties, e.g. density, temperature, pressure, and entropy, out to the overdensity radius $R_{500}$. 
The paper is organized as follows: in Sect.~\ref{sec:analysis}, we present the sample properties and the analysis of the \xmm\ and \chandra\ data of the sample; in Sect.~\ref{sec:results} we provide our results; the systematic uncertainties are discussed in Sect.~\ref{sec:syst} and we finally summarize our conclusions in Sect.~\ref{sec:conc}.

Throughout the paper we assume a flat $\Lambda$CDM cosmology with $\Omega_m=0.3$, $\Omega_\Lambda=0.7$ and $H_{0}=70$ km s$^{-1}$ Mpc$^{-1}$. All uncertainties quoted correspond to 68\%  single-parameter confidence intervals unless otherwise stated.

\begin{table*}[]
    \centering
\begin{tabular}{ c c c c  c  c c c }
\hline\hline
Cluster & redshift & R.A. & Dec. & t$_{CXO}$ & $t_{MOS1}$ & $t_{MOS2}$ & $t_{pn}$ \\
      & & [deg] & [deg] & [ks] & [ks] & [ks] & [ks] \\
\hline
SPT-CLJ0205-5829 & 1.322 & 31.4437 & -58.4855 & 57.8 &  69.4 &   70.2 &   52.7 \\
SPT-CLJ0313-5334 & 1.474 & 48.4809 & -53.5781 & 113.6 & 186.0 &  195.2 &  164.5 \\
SPT-CLJ0459-4947 & 1.70 & 74.9269 & -49.7872 & 136.2 & 461.9 &  471.6 &  410.3 \\
SPT-CLJ0607-4448 & 1.401 & 91.8984 & -44.8033 & 111.1 & 132.7 &  144.8 &   98.7 \\
SPT-CLJ0640-5113 & 1.316 & 100.0645 & -51.2204 & 173.4 & 127.7 &  131.9 &  114.0 \\
SPT-CLJ2040-4451 & 1.478 & 310.2468 & -44.8599 & 96.7 &  76.2 &   76.6 &   72.8 \\
SPT-CLJ2341-5724 & 1.259 & 355.3568 & -57.4158 & 112.4 & 107.7 &  107.7 &   93.0 \\
\hline\hline
\end{tabular}

    \caption{Properties of the sample; cluster name, redshift, coordinates of the centroid, \chandra\ clean exposure time, and \xmm\ (EPIC MOS1, MOS2, and pn) clean exposure times.}
    \label{tab:my_label}
\end{table*}
\section{Cluster Sample and Data analysis} \label{sec:analysis}
\subsection{Cluster Sample }

Our sample consists of 7 SPT-selected high redshift ($z > 1.2$) massive clusters of galaxies with S/N ratio greater than 6 and a total SZ  inferred mass greater than $3\times10^{14}\ M_{\odot}$ \citep{bleem15}.  The deep \xmm\ observations of these clusters have been performed in AO-16 (PIs E. Bulbul, and A. Mantz) and \chandra\ observations were performed in AO-16 through both XVP program (PI M. McDonald) and two guest observer (GO) programs (PI G. Garmire,  S. Murray). The total \chandra\ and \xmm\ clean exposure time used in this work is $\sim$2~Ms (see Table~\ref{tab:my_label}).

\subsection{Imaging Analysis} 
\label{sec:imaging}
\subsubsection{\xmm\ Imaging Analysis} \label{sec:xmm_analysis}
 We strictly follow the data analysis prescription developed by the \xmm\ Cluster Outskirts Project collaboration \citep[X-COP,][]{xcop} with their new background modeling method \citep{ghirardini18}. Thanks to the reduction of the systematic uncertainty on the background below 5\% through this method, we are able to measure thermodynamic properties of high redshift clusters out to $R_{500}$. We provide the summary of the analysis below. We use the \xmm\ Science Analysis System (SAS) and Extended Source Analysis Software \citep[ESAS, ][]{snowden08}, developed to analyze \xmm\ EPIC observations. In our analysis, we use XMM-SAS v17.0  and CALDB files as of January 2019 (XMM-CCF-REL-362). 
 
Filtered event files are generated using the XMM-SAS tasks \texttt{mos-filter} and \texttt{pn-filter}. 
The photon count images are extracted from the filtered event files from three EPIC detectors, MOS1, MOS2, and pn on board \xmm, in the soft and narrow energy band [0.7-1.2]~keV. The choice of this narrow band is to maximize the source to background ratio and minimize the systematic uncertainties in the modeling of the EPIC background \citep{ettori+11}. To create the total EPIC images, the count images from the three detectors are summed. Next, we use \texttt{eexpmap} to compute exposure maps by also taking the vignetting effect into account. The exposure maps are also summed using the scaling factors of 1:1:3.44 for MOS1:MOS2:pn detectors, i.e. the ratio between the effective area of MOS and pn in the [0.7-1.2]~keV energy band. These scaling factors are computed individually for each observation. 

The high-energy particle background images are generated using the background images extracted from the unexposed corners of the detectors, and rescaling them to the field-of-view  (FoV). After the light curve cleaning, residual soft protons still contaminate the FoV \citep[][]{salvetti17}. We measure the soft proton contamination in the FoV of each observation by calculating the fraction of count rates in the unexposed and exposed portions of the detector in a hard band ([7--11.5]~keV) \citep{lm08}. We then generate the 2D soft proton image \cite[][as described in their Appendix A]{ghirardini18}, to model the remaining soft proton contamination. We construct the total non-X-ray background (NXB) by summing the high energy particle background and the residual soft protons images. Thus, we obtain total photon images, exposure maps, and total non-X-ray background images for each observation.

To detect and excise point and extended sources in the FoV, we use the XMM-SAS tool \texttt{ewavelet} with a selection of scales in the range 1--32 pixels with S/N threshold of 5. We remove all the point sources found by \texttt{ewavelet} tool from the further analysis.
We also run CIAO point source detection tool \texttt{wavdetect} on \chandra\ images. The sources detected on \xmm\ and \chandra\ images are combined to remove missed point sources by \texttt{ewavelet}. See Sect.~\ref{sec:cxo_analysis} for details on the \chandra\ analysis.

\subsubsection{Point Spread Function Correction for \xmm}       
\label{sec:psf}

Due to the relatively large size of the Point Spread Function (PSF) of \xmm, some X-ray photons that originate from one particular region on the sky may be detected elsewhere on the detector. \xmm's 5~arcsec wide PSF (at aim point) needs to be taken into account to correct for this effect due to the small spatial scales of the clusters in our sample \citep{read+11}. To estimate the impact of the PSF on  surface brightness profiles, we first create a matrix, $PSF_{i,j}$, whose value is the fraction of photons originating in the i$^\text{th}$ annulus in the sky but detected in the j$^\text{th}$ annulus on the detector.

\begin{figure}
    \centering
    \includegraphics[width=0.5\textwidth]{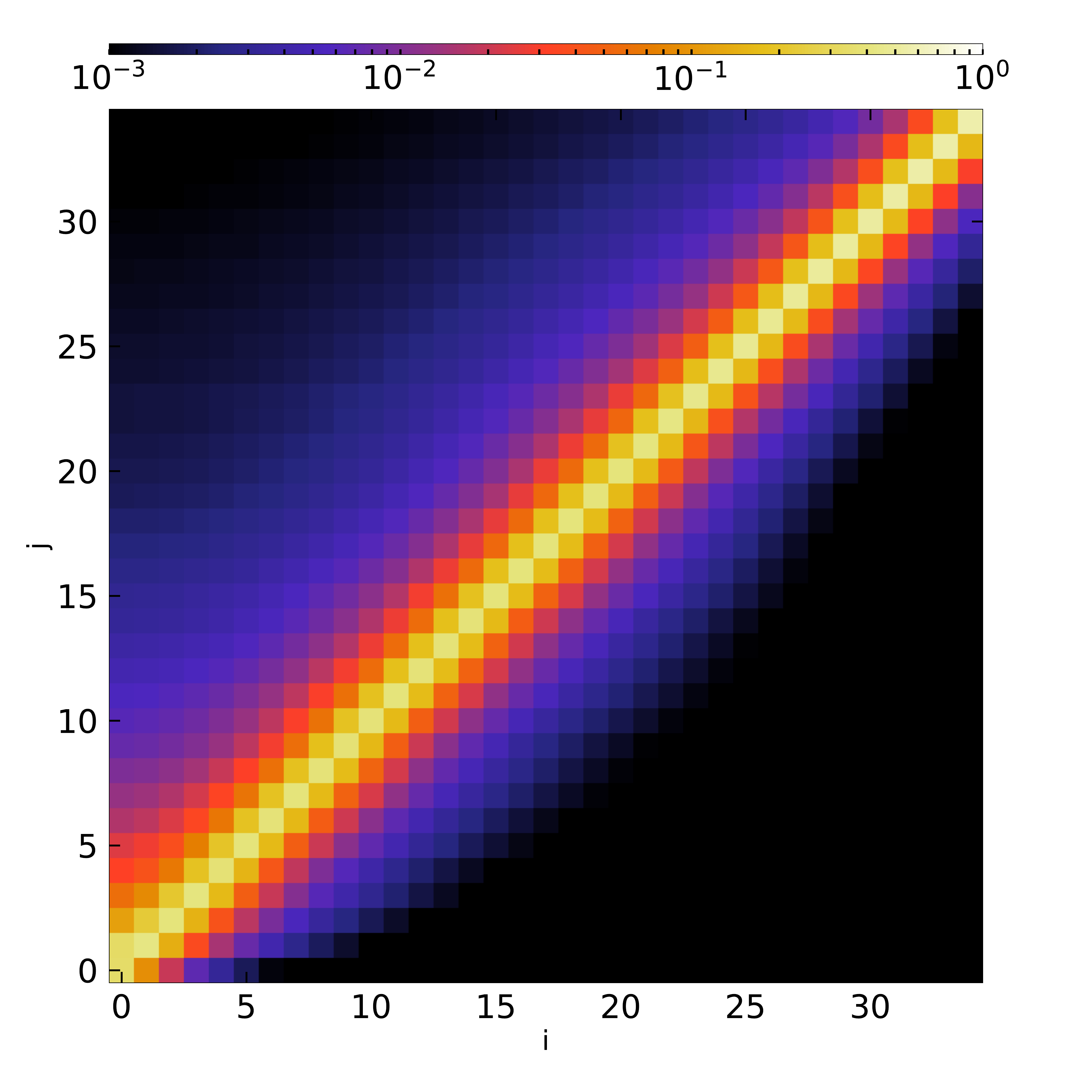}
    \caption{Example PSF matrix image used in our analysis. The matrix shows the contribution to the j$^{th}$ annulus from the i$^{th}$ annulus at each position $(i,j)$. The non-diagonal and asymmetric nature of the distribution shows that the contribution of the emission from the cluster center to the outskirts is not negligible and should be corrected for.}
    \label{fig:PSF}
\end{figure}

In practice, to model the PSF and build the PSF matrix, we, following \cite{eckert16}  and (2019), build an image of an annulus with a constant value inside the annulus itself and zero outside, with the constant chosen in such a way that the sum of all pixels is 1; this represents the probability density function (PDF) for the true photons generated in the annulus that represents their origin on the plane of the sky. The \xmm\ mirrors smear this annulus-limited PDF onto a larger fraction of the exposed CCDs.
We then use a functional form \citep[e.g., a King profile plus a Gaussian as in ][]{read+11} to model the instrumental PSF function in each location of the detector.
The observed photons are the result of the convolution of the original sky photons  by the PSF function, $S_{\rm b, obs} =  {\rm PSF} \circledast S_{\rm b, true}$.
The PDF is no longer limited to the annulus, but has spread to the surroundings.
The fraction of the PDF, originating from annulus `$i$', that now is present in annulus `$j$' is the value that we put in the corresponding line and row of the PSF matrix. 
An example image of the PSF matrix is given in Figure~\ref{fig:PSF}. While the majority of the photons which originate from a given annulus are detected in the same region (the largest values are on the diagonal), some fraction of them are detected in a different annuli.

\subsubsection{\chandra\ Imaging Analysis} \label{sec:cxo_analysis}

We process the \emph{Chandra} observations of the sample using the CIAO~4.11 \citep[\emph{Chandra} Interactive Analysis of Observations, ][]{ciao} and calibration files in CALDB 4.8.2. We filter the data for good time intervals, including the corrections for charge transfer inefficiency \citep{grant05}. We remove the photons detected in bad CCD columns and hot pixels, compute the calibrated photon energies by applying the ACIS gain maps, and correct for their time dependence. We also remove the time intervals that are affected by the background flares by examining the light curves.
We ran \texttt{wavdetect}, the standard CIAO tool to find point sources in \chandra\ observations, with scales in range 1-32 pixels and threshold for identifying a pixel as belonging to a source of $10^{-6}$. We merge point sources detected on \chandra\ images  with those detected on \xmm\ images as described in Sect.~\ref{sec:xmm_analysis}. All point sources detected in this process are excluded from further analysis.

We extract photon count images in the soft energy band [0.5-2.0]~keV, as it is routinely done when analyzing \chandra\ data.

For the instrumental background we use blank-sky background spectra which is rescaled based on the flux in the hard band [9.5 - 12]~keV to account for variations in the particle background. Exposure maps are generated to correct for vignetting effect. The particle background subtracted, vignetting corrected
images are shown in Figure~\ref{fig:raw_images}. Due to the small size of \chandra's PSF, 80\% of the total encircled counts are detected within 0.7 arcsec from its source. We, therefore, do not apply any PSF correction to \chandra\ data.

\subsubsection{Joint \chandra\  and \xmm\ Surface Brightness Analysis }
\label{sec:joint}

To compute the surface brightness profile, we first measure the number of photon counts ($N_{c,i}$) in concentric annuli around the cluster center. We find the cluster center by measuring the centroid in a 250--500 kpc aperture on \chandra\ images following the approach introduced by \cite{mcdonald+13}. This method allows us to find the center of the large scale distribution of the intra-cluster plasma independent of the core morphology. The widths of the annuli are required to be larger than 2~arcsec, increasing logarithmically, and with at least 30~counts contained within each annulus. For \xmm\ the width of these annuli is determined in such way that each has at least a total of 100 counts and the minimum width is larger than 5~arcsec. We then compute the mean exposure time $t_{exp,i}$ from the exposure map, and background counts using the total background map $N_{\rm NXB,i}$ for the two X-ray telescopes. The surface brightness in each annulus is calculated using the following relation:

\begin{equation}
    S_{B_i} = \frac{N_{\rm c,i} - N_{\rm NXB,i}}{t_{\rm exp,i} \cdot A_{\rm reg,i}}
    \label{eq:SB}
\end{equation}

\noindent where $A_{reg,i}$ is the area, in arcmin$^2$, of each annulus `$i$'.
\begin{table*}[]
\setlength{\tabcolsep}{2pt}
    \centering
\begin{tabular}{ c c c c c c c c c }
\hline
\hline
Cluster & $\log(n_0)$  &  $\log(r_c)$  &  $\log(r_s)$  &  $\alpha$  &  $\beta$  &  $\epsilon$  &  $\log (B_{XMM})$  &  $\log (B_{Chandra})$  \\
\hline
SPT-CLJ0205-5829  & $-4.87 \pm  0.28$ & $2.06 \pm  0.45$ & $5.62 \pm  0.09$ & $1.06 \pm  0.48$ & $-0.00 \pm  0.04$ & $6.17 \pm  0.35$ & $-10.99 \pm  0.02$ & $-10.81 \pm  0.02$\\
SPT-CLJ0313-5334  & $-4.83 \pm  0.32$ & $1.21 \pm  0.78$ & $5.62 \pm  0.08$ & $1.42 \pm  0.95$ & $0.03 \pm  0.03$ & $7.02 \pm  0.51$ & $-11.07 \pm  0.02$ & $-10.48 \pm  0.01$\\
SPT-CLJ0459-4947  & $-3.12 \pm  0.15$ & $2.08 \pm  0.24$ & $5.28 \pm  0.09$ & $1.09 \pm  0.46$ & $0.16 \pm  0.03$ & $4.69 \pm  0.22$ & $-10.78 \pm  0.00$ & $-10.09 \pm  0.01$\\
SPT-CLJ0607-4448  & $-3.16 \pm  0.22$ & $2.27 \pm  0.29$ & $5.16 \pm  0.24$ & $1.61 \pm  0.45$ & $0.20 \pm  0.05$ & $3.18 \pm  0.21$ & $-10.53 \pm  0.01$ & $-10.18 \pm  0.01$\\
SPT-CLJ0640-5113  & $-3.66 \pm  0.20$ & $2.02 \pm  0.39$ & $5.12 \pm  0.10$ & $1.21 \pm  0.47$ & $0.10 \pm  0.03$ & $4.31 \pm  0.19$ & $-10.78 \pm  0.01$ & $-10.38 \pm  0.01$\\
SPT-CLJ2040-4451  & $-5.21 \pm  0.32$ & $1.79 \pm  0.45$ & $5.75 \pm  0.11$ & $0.50 \pm  0.47$ & $0.01 \pm  0.04$ & $5.49 \pm  0.34$ & $-10.39 \pm  0.01$ & $-10.45 \pm  0.02$\\
SPT-CLJ2341-5724  & $-3.26 \pm  0.09$ & $2.72 \pm  0.10$ & $5.95 \pm  0.10$ & $0.45 \pm  0.38$ & $0.29 \pm  0.01$ & $3.24 \pm  0.24$ & $-10.48 \pm  0.01$ & $-10.61 \pm  0.01$\\
\hline
\hline
\end{tabular}
\caption{Best-fit parameters of the \cite{vikhlini+06} density model and measured background levels in the \chandra\ and \xmm\ observations are given.}
    \label{tab:density_bf}
\end{table*}
\begin{table*}
\setlength{\tabcolsep}{2pt}
    \centering

\begin{tabular}{ c c c c c c c }
\hline
\hline
Cluster & $T_0$  &  $r_{cool}$  &  $r_t$  &  $\frac{T_{min}}{T_0}$  &  $a_{cool}$  &  $\frac{c}{2}$ \\
 & keV & kpc & kpc \\
\hline
SPT-CLJ0205-5829  & $9.1 \pm  2.2$ & $23 \pm  14$ & $358 \pm  195$ & $0.45 \pm  0.19$ & $1.89 \pm  0.51$ & $0.36 \pm  0.18$\\
SPT-CLJ0313-5334  & $6.8 \pm  1.5$ & $24 \pm  15$ & $354 \pm  184$ & $0.44 \pm  0.19$ & $1.87 \pm  0.51$ & $0.40 \pm  0.18$\\
SPT-CLJ0459-4947  & $9.4 \pm  1.3$ & $22 \pm  12$ & $283 \pm  112$ & $0.46 \pm  0.19$ & $1.91 \pm  0.50$ & $0.44 \pm  0.14$\\
SPT-CLJ0607-4448  & $5.9 \pm  0.8$ & $25 \pm  16$ & $406 \pm  225$ & $0.44 \pm  0.19$ & $1.85 \pm  0.50$ & $0.28 \pm  0.18$\\
SPT-CLJ0640-5113  & $8.1 \pm  1.4$ & $23 \pm  13$ & $300 \pm  136$ & $0.45 \pm  0.19$ & $1.90 \pm  0.51$ & $0.45 \pm  0.16$\\
SPT-CLJ2040-4451  & $13.7 \pm  4.2$ & $22 \pm  13$ & $189 \pm  71$ & $0.46 \pm  0.19$ & $1.92 \pm  0.50$ & $0.62 \pm  0.14$\\
SPT-CLJ2341-5724  & $6.4 \pm  1.0$ & $23 \pm  14$ & $402 \pm  220$ & $0.44 \pm  0.19$ & $1.87 \pm  0.51$ & $0.30 \pm  0.18$\\
\hline
\hline
\end{tabular}
\caption{Best-fit parameters of the \cite{vikhlini+06} temperature model.}
    \label{tab:temperature_bf}
\end{table*}

From a theoretical point of view, the surface brightness profile is related to the number density through;
\begin{equation}
S_{B_i}\propto n_p (r)\, n_e(r)\, dl
\label{eq:tSb}
\end{equation}
\noindent where $n_p$ and $n_e$ are number densities of protons and electrons, and $dl$ is the integral along the line of sight. We fit the \cite{vikhlini+06} density model to the observed \chandra\ and \xmm\ surface brightness data, jointly. 
\begin{equation}
n_e^2(r) = \frac{n_0^2 \left( \frac{r}{r_c} \right)^{-\alpha}}{ \left( 1 + \left( \frac{r}{r_c} \right)^2 \right)^{3\beta-\alpha/2} \cdot \left(  1 + \left( \frac{r}{r_s} \right)^3 \right)^{\epsilon/3} } 
\end{equation}
The parameters of the ICM model are constrained by fitting the observed counts $N_{c,i}$ in each annulus against the predicted counts $\mu_i$ (see Equation~\ref{eq:PSF_multiscale}) using the following Poisson likelihood:

\begin{equation}    
    - \log \mathcal{L} = \sum_{i=1}^N \mu_i - N_{c,i} \log \mu_i
\label{eq:poisson_like}
\end{equation}
The net number of counts $\mu_i$ inferred by the ICM model in the $i$th annulus is calculated using the predicted surface brightness, Eq.~\ref{eq:tSb}, convolved with the PSF matrix, considering the exposed area and time for each annulus, as well as both sky and particle background.

%
\begin{equation}
    \mu_i = \left\{ \sum_j {\rm PSF}_{i,j} \cdot \left( S_{\rm b,ICM,i} + {\rm B_{sky}} \right) \right\} \cdot t_{\rm exp,i} \cdot A_{\rm reg,i} + N_{\rm NXB,i}
    \label{eq:PSF_multiscale}
\end{equation}

\noindent where $t_{\rm exp,i}$ and $A_{\rm reg,i}$ are respectively the exposure time and area of the annulus `i', $B_{sky}$ is the cosmic X-ray background, and $N_{\rm NXB,i}$ are the detector background counts. 
 
The sum of \xmm\ and \chandra\ likelihoods is used as total likelihood for the fit. We first minimize the $\chi^2 = - 2 \log \mathcal{L} $ using the Nelder-Mead method \citep{Gao2012}. Then we fit using the Bayesian nested sampling algorithm MultiNest \citep{multinest} using shallow gaussian priors centered around the Nelder-Mead method best-fit results and with a standard deviation of 1 (or 2.3~dex) in order to ensure that the fit is not stuck in a local minimum.

The surface brightness profiles and best-fit models are shown in Figure~\ref{fig:gallery_sb}, while the best-fit parameters of the ICM model are given in Table~\ref{tab:density_bf}.
We note that the emissivity measurements of \chandra\ and \xmm\ observatories are consistent with each other within 3\%, therefore, calibration differences are irrelevant in the measurements of emissivity and number density \citep[as also shown in ][]{bartalucci17}.

\subsection{\xmm\ Spectral Analysis}
\label{sec:spec_xmm}
We extract spectra using the XMM-ESAS tools \texttt{mos-spectra} and \texttt{pn-spectra} \citep{snowden08}. Redistribution matrices (RMFs) and ancillary response files (ARFs) are created with {\it rmfgen} and {\it arfgen}, respectively. The point sources (see Section \ref{sec:xmm_analysis} for details) are excluded from the spectral analysis. The spectral fitting package \textsc{Xspec} v12.10 \citep{xspec} with ATOMDB v3.0.9 is used in the analysis \citep{foster+2012}. The Galactic Column density is allowed to vary within 15\% of the measured LAB value in our fits \citep{kalberla05}. The extended C-statistics are used as an estimator of the goodness-of-fit \citep{cash79}. The abundances are normalized to the \cite{aspl} solar abundance measurements with the mean molecular weight $\mu = 0.5994$ and the mean molecular mass per electron $\mu_e = 1.1548$, and the ratio between number density of protons to electrons equal to $n_p/n_e = 0.8527$.
The MOS spectra are fitted in the energy band of 0.5--12~keV, while we use the 0.5--14~keV energy band for pn. We ignore the energy ranges between 1.2--1.9~keV for MOS, and 1.2--1.7~keV and 7.0--9.2~keV for pn due to the presence of bright and time-variable fluorescence lines. The energy band below 0.5~keV, where the EPIC calibration is uncertain, is eliminated from spectral fits. The source spectrum is modeled with an absorbed single-temperature thermal model {\it apec} with varying  temperature,  metallicity, and normalization. For the clusters with multiple observations the model parameters are tied between multiple spectra and fitted jointly. 

The particle background is determined using the rescaled filter-wheel-closed spectra, that allows us to measure the intensity and the spectral shape. On top of this, we include an additional model component for the residual soft protons \citep{salvetti17}, modeled as a broken power law with shape fixed \citep[slopes 0.4 and 0.8 and break energy 5 keV][]{lm08} and normalization free. 
Regarding the sky background, we model it as the sum of three components: (i) the cosmic X-ray background (CXB) with an absorbed power law with photon index fixed to 1.46, (ii) the galactic halo (GH) with an absorbed APEC model with temperature free to vary in the range [0.1-0.6] keV, and (iii) the local bubble (LB) with an APEC model with temperature fixed to 0.11 keV\citep{snowden08,lm08}. The normalizations of the CXB, LB, and GH background components are set free.
To find the sky parameters we fit the background region, by extracting a spectrum 5 arcmin ( $\sim 5 R_{500}$ ) away from the core.
We impose gaussian priors on these parameters with width equal to the parameter uncertainty found in the fitting of the background region.

\begin{figure}
    \centering
    \includegraphics[width=0.5\textwidth]{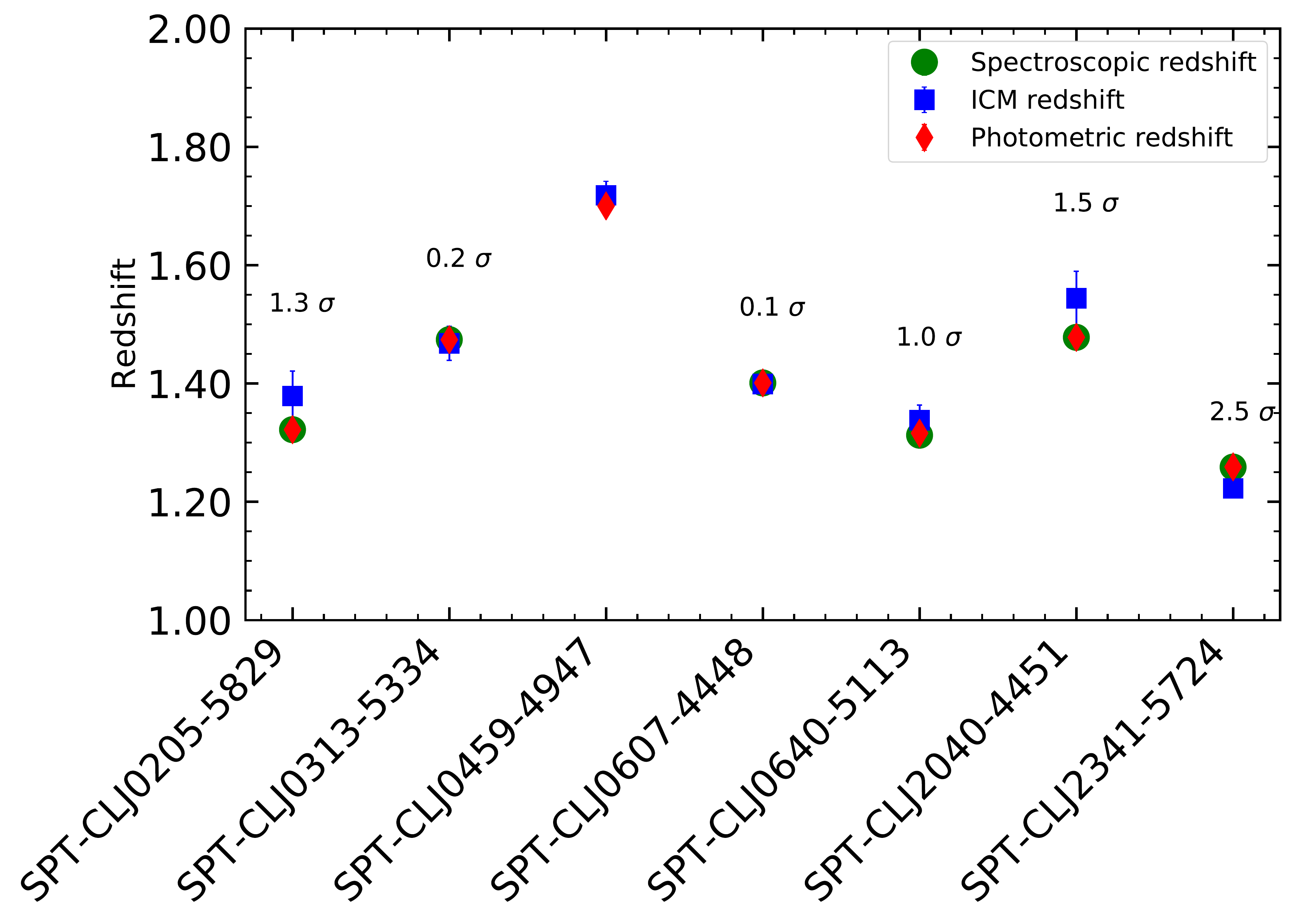}
     \caption{Comparisons of X-ray redshifts (in blue) with the photometric redshifts in red  \citep{bleem15}, spectroscopic redshifts in green \cite{Bayliss2040,Stalder0205,khullar19}. The error bars indicate the sum of statistical and systematic uncertainties at 1$\sigma$ level.
    }
    \label{fig:redshift}
\end{figure}

We first extract the \xmm\ spectra within $R_{500}$ to measure the redshifts of the clusters from the X-ray data. Fitting the spectra 
within $R_{500}$, so that the statistics are of high quality to determine an accurate X-ray redshift. We fit these spectra using an absorbed single temperature thermal model with free temperature, metallicity, redshift, and normalization.
Taking into account the gain calibration uncertainty of \xmm\ pn at 3 keV (the redshifted position of the Fe-K line) of 12~eV (private communication with the \xmm\ calibration team), we find that the redshifts are consistent with the previously reported photometric \citep{bleem15} and spectroscopic redshifts \citep{Bayliss2040, khullar19, Stalder0205} within $2\sigma$ confidence level for these clusters. 
A comparison of redshifts based on X-ray data with photometric and spectroscopic redshifts are shown in Figure~\ref{fig:redshift}. We point out that for SPT-CLJ0459-4947 the previously reported redshift \citep{Bocquet+19} is measured using the position of the Fe-K line from \emph{XMM-Newton} data from LP by A. Mantz.

To examine the radial profiles of thermodynamic properties, we next extract the spectra from concentric annuli with sizes increasing logarithmically around the cluster centroid. The minimum width of annuli is set to be $\sim$15~arcsec to minimize the effect of the \xmm's PSF, but still having a large enough statistic to determine the projected temperature. We group the output spectra to ensure having a minimum of 5 counts per bin.  The \xmm\ PSF is taken into account 
using the cross-talk ARFs generated by the SAS task \texttt{arfgen}. This method allows the spectra to be co-fitted by taking into account the cross-talk contribution to an annulus from another region \citep{snowden08,ettori+10}. The use of flat constant priors on the temperature, metallicity, and  the use of the `jeffreys' prior on the normalizations (i.e. $ {\rm Prior (K_{apec)}} =  {\rm K_{apec}^{-1}}$) allows us to account for  the uncertainty on the sky background as well as the uncertainty in their free parameters.
The spectra are fit using the Monte Carlo Markov Chain (MCMC) implementation in $\texttt{Xspec}$ of the Goodman-Weare algorithm  \citep{Goodman+10}, with 50000 steps and 1000 burn-in period to ensure we investigate the parameter space and derive the uncertainties on free parameters (temperature, metallicity, and normalization) in our fitting software. 
At the end of this process, we obtain the best-fit projected temperatures and their covariance matrix, which are easily computed using the MCMC chain.

To obtain the three-dimensional deprojected temperature profile of each cluster, we project the ICM temperature model on the plane of the sky by taking into account emission weighting to determine spectroscopic-like temperature \citep{mazzotta+04}, 
\begin{equation}
T_{\rm 2D,sl,i} = \frac{\int n_e^2 T_{\rm 3D}^{1-\alpha} dV}{\int n_e^2 T_{\rm 3D}^{-\alpha} dV}.
\label{eq:Tspeclike}
\end{equation}
\noindent where $\alpha = 3/2$ and $n_e$ is the electron number density, $T_{sl}$ is the predicted 2D spectral temperature, and the temperature model $T_{3D}$ is a widely used phenomenological model to describe the temperature profiles \citep{vikhlini+06}:

The 3D ICM model we used in this work is  

\begin{equation}
T_{3D} (r) = T_0 \frac{\frac{T_{min}}{T_0} + \left( \frac{x}{r_{cool}} \right)^{a_{cool}}}{1+\left( \frac{x}{r_{cool}} \right)^{a_{cool}}}\frac{1}{\left(1+\left( \frac{x}{r_t} \right)^2 \right)^{\frac{c}{2}}}
\label{eq:T_vikh}
\end{equation}

We first minimize the $\chi^2 = - 2 \log \mathcal{L} $ using the Nelder-Mead method \citep{Gao2012}. Then we fit using the Markov Chain Monte Carlo method using the code \texttt{emcee} \citep{emcee} using gaussian priors centered around the Nelder–Mead method results and with a sigma of 0.5 (or 1.15 dex).
 We use 10000 steps with burn-in length of 5000 steps to have resulting chains independent of the starting position and thinning of 10 to reduce the correlation between consecutive steps.
The likelihood adopted in the fit is;

\begin{equation}
    \log \mathcal{L}= -  \left(\log T_{obs} - \log T_{sl} \right) \Sigma_{i,j} \left(\log T_{obs} - \log T_{sl} \right)^{\rm \mathbf{T}}
    \label{eq:loglikeT}
\end{equation}
\noindent where $T_{obs}$ and $T_{sl}$ are the arrays of the measured spectral temperatures and of the spectroscopic-like projected temperatures as of Equation~\eqref{eq:Tspeclike} respectively, and $ \Sigma_{i,j}$ is the spectral log-temperature covariance matrix, see Sect.~\ref{sec:spec_xmm}.
 Thus a $\chi^2$-like log-likelihood, where the temperature distribution in each annulus is assumed to be a log-normal \citep{Andreon+12} and the full covariance between the annuli is considered.
The best-fit parameters for the temperature profile are given in Table~\ref{tab:temperature_bf}.

\section{Results}
\label{sec:results}
In this section we explore thermodynamic properties (e.g., density, temperature, pressure, and entropy) of the high redshift SPT clusters in our sample taking advantage of the SPT SZ survey's clean selection function and its high sensitivity. We further compare the thermodynamic properties  of the ICM of the clusters in our sample with the X-COP sample to investigate their evolution with redshift. The X-COP sample is selected based on the \emph{Planck} signal-to-noise ratio including only low redshift clusters with $z < 0.1$ \citep[][ G18 hereafter]{ghirardini18_b}. In G18, the authors were able to recover ICM properties out to the Virial radius using the joint X-ray and SZ analysis, adding on to the previous studies which probe the region within $R_{500}$ by joining X-ray and SZ observations \citep[e.g.][]{ameglio+07,hasler2012, bonamente2012,eckert13a,eckert13n,Shitanishi+18}. We further remark that the analysis done for the high redshift clusters is almost identical to the analysis applied in G18 for the X-COP cluster sample, allowing us for controlled measurement of the evolution in the thermodynamic quantities.

The self-similar model \citep{Kaiser+86}, which assumes purely gravitational collapse, predicts a particular evolution with redshift of the cluster properties once they are scaled based on their common quantities, e.g. mass within an overdensity radius \citep{voit+05}. 
We, therefore, measure the mass of our clusters and rescale our thermodynamic quantities with this mass within $R_{500}$. In the next section  we describe our method for the mass reconstruction under the assumption of hydrostatic equilibrium (HE), and then show the thermodynamic profiles and describe their properties.

\begin{figure*}
    \centering
    \includegraphics[width=0.49\textwidth]{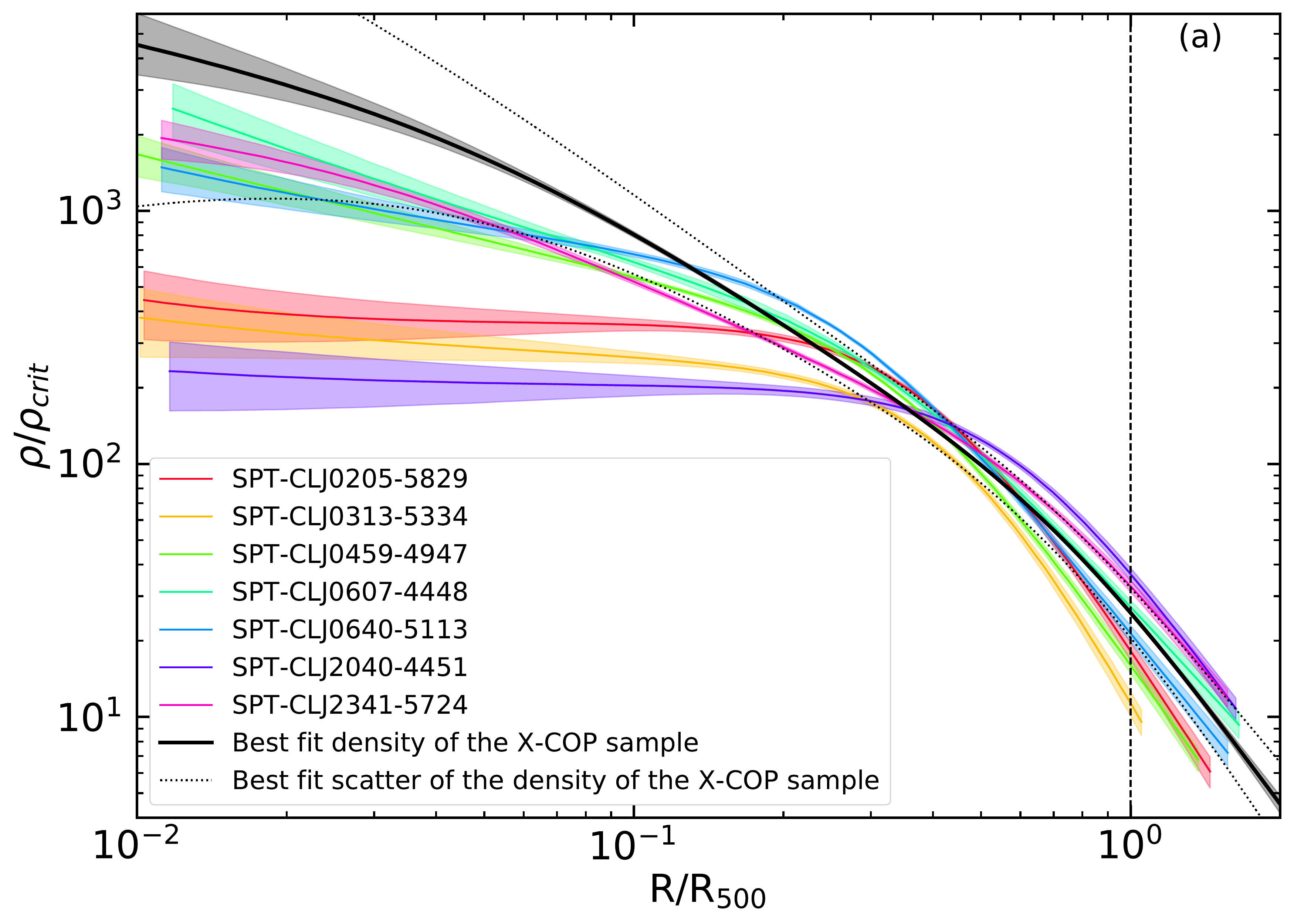}~
    \includegraphics[width=0.49\textwidth]{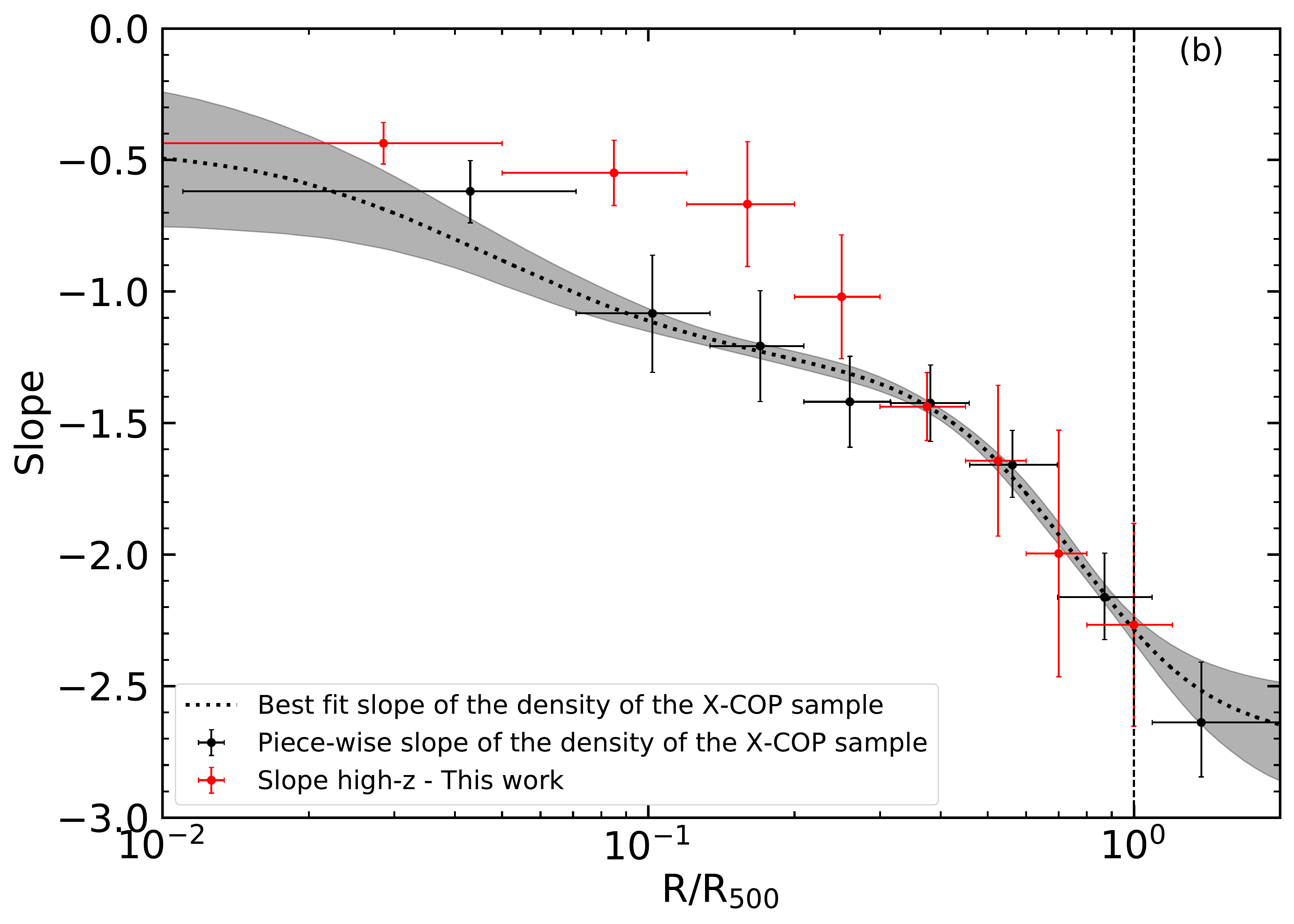}

    \includegraphics[width=0.49\textwidth]{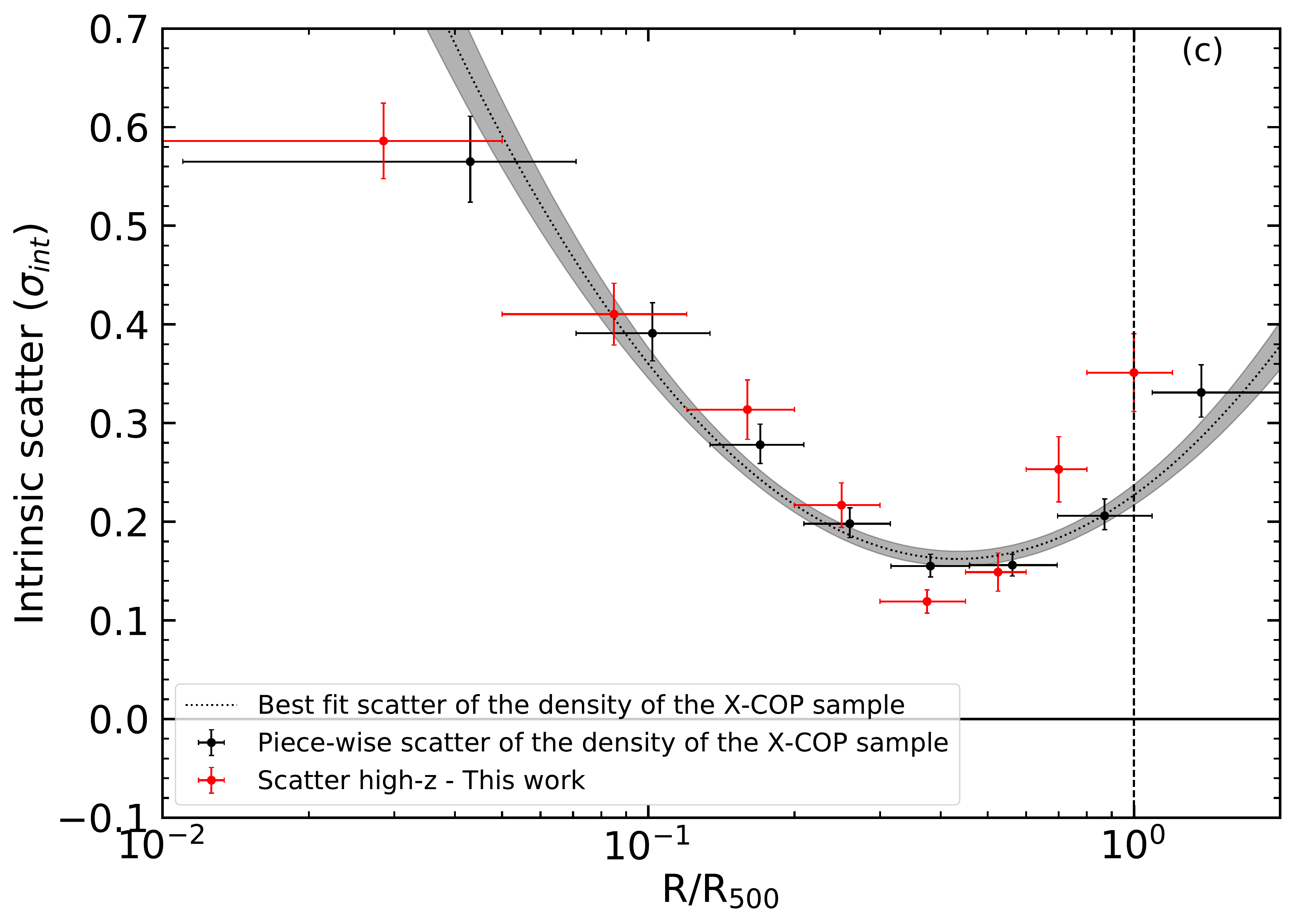}~

    \caption{
\emph{Top Panel}: (a) Density profiles from the sample. The solid black line represents the average density profile of the X-COP clusters. (b) Slope of the density profile of the high-z SPT-selected clusters obtained by the piece-wise power-law fitting technique is shown with red crosses compared to those of the X-COP clusters (shown in the dotted black line and the black crosses).
\emph{Bottom Panel}: (c) Scatter in the density profiles of the high-z SPT-selected clusters (red crosses) and the X-COP clusters (black crosses). 
The vertical dashed line represents the location of $R_{500}$ in all panels. }
    \label{fig:density}
\end{figure*}
\subsection{Total Cluster Mass Reconstruction}
\label{sec:mass}

A common way to measure the total mass \Mfiveoo\ is to use mass proxies calibrated with an X-ray or SZ observable, e.g., $L-M$ or $\xi-M$ scaling relations \citep[e.g.][]{pratt+09,Bocquet+19,bulbul19}. However, to avoid introducing a bias in our results by using the evolution in a specific scaling relation, we directly measure the cluster total mass using X-ray observations. 
The direct measurements based on X-ray data can be obtained from the thermodynamic properties using the assumption of hydrostatic equilibrium and spherical symmetry, i.e.

\begin{equation}
M(<R) = - \frac{R\, k_B\, T}{G\, \mu m_p} \left[ \frac{d \log \rho_g}{d \log R} + \frac{d \log T}{d \log R} \right].
    \label{eq:HEE}
\end{equation}

\noindent where $G$ gravitational constant, $m_{p}$ mass of the proton, and  $\rho_g$ is the gas density. There are several methods that are used in the literature to solve the previous equation \citep[see][for a review]{ettori+13}. Throughout this work, we adopt a ``forward'' modeling approach to obtain a measurement of $M_{500}$, the total cluster mass within $R_{500}$. 
We make use of the best-fitting density and temperature profiles as recovered in Sect.~\ref{sec:imaging} and \ref{sec:spec_xmm} respectively, propagating them through the HE equation to recover the mass profile. 

This method has the advantage of starting from smooth thermodynamic profiles, where the large number of parameters in these functional forms allow us to reproduce the density and temperature profiles over a large radial range. We direct the reader to the Appendix \ref{app:mass} for comparison with literature results, and with other mass reconstruction techniques we have employed to solve Equation~\eqref{eq:HEE}.

\subsection{Density, Temperature, Pressure, and Entropy Profiles}
\label{sec:thermoprop}
   \vspace{2mm}

The deprojected electron density profile $n_e(r)$ (see Sect.~\ref{sec:joint}) obtained from surface brightness analysis is first converted into gas density $\rho(r) = \mu_e m_p n_e(r)$ and then rescaled by the critical density of the Universe $\rho_c = \frac{3 H^2(z)}{8 \pi G}$, where $H(z) = H_0 E(z)$ and $E^2(z) = \Omega_\Lambda + \Omega_m (1+z)^3$. Panel (a) of Figure~\ref{fig:density} shows the gas density profiles of the sample. 
We notice that, in the outskirts, the profiles of the SPT-selected high-z and the Planck selected nearby X-COP clusters are fully consistent with each other, while in the core the SPT-selected high-z profiles are factor of a few smaller.
In the core, the observed scatter (measured as in Eq.~6 in G18) is an order of magnitude both in the SPT-selected high-z and the Planck selected nearby X-COP clusters, due to the cool core/non-cool core states in both samples, i.e. the effect of this dichotomy mostly dominates the scatter near the core. The scatter becomes minimal around 0.4 $R_{500}$, at the same location where X-COP clusters reach their minima in the scatter. Towards $R_{500}$ in the outskirts the scatter increases again in both samples.
The increase in the high redshift sample is faster, reaching the value of about 0.35 at $R_{500}$, while the scatter in the X-COP sample remains at 0.2 at the same radius. A comparison of the scatter is seen in panel (c) of Figure~\ref{fig:density}.

To be able to measure the slope of the density profiles, we perform a piece-wise power-law fitting technique as described in detail in G18. Comparing our sample with the nearby X-COP clusters, we find that in the core, the slope in our sample is flatter compared to the X-COP clusters, while in the outskirts ($>0.3R_{500}$) the mean slopes are consistent with each other (see panel (b) of Figure~\ref{fig:density}). 

\begin{figure*}
    \centering
    \includegraphics[width=0.49\textwidth]{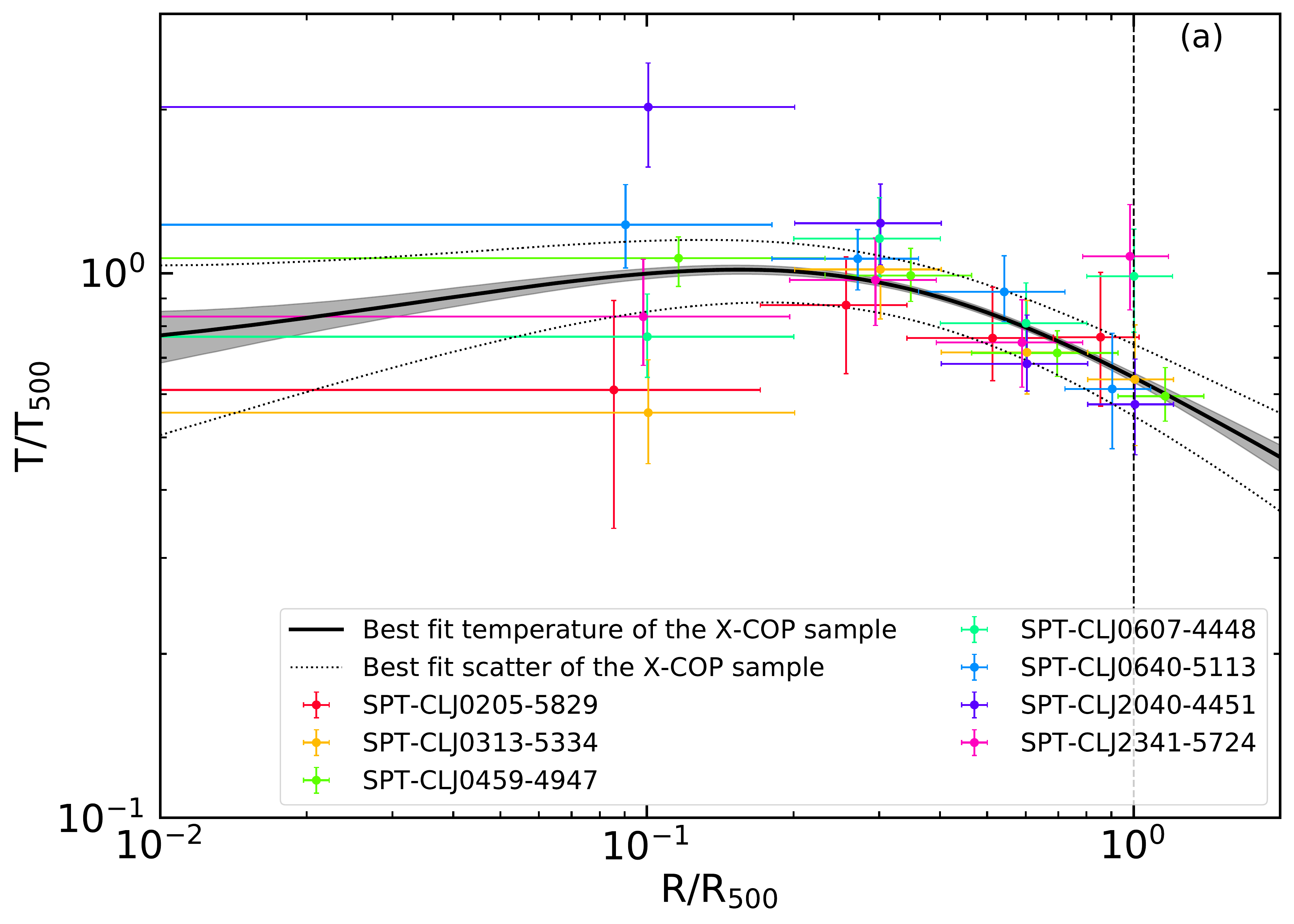}~
    \includegraphics[width=0.49\textwidth]{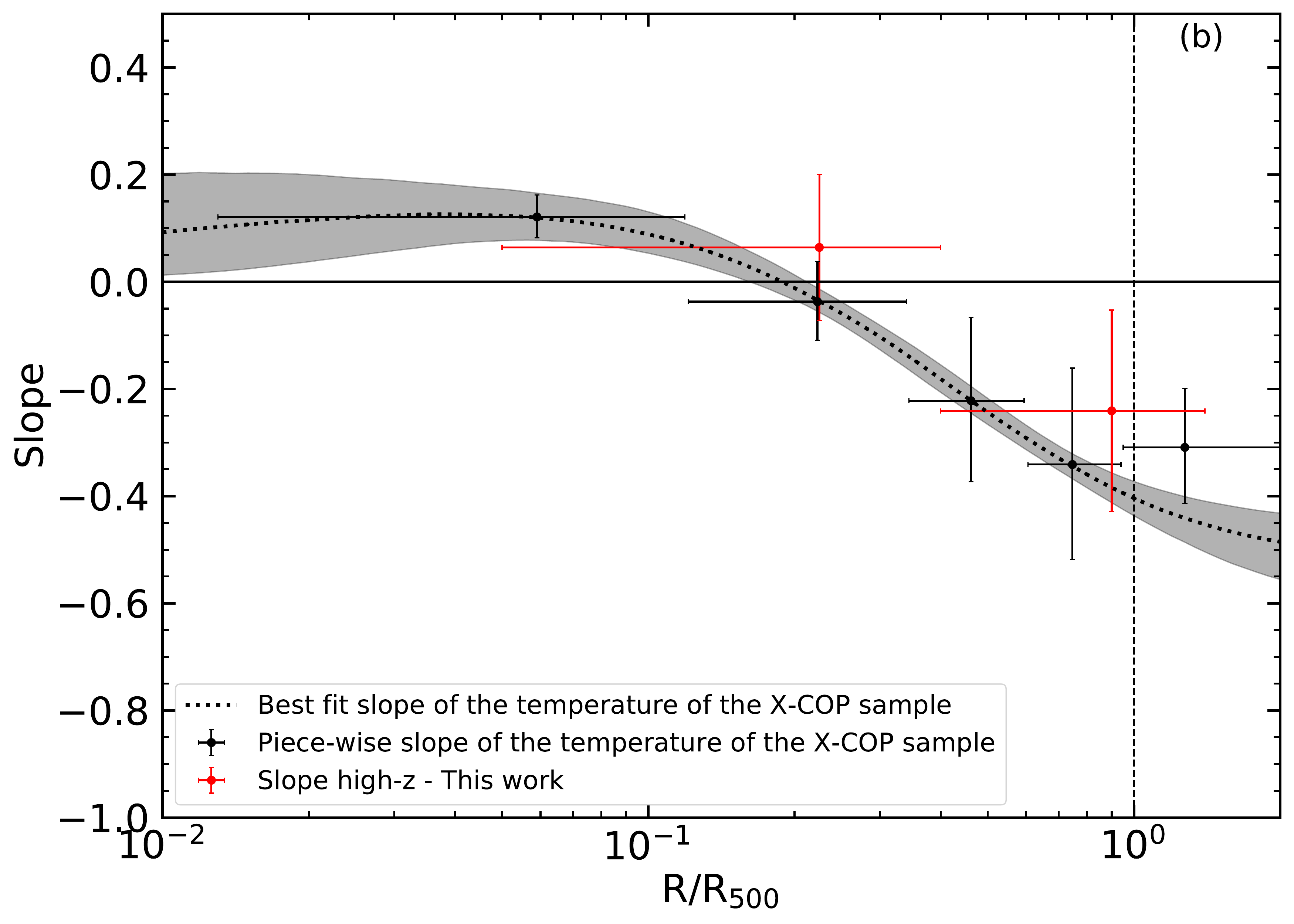}

    \includegraphics[width=0.49\textwidth]{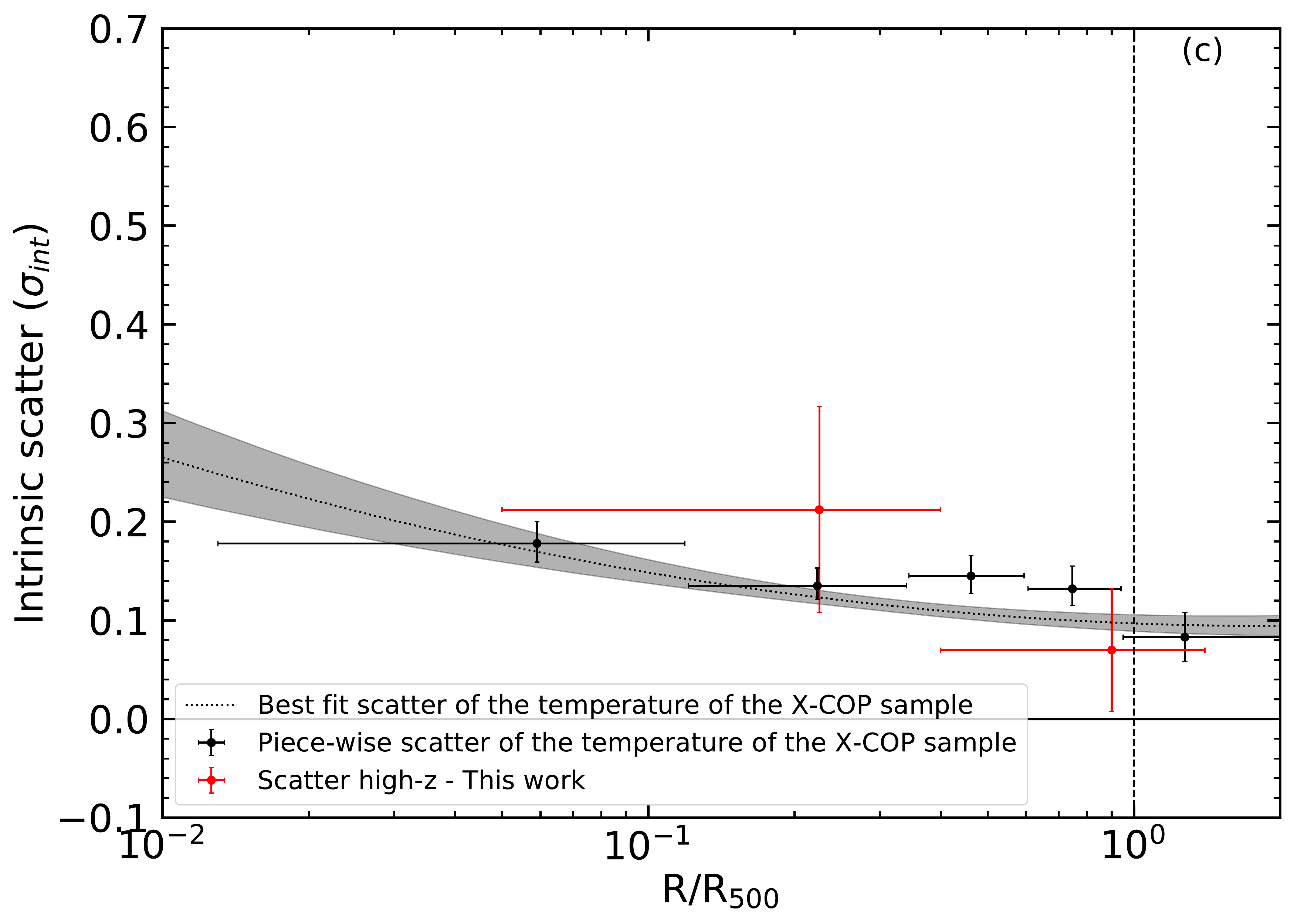}

    \caption{Same as Figure~\ref{fig:density}, but for the temperature profiles. }
    \label{fig:temperature}
       \vspace{2mm}
\end{figure*}

 Next, we study the temperature profiles of the SPT clusters and compare them with the nearby X-COP clusters. For this comparison, the spectroscopic temperature profiles (see Sect.~\ref{sec:spec_xmm} for details) are scaled by the self-similar $T_{500}$, also used in Equation~10 of G18 for the X-COP clusters \citep{voit+05}:

\begin{equation}
T_{500} = 8.85 \text{\ keV} \left(\frac{ M_{500} }{ h_{70}^{-1} 10^{15} M_\odot } \right)^{2/3} E(z)^{2/3} \left( \frac{\mu}{0.6} \right)
\label{eq:T500}
\end{equation}

\noindent where the total mass M$_{500}$ is measured in Sect.~\ref{sec:mass}, and used self-consistently in calculations of R$_{500}$ and $T_{500}$. In panel (a) of Figure~\ref{fig:temperature}, we compare the rescaled temperature profiles of the SPT high-z clusters with the nearby X-COP clusters. We find that the scaled temperature profiles in the two samples are consistent with each other in the entire radial range out to $R_{500}$. The size of the PSF of \xmm\ is comparable to the size of the core of these high-redshift clusters, therefore we cannot resolve well temperatures within $<$150~kpc, or 0.1 $R_{500}$. Performing a piece-wise power-law analysis in two radial bins, we obtain similar slopes and the intrinsic scatter in the temperature profiles when comparing them with the X-COP clusters results.

The pressure profiles are obtained by combining the deprojected density and temperature profiles as $P = n_e T_e$. Pressure profiles can be constrained from both X-ray and SZ observations and used for constraining astrophysical properties and the total mass of clusters out to their Virial radius \citep{bonamente2012, ghirardini18}. 

\begin{figure*}
    \centering
    \includegraphics[width=0.49\textwidth]{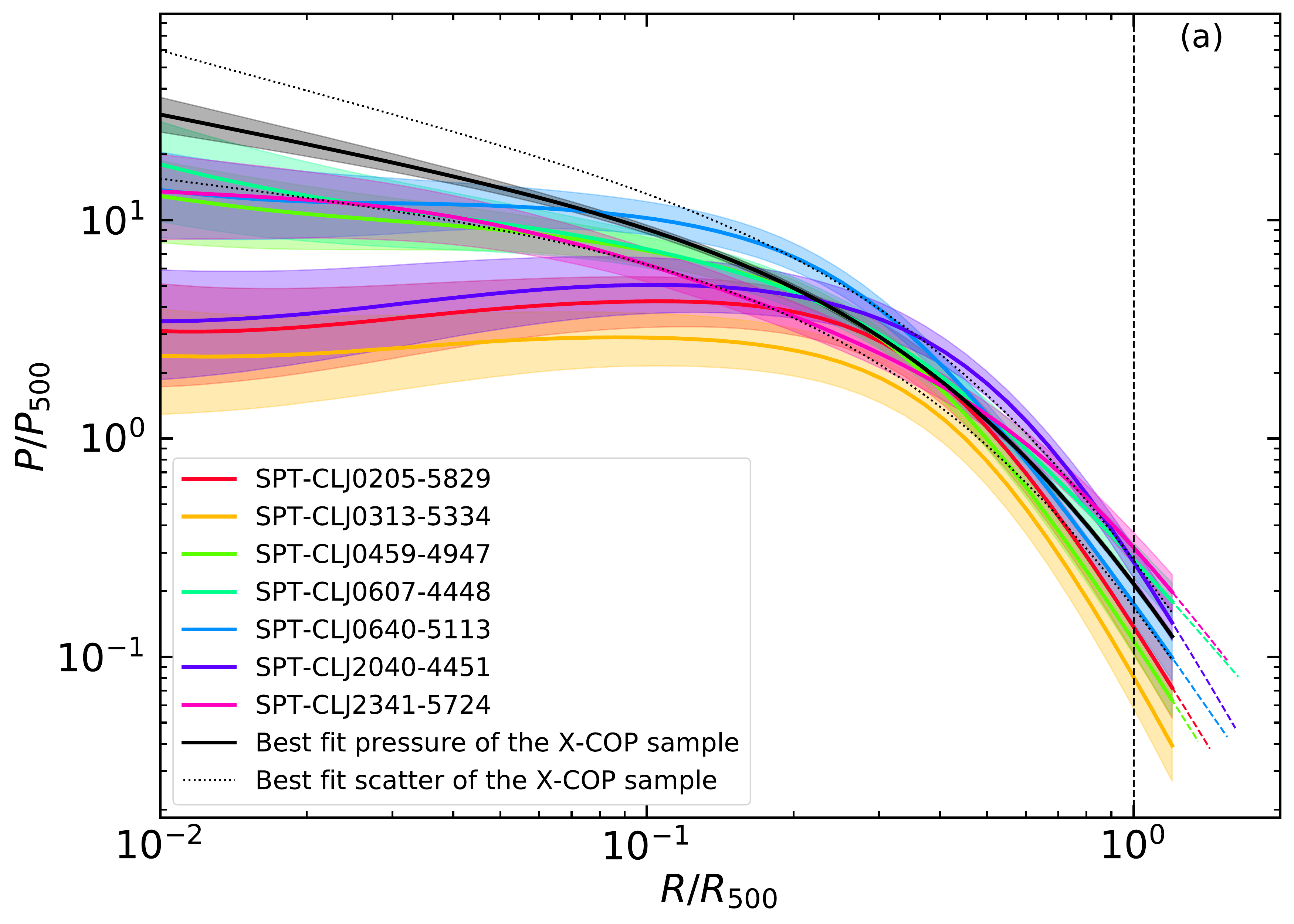}~
    \includegraphics[width=0.49\textwidth]{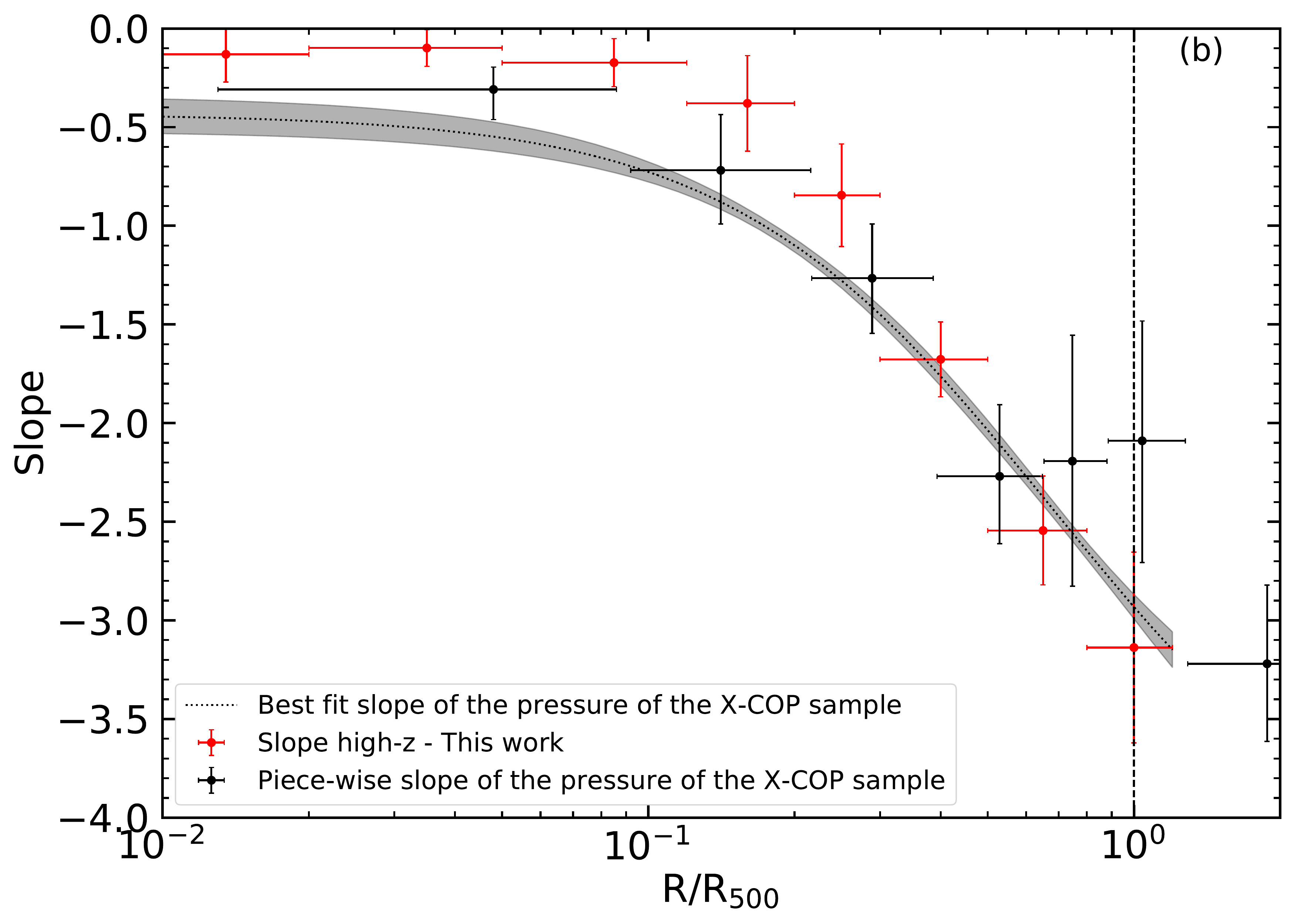}
    \includegraphics[width=0.49\textwidth]{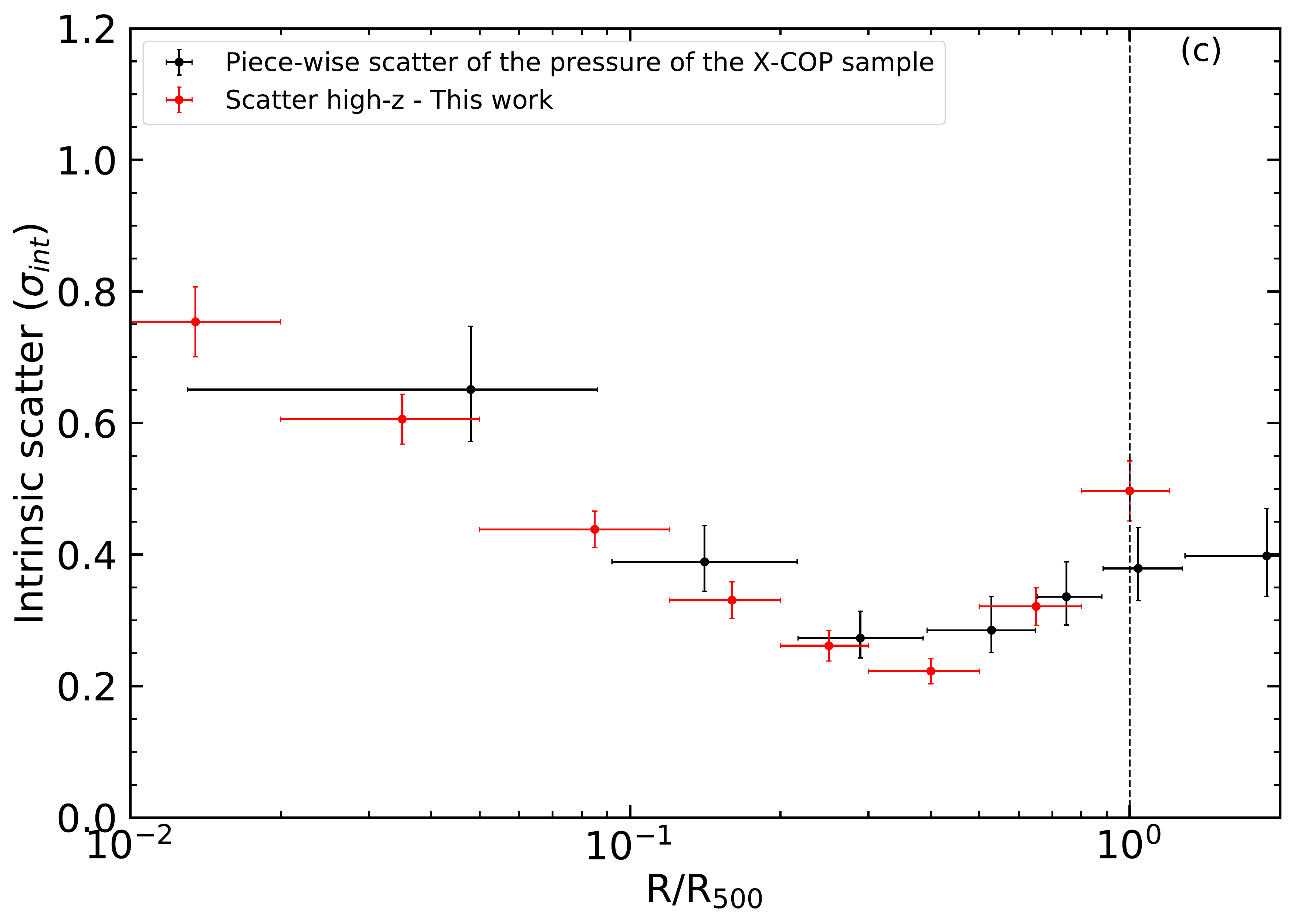}~
    \caption{Same as Figure~\ref{fig:density}, but for the pressure profiles. }
    \label{fig:pressure}
\end{figure*}

We rescaled the pressure using the self-similar pressure $P_{500}$ as described in \citet{nagai+07}:

\begin{multline}
P_{500} = 3.426 \times 10^{-3} \text{\ keV} \text{\ cm}^{-3} \left(\frac{ M_{500} }{ h_{70}^{-1} 10^{15} M_\odot } \right)^{2/3} E(z)^{8/3} \cdot \\ \cdot \left(\frac{f_b}{0.16} \right) \left( \frac{\mu}{0.6} \right) \left( \frac{\mu_e}{1.14} \right)
\label{eq:P500}
\end{multline} 
Panel (a) of Figure~\ref{fig:pressure} shows a comparison of the rescaled pressure profiles of our sample of high-z clusters with the X-COP sample. 
We find that in the core of SPT high-z clusters the rescaled pressure is on average lower and flatter compared with  what is measured in nearby clusters.
In the outskirts, pressure becomes consistent with the finding of low redshift X-COP clusters. The scatter is also fully consistent between high and low redshift clusters in all our radial points except the outermost at $R_{500}$, when at high redshift it is 20\% higher.

\begin{figure*}
    \centering
    \includegraphics[width=0.49\textwidth]{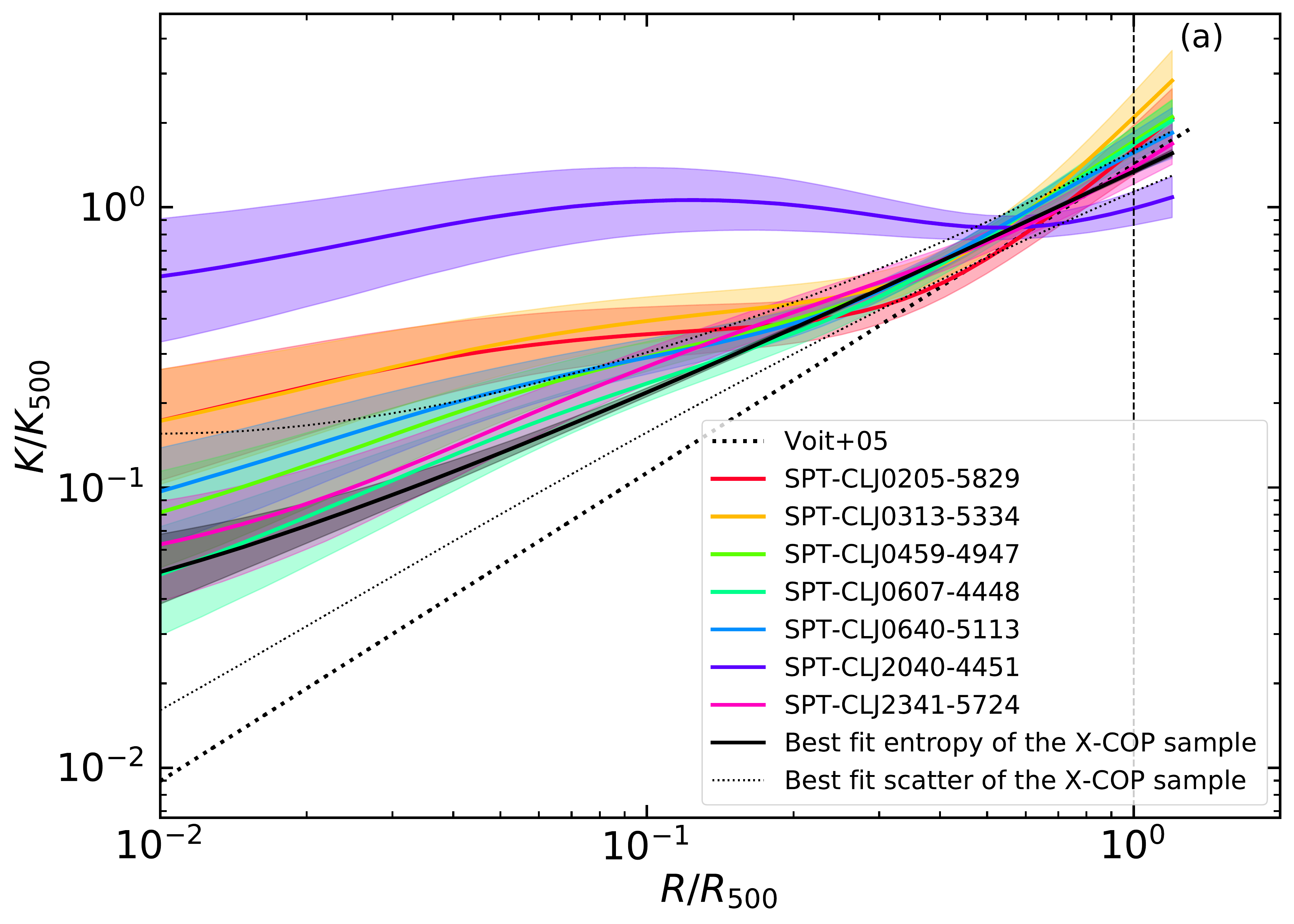}~
    \includegraphics[width=0.49\textwidth]{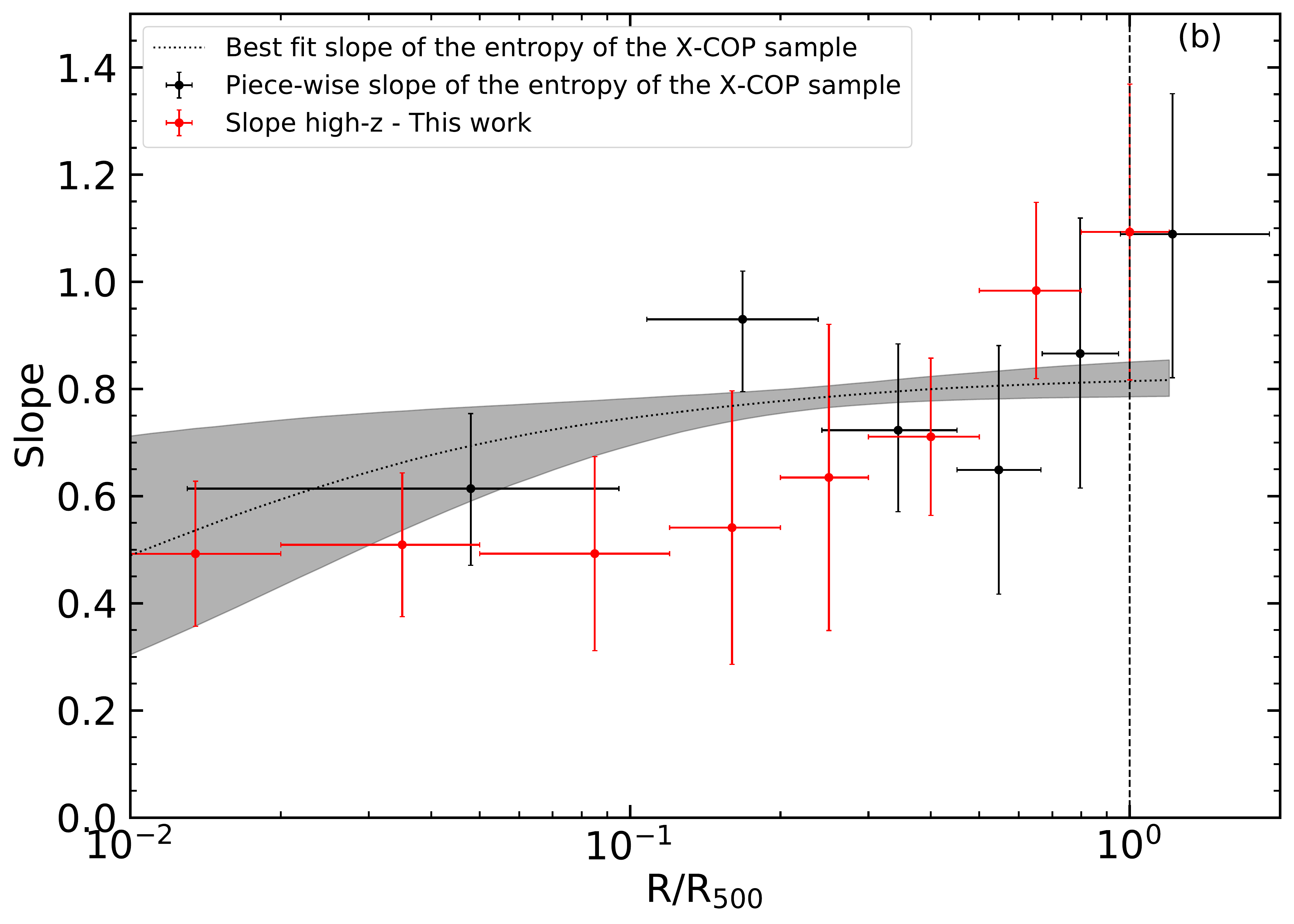}

    \includegraphics[width=0.49\textwidth]{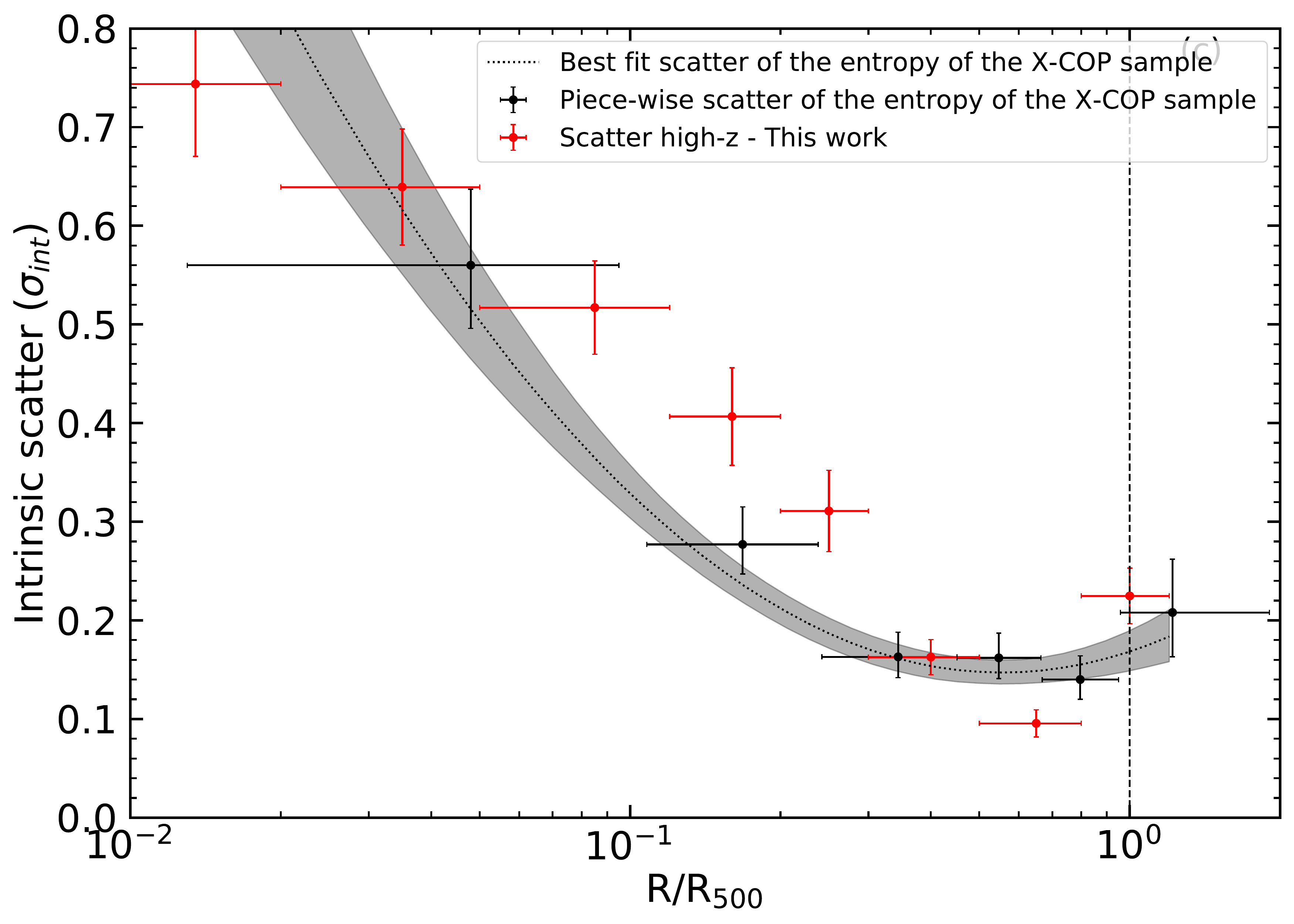}~

    \caption{Same as Figure~\ref{fig:density}, but for the entropy profiles. }
    \label{fig:entropy}
    \vspace{2mm}
\end{figure*}

Another thermodynamic property that could be extracted from X-ray observations is the entropy. Entropy is often used to constrain the clumpiness and self-similarity in cluster outskirts \citep{walker12a, Urban+11, bulbul2016}. The entropy profiles are obtained using the relation, $K = T n_e^{-2/3}$.
Similarly, the entropy is rescaled with the self-similar value $K_{500}$ for comparison \citep[see][]{voit+05}:

\begin{multline}
K_{500} = 1667 \text{\ keV cm}^2 \left(\frac{ M_{500} }{ h_{70}^{-1} 10^{15} M_\odot } \right)^{2/3} E(z)^{-2/3} \cdot \\ \cdot \left(\frac{f_b}{0.16} \right)^{-2/3} \left( \frac{\mu}{0.6} \right) \left( \frac{\mu_e}{1.14} \right)^{2/3}
\label{eq:K500}
\end{multline}
\begin{figure*}
    \includegraphics[width=0.49\textwidth]{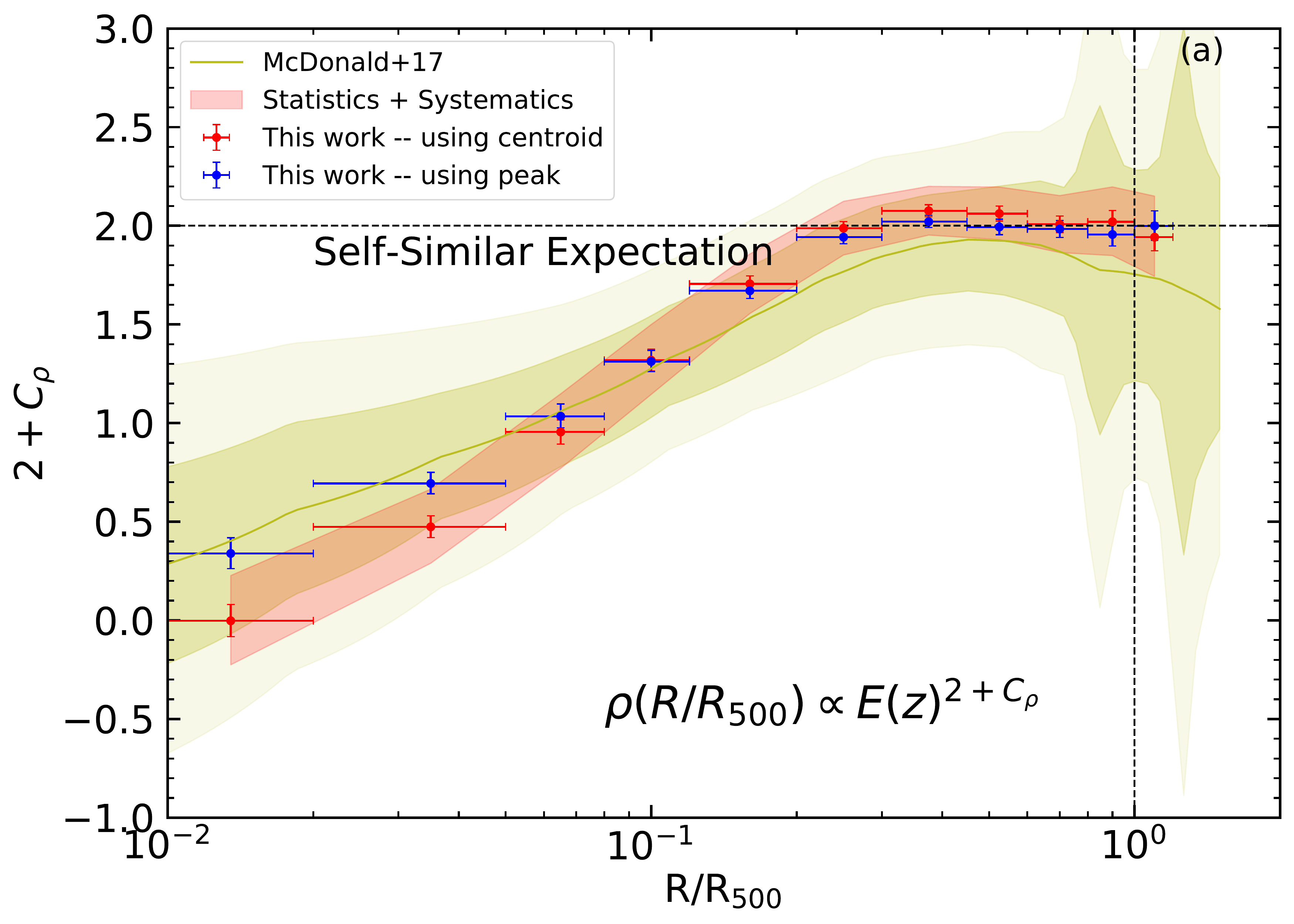}
    \includegraphics[width=0.49\textwidth]{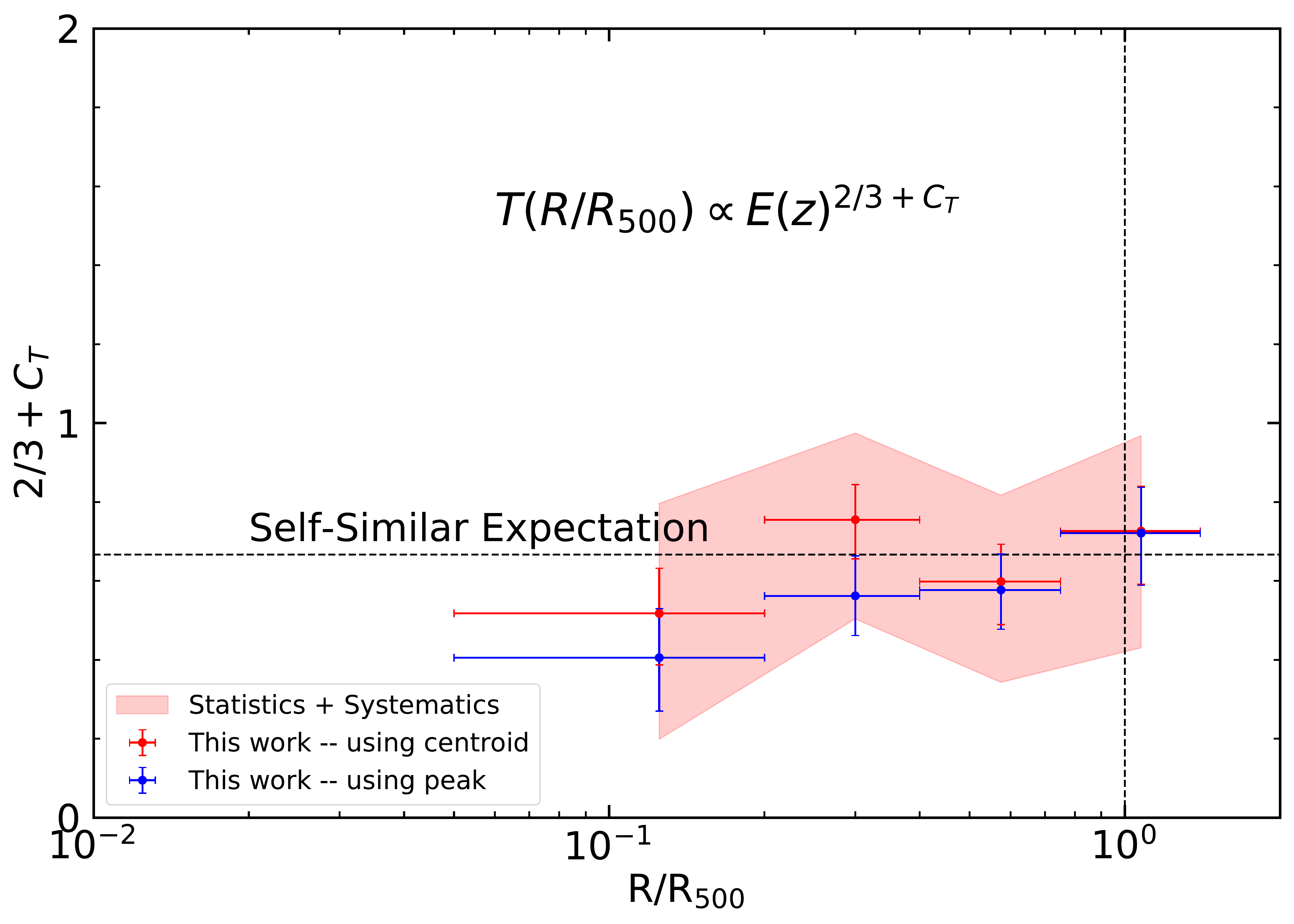}
    \includegraphics[width=0.49\textwidth]{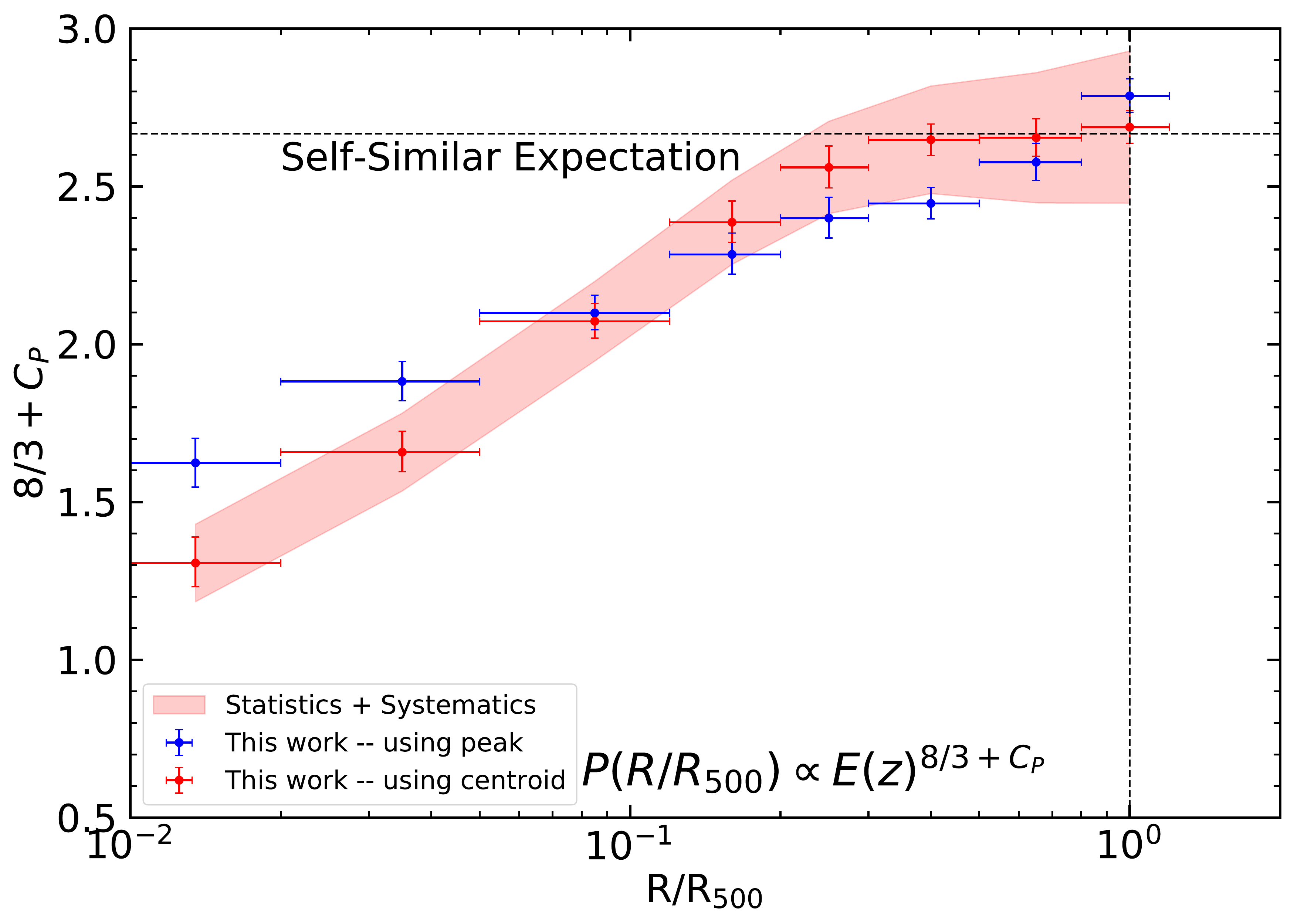}
    \includegraphics[width=0.49\textwidth]{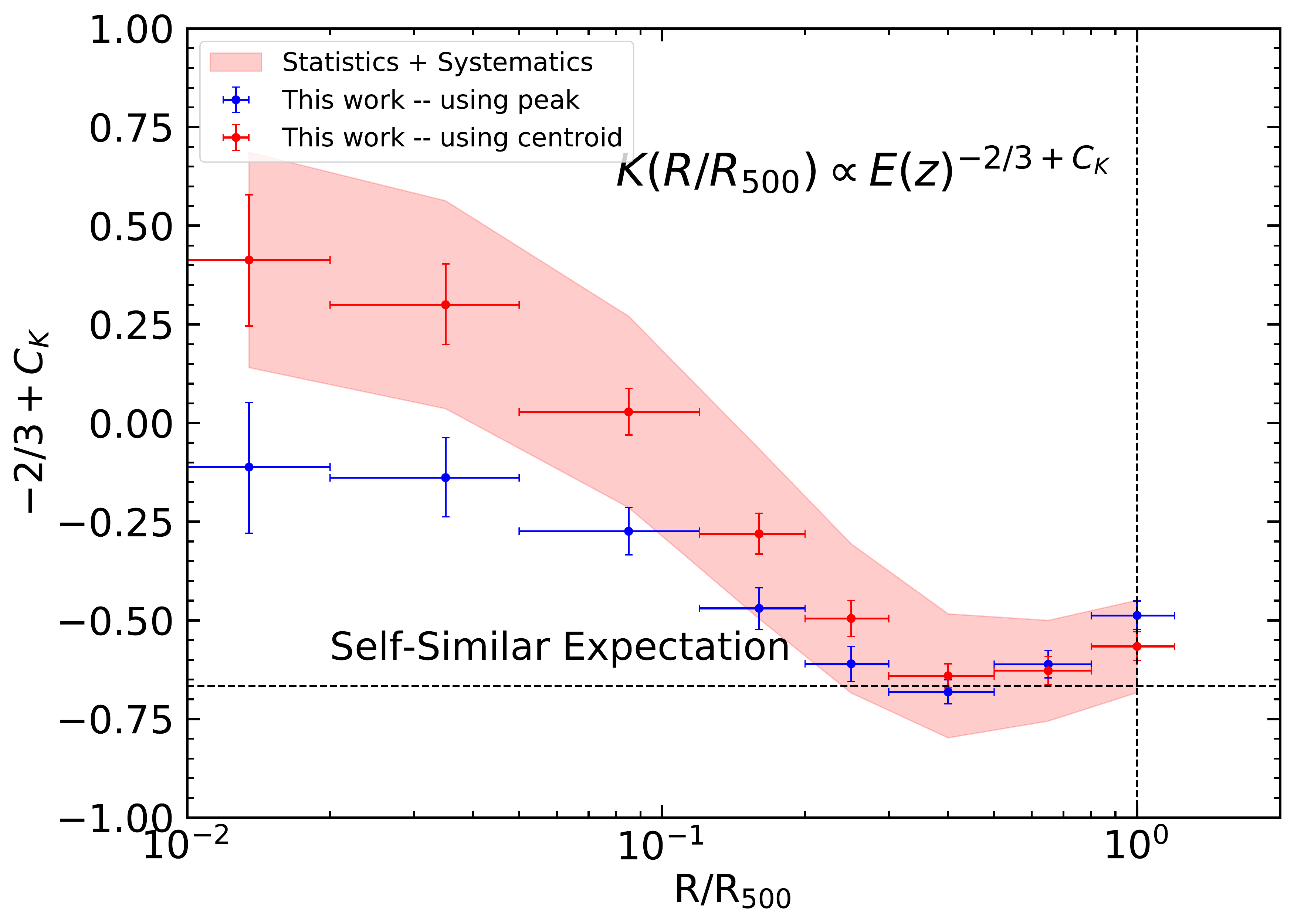}

\caption{\emph{Top Panel}: (a) Evolution in the density profiles as a function of redshift obtained using the centroid (in red) and X-ray peak (in blue). The red shaded region around our data points represents the sum in quadrature of the statistical and systematic uncertainties (see Sect.~\ref{sec:syst} for details). The yellow shaded area represents the same result as found by MD17. Zero values of $2+C_\rho$ indicate no evolution with redshift. The self-similar evolution of $C_\rho = 0$ (corresponding to $\rho \propto E(z)^2$) is represented by an horizontal dashed line. 
The other panels are the same but for temperature (b) with self-similar predicted evolution corresponding to $T \propto E(z)^\frac{2}{3}$, 
for pressure (c) with self-similar predicted evolution corresponding to $P \propto E(z)^\frac{2}{3}$, and entropy (d)  with self-similar predicted evolution corresponding to $K \propto E(z)^{-\frac{2}{3}}$.
Moreover for pressure and entropy, below 0.1 $R_{500}$ the values of the evolution are extrapolated because temperature measurements are not resolved on smaller scales.
The vertical dashed line represents the location of $R_{500}$ in all panels. 
 }
\label{fig:evo}
\vspace{2mm}
\end{figure*}

In Figure~\ref{fig:entropy} we show the entropy profiles of the sample, the slope of the entropy, and the intrinsic scatter. An excess is observed in the entropy compared to self-similarity within (0.3$R_{500}$) near the core. We attribute this excess to non-gravitational processes (e.g., AGN feedback, infalling substructures,  merging activities) in the cores. A similar entropy excess in the core was reported in nearby low redshift clusters \citep{urban14, bulbul2016, ghirardini18_b, walker2019}, however smaller than the entropy excess observed in these high-z clusters. The high-z entropy excess may be due to
the increased incidence of non-gravitational effects, e.g. galaxy and cluster formation, minor mergers at higher redshifts that triggers AGN
activity \citep{birzan17, Hlavacek-Larrondo+12,mcdonald16}. 

The entropy profiles are flat in the cores, steepen and become consistent with the self-similar model beyond $\sim$~0.2$R_{500}$, similarly and fully consistent with the entropy profiles in the outskirts of nearby clusters \citep[see][for a review and references therein]{walker2019}.
 The intrinsic scatter is comparable for both samples.

\subsection{Evolution of thermodynamic Properties with Redshift}

In this section, we investigate the redshift evolution of thermodynamic properties of the ICM and measure the deviation from self-similarity of our sample. Following a similar approach described in MD17, we 
determine the evolution of the density in different radial bins. 
We characterize the evolution of the thermodynamic quantities using the functions given below

\begin{equation}
    \begin{cases}
    \left( \rho \right)_z  = \left( \rho \right)_{z=0} \cdot E(z)^{2+C_\rho}\\
    \left( T \right)_z = \left( T \right)_{z=0} \cdot E(z)^{2/3+C_T}\\ 
    \left( P \right)_z = \left( P \right)_{z=0} \cdot E(z)^{8/3+C_P}\\
    \left( K \right)_z = \left( K \right)_{z=0} \cdot E(z)^{-2/3+C_K}\\
    \end{cases}
    \label{eq:evolution}
\end{equation}
\noindent where $C_{\rho, T, P, K}$ represent the deviation with respect to self-similar values for the evolution \citep{Kaiser+86} of density, temperature, pressure, and entropy. Starting from the density, temperature, pressure, and entropy profiles of the nearby X-COP sample, we infer the expected profiles at the redshifts of the SPT high-z sample assuming a simple deviation from the self-similar evolution, as indicated in Equation~\eqref{eq:evolution}. We, then, compare these profiles with the thermodynamic profiles of the SPT clusters using a log-likelihood $\log \mathcal{L} = -\chi^2/2$ to fit and to determine the best-fit evolution parameters $C_{\rho, T, P, K}$. The best-fit parameters of these fits are given in Table~\ref{tab:evo}.  The uncertainties of the X-COP profiles as well as their measured scatter, and the uncertainties on $R_{500}$ and $Q_{500}$, see Eq. \eqref{eq:T500}, \eqref{eq:P500}, and \eqref{eq:K500}, are propagated through the fit. We also include the systematic uncertainties related to our observations in our measurements (see Section~\ref{sec:syst} for details). The systematic and statistical uncertainties are summed in quadrature to estimate the total uncertainty.

 We note that the cluster centers are determined from the \chandra\ data and initial results are obtained using the centroid of the large scale ICM emission in this analysis. The choice of cluster center plays an important role especially when measuring the evolution of the central cluster properties \citep{Sanders+18}. To investigate the effect of the center location, we determine the evolution in density using both the centroid and the X-ray peak. The evolution in density, temperature, pressure, and entropy profiles obtained using both the centroids (red) and X-ray peaks (green) are shown in Figure~\ref{fig:evo}.

We find no evolution in the density at small radii ($\sim0.3R_{500}$) using large scale centroids. The self-similar evolution in cluster cores is excluded significantly by $\sim$11$\sigma$. Using of the X-ray peaks instead of the centroids, the evolution values move slightly towards self-similarity in the core. However, the departure from self-similarity is still significant at a $\sim$9$\sigma$ confidence level. We also note that the intrinsic scatter in density of high redshift clusters, shown in Figure \ref{fig:density} panel (c), at small radii, is similar to that of the low X-COP redshift clusters.
Non-gravitational phenomena, e.g., Active Galactic Nuclei feedback, sloshing, dominate the physical processes in cluster cores and can affect the evolution in the core of the clusters. Thus, our finding may point that non gravitational physical processes that regulate cluster cores were already in place since a redshift of 1.8 (with look back time of $\sim$10 Gyr). Our results in cluster cores are consistent with the results in MD17 at the 1$\sigma$ confidence level. However the uncertainties in the measurements are reduced at least by a factor of two. \citet{Sanders+18} suggests that use of the X-ray peak instead of centroids could mimic a potential evolution in cluster cores and bias the results in evolution studies. Changing the cluster center does not significantly affect our results. 

At large radii, the evolution in density becomes consistent with the self-similar expectation around 0.1$R_{500}$ and remains fully consistent out to $R_{500}$. MD17 also reported the best-fit evolution consistent with the self-similarity, however  due to the limited statistics the authors could not rule out no evolution scenario. We tightly constrain self-similarity in cluster outskirts and confirm it with a higher significant level.  We also observe an increase of the scatter on cluster density profiles (see Figure~\ref{fig:density}) in cluster outskirts. This may imply that although the cluster-to-cluster variance in the outskirts increases because of larger mass accretion rates and merger activity at higher redshifts \citep{Wechsler02,Fakhouri09,Tillson11,avestruz16}, the average evolution in density, however, remains consistent with this self-similarity.

In the case of temperature profiles, we don't measure any significant deviation from self-similarity from the cluster cores out to $R_{500}$. The intrinsic scatter is also consistent with that of the low redshift clusters within uncertainties (see Figure~\ref{fig:temperature}). Therefore, the cluster temperature evolution and the cluster to cluster variance do not seem to change from low to high redshifts. The change of the cluster centre makes a very small difference and does not change the results. This is not surprising considering the large uncertainties on temperature measurements.

We observe a mild evolution in pressure profiles in cluster cores. Similarly, the evolution becomes consistent with self-similarity at $\sim$0.1~$R_{500}$ and larger scales. At small scales, pressure profiles deviate significantly from self-similar evolution at a 9$\sigma$ level. Using the X-ray peak as the cluster center does not change the results significantly.  

Interestingly, in the core, a mildly significant ($\sim3\sigma$ confidence) evolution is observed for the entropy, if we use the centroid as the cluster center. Changing the cluster center to the X-ray peak reduces significantly the observed evolution. In the outskirts the evolution becomes fully consistent with self-similarity, regardless of the center used.

It's important to remind the reader that the evolution measured in cluster cores for pressure and entropy is quite dependent upon the adopted cluster temperature model, because the first temperature bin is very large, encapsulating the entire cluster core, $<0.1 R_{500}$.

\subsection{Polytropic index}
\label{sec:poly}
\begin{figure}[t]
\includegraphics[width=0.5\textwidth]{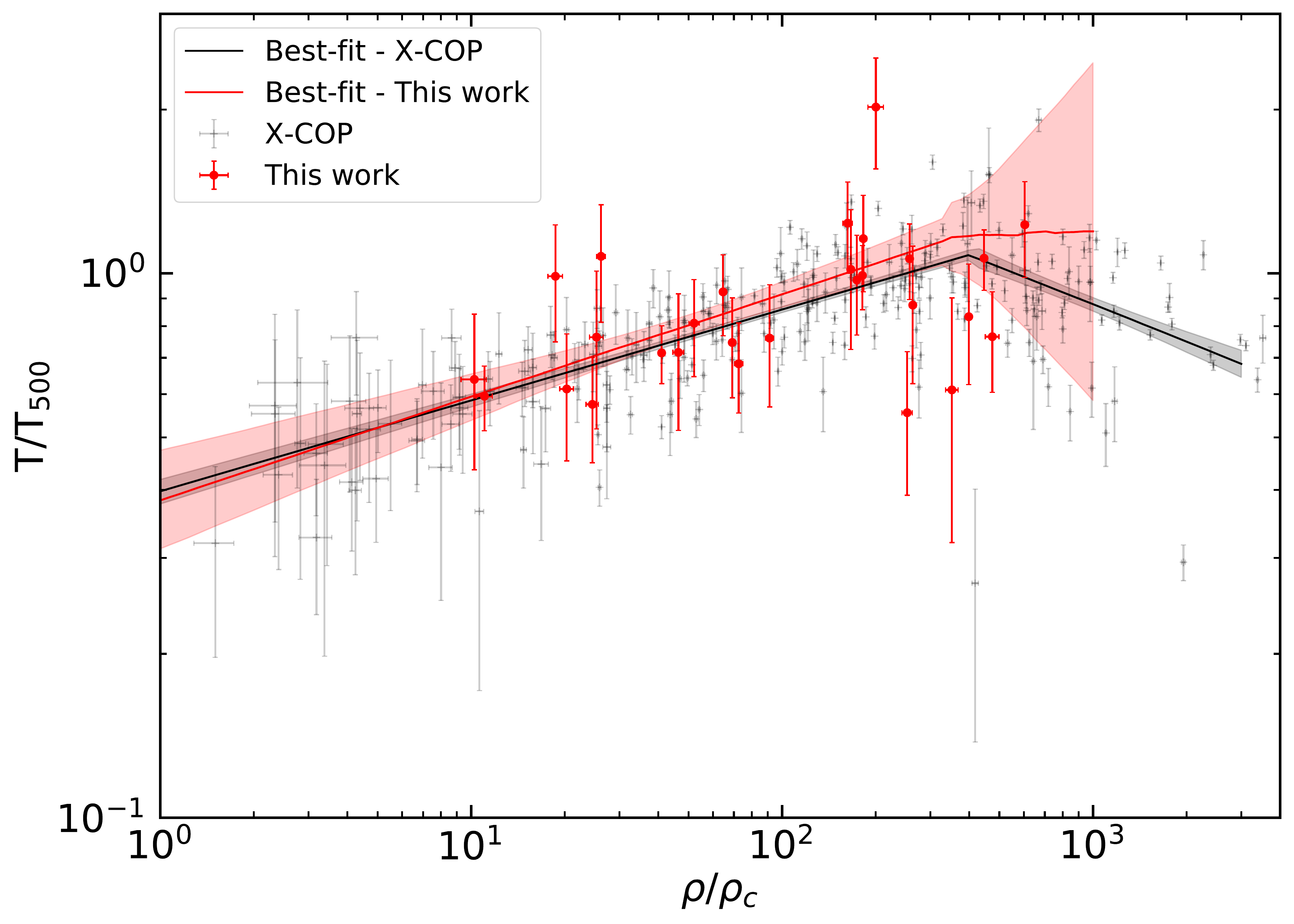}
\caption{Rescaled temperature against rescaled density in high redshift cluster sample (in red) and in low redshift clusters \citep[in black,][]{Ghirardini+19}. The lines represent the best-fit broken power-law to the data. In particular, we find that the slope in the relation is consistent in low and high redshift clusters in the low density regime, i.e. in cluster outskirts, supporting again the self-similar model of cluster evolution.}
\label{fig:poly}
\end{figure}
The global structure of the ICM can be effectively described by a polytropic equation of state $P_e = K \rho^\Gamma$, where the polytropic index is indicative of stratification of the ICM \citep{shaw+10}.
Both simulations \citep{komatsu+01,Ostriker+05,Ascasibar+06,capelo+12} and observations \citep{Markevitch+98,Sanderson+03,bulbul+10, eckert15,Ghirardini+19} find that the stratific	ation of the ICM, especially in the outer part, is well represented by a polytropic equation of state with $\Gamma$ in the range 1.1-1.3. In particular, the X-COP collaboration reports that the value of $\Gamma$ in cluster outskirts, where $\rho / \rho_c \lesssim 400$, is $\Gamma = 1.17 \pm 0.01$ at redshifts below 0.1. However, the polytropic index in the high redshift Universe, or its evolution, has never been investigated. We find that the polytropic index, see also Figure~\ref{fig:poly}, is $1.19 \pm 0.05$ in low density regions, i.e. in the cluster outskirts. This value is fully consistent with the value measured at low redshifts in the X-COP clusters, indicating that there is no significant evolution with redshift, i.e. the ICM stratification is the same at low and high redshift. In high density regions, i.e. in the core, we are not able to resolve the index due to the large size of the \xmm\ PSF.

\section{Systematics}
\label{sec:syst}
\begin{table*}
\centering
\begin{tabular}{ c | c c | c | c c | c c }
\hline
\hline
Thermodynamic & \multicolumn{2}{c|}{HE} & Clumping & \multicolumn{2}{c|}{Progenitor} & \multicolumn{2}{c}{Calibration}\\
At & 0.01 $R_{500}$ & $R_{500}$ & - & 0.01 $R_{500}$ & $R_{500}$ & 0.01 $R_{500}$ & $R_{500}$\\
\hline
Density & 0.02 & 0.10 & 0.058 & 0.14 & 0 & 0.03 & 0.09\\
Temperature & 0.08 & 0.10 & 0.061 & 0.12 & 0 & 0.07 & 0.04\\
Pressure & 0.09 & 0.19 & 0.061 & \multicolumn{2}{c|}{-} & 0.04 & 0.13\\
Entropy & 0.10 & 0.11 & 0.025 & 0.25 & 0 & 0.03 & 0.01\\
\hline
\hline
\end{tabular}
\caption{
Range in which each systematic bias discussed in Sect.~\ref{sec:syst} affects each thermodynamic quantity in the core and at  $R_{500}$. The thermodynamic biases are, from left to right: 1) hydrostatic bias caused by how the profiles are rescaled, 2) clumping bias caused by the presence of unresolved clumps, 3) bias caused by the fact that SPT high-z clusters are not exactly the progenitors of the redshift 0 clusters we are comparing them with, and 4) calibration bias caused by difference between \chandra\ and \xmm temperatures. 
}
\label{tab:sys_rad}
\end{table*}

In this section, we examine several systematic uncertainties that affect our results on the evolution of the thermodynamic properties of clusters, evaluating their magnitudes. The variation of the thermodynamic property $\mathrm{Q}$ can be converted into the systematic uncertainty on the evolution following the formula below.

\begin{equation}
  \mathrm{Q} + \Delta \mathrm{Q} = E(\hat{z})^{k} \cdot \mathrm{Q}
    \label{eq:Qvar}
\end{equation}

\noindent where $\hat{z}$ is the average redshift of our sample, and $k$ is the systematic uncertainty on the evolution of each thermodynamic property Q. Solving this equation for the systematic uncertainty $k$ gives the following equation:

\begin{equation}
     k = \frac {\log \left( 1 + \frac{\Delta \mathrm{Q} }{\mathrm{Q} } \right)}{ E(\hat{z})}.
     \label{eq:sys_evo}
\end{equation}
We consider the systematic uncertainties related to hydrostatic mass bias, clumping factor, cluster progenitors, and calibration differences between \xmm\ and \chandra\ below.
In Table~\ref{tab:sys_rad} we show the amplitude of the mass bias on each thermodynamic quantity in the core and at $R_{500}$.

\subsection{Mass bias}

The evolution measured on a thermodynamic property is affected by a potential bias due to the assumption of hydrostatic equilibrium. If the hydrostatic masses we use in this work are biased by a factor of $(1+b)$, this bias translates to a bias in the fiducial radius that can be written as;
\begin{equation}
    \left( \frac{R}{R_{500}} \right)_z = \left( \frac{R}{R_{500}} \right)_{z=0} \times (1+b)^{-\frac{1}{3}}.
    \label{eq:biasR}
\end{equation}

\noindent And also it translates into an uncertainty on a rescaled thermodynamic property  Q as;
\begin{equation}
    \left( \frac{\mathrm{Q}}{\mathrm{Q_{500}}} \right)_z = \left( \frac{\mathrm{Q}}{\mathrm{Q}_{500}} \right)_{z=0} \times (1+b)^{-\frac{2}{3}} 
    \label{eq:biasQ}
\end{equation}
\noindent where $Q=T,P,K$

A bias, which is introduced by mass, can mimic evolution on these thermodynamic properties. Unfortunately, evolution measurements and mass bias are highly degenerate, thus it is not possible to fit Equation~\eqref{eq:evolution} and Equation~\eqref{eq:biasR}-\eqref{eq:biasQ} simultaneously to constrain   the mass bias and the evolution with redshift.
Given that the low redshift X-COP sample and high redshift SPT sample have very similar selection criteria, i.e. a selection based on SZ S/N, and the masses are obtained in a similar way, i.e. assuming HE, the mass bias is expected to affect the results from the two samples by the same amplitude and direction.

An estimate of the mass bias can be obtained by measuring the average ratio between several mass measurements. In Sect.~\ref{sec:mass} we have described our reference method of solving the HE equation to measure $M_{500}$, and, in Appendix~\ref{app:mass}, we compare this measure with other techniques, and other masses in the literature obtained from scaling relations. Figure~\ref{fig:mass} shows the cluster masses obtained through these methods. 
To estimate this bias we measure the average ratio between the measured masses and the mass obtained using the method described in Sect.~\ref{sec:mass}. Since the error bars are not homogeneous, we apply a bootstrap method through all the mass measurements for each cluster. In practice, we measure the mass bias $10^6$ times where each time a new distribution of masses is drawn from the masses shown in Figure~\ref{fig:mass}.
This method yields a mass bias of $1+b = 1.12 \pm 0.01$. The result implies that high redshift clusters have 12\% higher mass bias compared to the nearby clusters. Given that clusters at high redshifts are still forming and not yet thermalized, an increase in the non-thermal pressure support due to gas motions in their outskirts and elevated AGN activity in their cores resulting in an increase in mass bias with redshift is expected. 

Using the mass bias obtained above, we then estimate the corresponding systematic bias in the evolution. This bias affects both \emph{x} and \emph{y} axes, except for density where the rescaling on the \emph{y}-axis is independent of mass. The bias is translated into

\begin{equation}
    \Delta Q = \Delta R \cdot \frac{dQ}{dR}  = (\Delta M)^{1/3} \cdot \frac{dQ}{dR} 
\end{equation}

on the \emph{x}-axis, and 
\begin{equation}
    \Delta Q = (\Delta M)^{2/3} 
\end{equation}
on the \emph{y}-axis, then by summing up in quadrature these two values and applying Equation~\eqref{eq:Qvar} and Equation~\eqref{eq:sys_evo} we measure the systematic uncertainty on the evolution of the thermodynamic quantities caused by the mass bias.

\subsection{Clumping factor}
Unresolved clumps in our observations can lead to higher local densities measured and can bias the observed thermodynamic quantities. In G18 the authors correct the density for the presence of these clumps by both removing the extended sources  contaminating the FoV, and by also computing the median of surface brightness distribution in each annulus, which has been shown to be unbiased by the presence of high density unresolved substructures \citep{zhuravleva13,roncarelli+13}. In particular, to compute the median, a Voronoi tessellation algorithm needs to be performed \citep{voronoi} to produce cells containing surface brightness elements. In this work, we eliminate the detected  point and  extended sources from our analysis. Due to low counts observed and the small extension of the clusters on the sky, the cells produced via the Voronoi tessellation algorithm would be very few and highly correlated with each other. Therefore, it's not possible to compute the median of the surface brightness distribution in the same way as applied to the X-COP sample. Instead, we estimate this bias by adopting the upper limit of 10\% within $R_{500}$ measured in a sample of ROSAT clusters in \citet{eckert15}. We find that the density profiles are biased by a systematic uncertainty of $\Delta \rho / \rho = 0.10$. This translates into a systematic on the density measurements of $\sim$0.06 (see Equation~\ref{eq:sys_evo}).

For the other thermodynamic properties, we combine the effect aforementioned with the bias of 5\% in the pressure arising by the presence of clumps \citep[as measured in simulations by ][ where the 5\% refers to the upper limit within $R_{500}$]{khedekar13}. This translates into a bias of 5\% on the temperature, consistent with the predicted theoretical bias by \citet{avestruz16}, and -2\% bias on the entropy.

\subsection{Progenitors}
It is possible that these SPT-selected high redshift clusters are not the progenitors of the low redshift clusters in X-COP. In fact, the predicted mass of the SPT clusters is expected to be greater than $10^{15} M_\odot$ at redshift zero when the mass growth curve is taken into account \citep[][]{Fakhouri10}. Therefore, the SPT-selected clusters are more massive than the X-COP clusters \citep{ettori+18}, where the reported masses are less than $ 10^{15} M_{\odot}$. We treat the effect due to the fact that the X-COP clusters could be evolved from a different population of clusters than the SPT clusters as a systematic uncertainty. 

To estimate this bias we assume that the gas density follows the dark matter density as a first approximation. We then use a concentration-mass-redshift relation in \citet{amodeo+16} to calculate the relative change in the density from a cluster with mass of $15\cdot10^{14}M_\odot$, i.e. the expected mass of SPT clusters at redshift of 0 \citet{Fakhouri10}, to a mass of  $7\cdot10^{14}M_\odot$, i.e. the average mass of X-COP clusters. Assuming that pressure follows the Universal pressure profile, we estimate the thermodynamic quantities. These values are then propagated as systematic errors as shown in Equation~\eqref{eq:sys_evo}. The results in the core and in the outskirts are given in 
Table~\ref{tab:sys_rad}. We note that the self-similar model predicts an evolution which is independent of mass. Therefore, once the thermodynamic quantities are rescaled with their self-similar value, the fact that they are too massive to be the progenitors of the X-COP clusters is of minor importance, especially at large radii.

\subsection{Calibration Difference between \chandra\ and \xmm}

Calibration differences between \chandra\ and \xmm\ are described in the literature. Temperature measurements can be biased up to 40\% depending on the energy band used and cluster temperature \citep[e.g.][ and references therein]{schellenberger+15}. On the other hand, density measurements by \chandra\ and \xmm\ are fully consistent within the uncertainties \citep[see ][ and also Sect.~\ref{sec:joint}]{bartalucci+17_general}.

To quantify the bias due to calibration differences, we extract both \xmm\ and \chandra\ spectra of the region within $R_{500}$, and fit the spectra using a single temperature thermal apec model. We note that 
in the case of the SPT high redshift clusters, it is not possible to measure temperature profiles using \chandra\ observations in several radial bins due to the limited statistics. A comparison of measured single temperatures is shown in Figure~\ref{fig:T_cxo_xmm}. We find the temperature measurements are consistent with each other within statistical uncertainties. However, we note that  the uncertainties on the \chandra\ measurements are large because of limited statistics.

To estimate the systematic uncertainty on each thermodynamic quantity $Q$ caused by this discrepancy in the temperature is not trivial. The increase of the temperature would lead to an increase in the total mass by the same amount, if the  slope of temperature profile does not change. 
\cite{schellenberger+15} report that temperature measurements based on \chandra\ data are, on average, 17\% higher temperatures than those derived from \xmm\ for the average mass of the clusters in our SPT sample. 
 Thus, a systematic of 17\% on the temperature becomes a 17\% systematic on the mass, thus a 5.7\% bias on $R_{500}$ (a third considering the propagation of uncertainty), and 11.3\%  bias on $Q_{500}$ (two thirds considering that all self-similar quantities depend on mass with power of 2/3).  Thus the variation on each thermodynamic quantity is

\begin{equation}
    \frac{\Delta Q/Q_{500}}{Q/Q_{500}} = \underbrace{17\%}_{\textrm{ on Q}}-\underbrace{11.3\%}_{\textrm{ on Q}_{500}}-\underbrace{5.7\% \cdot \frac{dQ}{dR}}_{\textrm{ on R}_{500}}  .
    \label{eq:evo_Tbias}
\end{equation}
We point out that, for the last two terms, the variation on the rescaled thermodynamic quantity from the radial and the $Q_{500}$ rescaling is in the opposite direction with respect to the systematic bias on the quantity $Q$.
Thus by computing the slope at each radius we get the relative rescaled thermodynamic variation at each radii, and finally using Equation~\eqref{eq:sys_evo} we obtain the systematic bias affecting the evolution of each thermodynamic quantity as given Table~\ref{tab:sys_rad}.

\section{Conclusions}\label{sec:conc}

In this paper we have studied the thermodynamic profiles for the 7 most massive clusters at redshift above 1.2 in the SPT-SZ survey. These clusters were observed by both \chandra\ and \xmm\ for a total clean exposure time of about 2 Ms. We combine the data from these two telescopes to recover density, temperature, pressure, and entropy profiles and examine their evolution with redshift from cluster cores to outskirts. Furthermore, we measure the temperature profiles of a complete set of SPT-selected high redshift clusters for the first time allowing us to reconstruct the total cluster masses under the assumption of hydrostatic equilibrium. Our results include the systematic uncertainties that are  extensively studied in Section~\ref{sec:syst}.

Deep \xmm\ observations of the SPT-selected clusters have sufficient statistics to determine the redshifts from the X-ray data alone. The Fe-K line at 6.7 keV (rest frame) is clearly detected in the spectrum of each cluster in the sample. The centroids of these emission lines are used to measure the redshifts. We show that the redshifts obtained from the X-ray data of the SPT high-z clusters are consistent with the previously reported redshifts obtained through optical photometry and spectroscopy \citep{Bayliss2040, bleem15, khullar19}.

Combination of \chandra's high spatial resolution and \xmm's large FoV and effective area is the most powerful way to measure thermodynamic profiles of clusters at high redshifts, $z > 1.2$ from their cores (0.01 $R_{500}$) to the outskirts ($R_{500}$). Accurate measurements of temperature profiles enable a few key measurements for these clusters, e.g. total mass, pressure, and entropy. We are able to measure their total mass though the hydrostatic equilibrium assumption with relatively small uncertainities (10 - 20 \%) at these redshifts. The hydrostatic masses are generally in good agreement with reported masses in the literature obtained through SZ signal-to-noise and scaling relations \citep{bleem15}. 

We further measure the density, temperature, pressure, and entropy profiles of the high-z SPT cluster sample and compare their distributions with the previously reported thermodynamic properties of the nearby X-COP clusters. The scatter of all the thermodynamic quantities are similar in low and high redshift clusters in small spatial scales near the cores. At large radii, the scatter increases more steeply in the sample of  high redshift clusters. This may be due to an increased frequency of merger events and higher mass accretion rate at high redshifts \citep{Wechsler02, Fakhouri09, Tillson11}.

The average profiles of density, temperature, pressure, and entropy of high-z clusters are self-similar and consistent with those of the X-COP clusters at large spatial scales near $R_{500}$. Temperature profiles of high redshift clusters are self-similar at all radii. We also report that the polytropic index (1.19~$\pm$~0.05) is fully consistent with that measured at low redshift clusters indicating that there is no significant evolution with redshift. The high observed scatter in density, pressure, and entropy in cluster cores is due to cool-core/non-cool core dichotomy in these cluster samples. The scatter in the thermodynamic properties becomes minimal at $0.4R_{500}$ and increases towards $R_{500}$ in the SPT-selected high-z clusters. The increase in the merger frequency and mass accretion rate in high-z clusters may contribute to high scatter in cluster outskirts \citep{Wechsler02,Fakhouri09,Tillson11}. 

We are also able to constrain the evolution in density and temperatures profiles of the cluster. Measurements of the evolution in entropy, pressure profiles with redshift also becomes available owing to precise temperature constraints for the first time. We find that the evolution in thermodynamic profiles deviates significantly from the self-similar evolution in cluster cores, while in the outskirts, the profiles are on average are in agreement with the prediction from the self-similar model. We find no evidence for evolution in the density in cluster cores, confirming the results in MD17. We point out that the analysis performed in this paper, and the one in MD17 are different in how self-similarity has been probed. We have considered two high-S/N cluster samples at low and high redshift, while in MD17 the authors have considered $\sim$100 low-S/N clusters. Therefore it is striking that two different analysis on two different samples yield the same results on the evolution of cluster density profiles.
We observe only mild evolution in pressure and entropy profiles in cluster cores. On the other hand, the evolution of temperature profiles is in agreement with self-similarity. Utilization of the X-ray peak instead of the centroid of the large scale emission does not significantly affect our results.

Planned and future X-ray telescopes with sufficiently small spatial resolution and large effective area, e.g. {\it Athena, Lynx} will provide sufficient statistics to precisely measure temperature and density profiles down to kpc scales in the cores of a large sample of clusters \citep{nandra+13, gaskin2019}. These measurements will allow us to probe in detail the role of AGN feedback in the first clusters formed and to shed light onto the accretion processes in cluster outskirts and the structure formation in the Universe.

\section*{Acknowledgements}
\emph{Facilities}:
10m South Pole Telescope (SPT-SZ) --
\emph{Chandra} X-ray Observatory --
\emph{XMM-Newton}

This work was performed in the context of the South Pole Telescope scientific program. SPT is supported by the National Science Foundation through grant PLR- 1248097. Partial support is also provided by the NSF Physics Frontier Center grant PHY-0114422 to the Kavli Institute of Cosmological Physics at the University of Chicago, the Kavli Foundation and the Gordon and Betty Moore Foundation grant GBMF 947 to the University of Chicago. This work is also supported by the U.S. Department of Energy.  Work at Argonne National Lab is supported by UChicago Argonne LLC, Operator of Argonne National Laboratory (Argonne). Argonne, a U.S. Department of Energy Office of Science Laboratory, is operated under contract no. DE-AC02-06CH11357. BB is supported by the Fermi Research Alliance LLC under contract no. De-AC02- 07CH11359 with the U.S. Department of Energy. 




\appendix
\section{Mass reconstruction}
\label{app:mass}
In this Appendix we solve the hydrostatic equilibrium equation with other approaches besides the one used throughout this paper, see Sect.~\ref{sec:mass}. 

\subsection{Forward Modeling Approach}
In Sect.~\ref{sec:mass} we have used a ``forward'' modeling where a temperature model is combined with a density model to solve hydrostatic equilibrium equation and recover the mass profile. However it is possible to do the same using a pressure model in combination with the density model, because it would be equivalent of doing the same but using pressure divided by density as the temperature model.
We use the five-parameters functional form \citep{nagai+07} to model the pressure, and then recover the three-dimensional temperature profile by dividing it by the density profile. Then, everything goes like in Sect.~\ref{sec:mass}. We indicate this method ``forward P'' distinguishing it from the one used in  Sect.~\ref{sec:mass}, indicated as ``forward T''

\subsection{Backward Modeling Approach}
A popular model used in the literature is the ``backward'' modeling, which assumes a dark matter distribution, e.g. the Navarro-Frenk-White (NFW) model \citep{nfw+97}, and then the observed temperature profiles are fitted against their profiles as predicted by the combination of the mass model with the density profile. Only two parameters are required to fully characterize the NFW mass model, scale radius and concentration \citep[see ][ for details]{ettori+10}.

\begin{equation}
M_{tot,NFW} = \frac{4}{3} \pi \rho_c(z) 500 \frac{r_s^3 c_{500}^3}{\log (1+c_{500}) - \frac{c_{500}}{1+c_{500}}} \left( \log (1+\frac{r}{r_s}) - \frac{\frac{r}{r_s}}{1+\frac{r}{r_s}} \right) 
\end{equation}
or using the equation $R_{500} = r_s c_{500}$
\begin{equation}
M_{tot,NFW} = \frac{4}{3} \pi \rho_c(z) 500 \frac{R_{500}^3}{\log (1+c_{500}) - \frac{c_{500}}{1+c_{500}}} \left( \log (1+c_{500}\frac{r}{R_{500}}) - \frac{c_{500}\frac{r}{R_{500}}}{1+c_{500}\frac{r}{R_{500}}} \right) 
\end{equation}

Since the large binsize of the annuli caused by the large \xmm\ PSF of about 15 arcsec that corresponds to a physical size of 150 kpc, the constraints on the concentration parameter are very weak, meaning that the concentration is almost unconstrained.
Thus we apply this technique two times, once leaving concentration completely free, i.e. with flat priors, and once choosing a gaussian prior on the concentration parameter, centered on the concentration--mass relation provided by \cite{diemer18}\footnote{as implemented in the code {\tt COLOSSUS} \citep{Diemer+17}, with $\Omega_{\rm m}=0.3, \Omega_{\Lambda}=0.7, \sigma_8=0.8, H_0 = 70$ km s$^{-1}$ Mpc$^{-1}$}: $\log c_{500} = 0.885 -0.049 \log (M_{500}/5 \, 10^{14} M_{\odot})$, and with an instrinsic scatter of $\sigma_{log_{10}(c_{500})} = 0.1$ \citep[from][]{Neto+07} propagated throught our analysis.

The fit is done using the code \texttt{emcee} \citep{emcee}, starting from a standard maximum likelihood fit, $\chi^2$ minimization using the Nelder-Mead method \citep{Gao2012}, using 10000 steps with burning length of 5000 steps to have resulting chains independent from the starting position, and thinning of 10 in order to reduce the correlation between consecutive steps.

\begin{figure*}
    \centering
    \includegraphics[width=\textwidth]{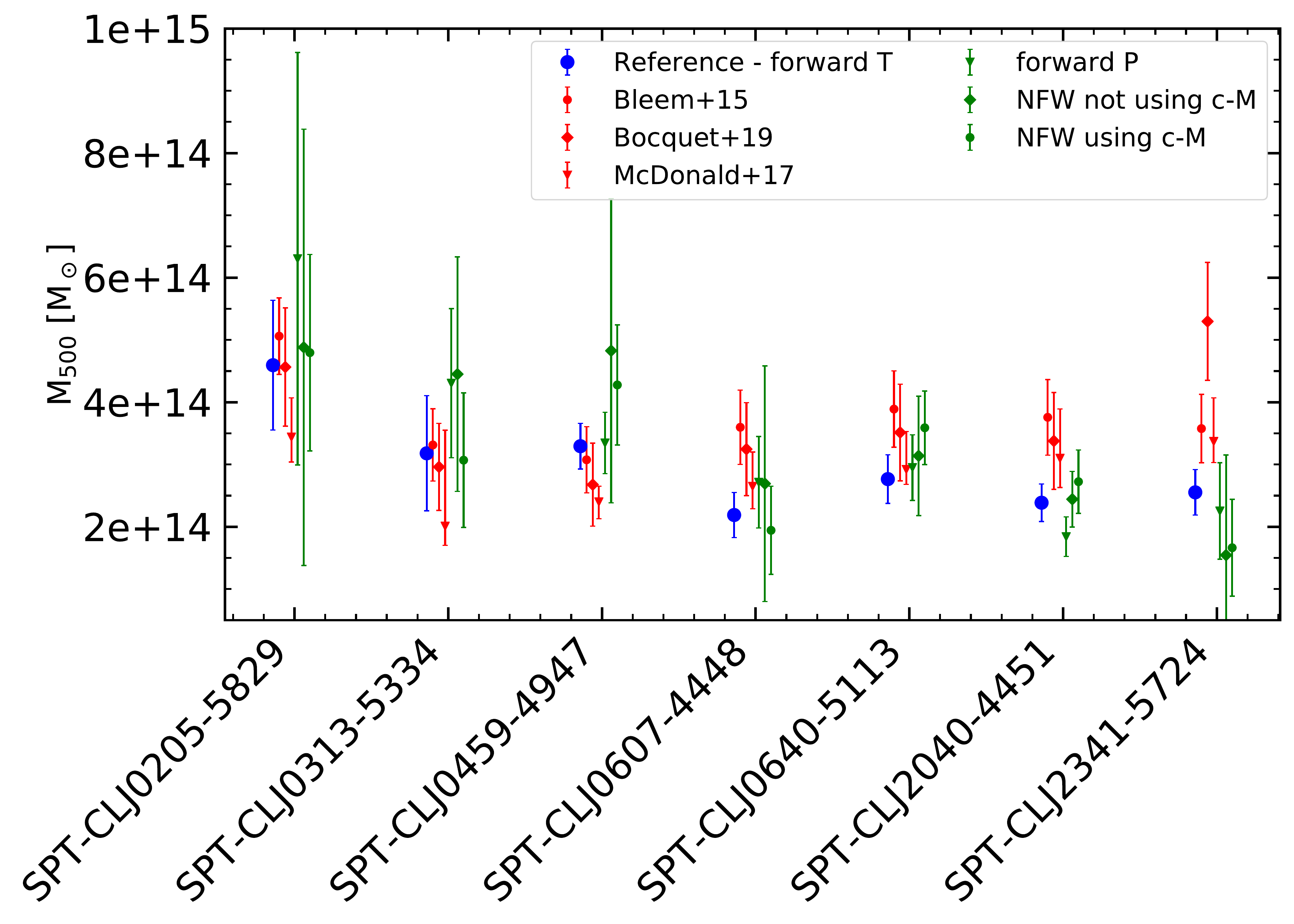}
    \caption{Mass comparison for the object in our sample. In red are the masses from the SPT catalog \citep{bleem15}, and the masses from SPT cosmological analysis \citep{Bocquet+19}. In black are the masses recovered by MD17 using the $M_{gas}-M_{tot}$ scaling relation. In green are the NFW best-fit masses in the two cases described in the text. In blue are the forward best-fit $M_{500}$ computed using functional form to fit the temperature and density profiles.}
    \label{fig:mass}
\end{figure*}

\subsection{Reconstructed mass}
Our reconstructed $M_{500}$, using the method described above and in Sect.~\ref{sec:mass}, are shown in Figure~\ref{fig:mass}. We compare our mass reconstruction among themselves, and with the SPT masses as calculated in the catalog \citep{bleem15} using $M-\zeta$ fixed scaling relation, with the masses calculated from the scaling relations obtained for the SPT cosmological results \citep{Bocquet+19}, and with the masses used in MD17 which come from the $M_{gas}-M_{tot}$ scaling relations \citep{vikhlinin+09}. Overall the masses we measure are consistent with all the other masses we are comparing with, with two peculiar cases:
1) SPT-CLJ0459-4947, for which the masses coming from the \emph{forward} reconstruction agree with the other masses in the literature, i.e. the two SPT masses and the masses in MD17, but the NFW reconstruction prefers a higher mass. This can potentially indicate that the NFW mass model could not be the best model 	to describe the dark matter potential for this object.
2) SPT-CLJ2341-5724, which has all the masses coming from our analysis consistent within 1$\sigma$, however when comparing with the literature masses, we find that these are much higher than what we measure, indicating the possibility that this cluster does not fall on the scaling relations used to determine the literature masses.
The recovered mass of SPT-CLJ0205-5829 has very large uncertainties. This is because the \xmm\ 55 ks observation 0803050201 is highly flared, with only about 10 ks remaining after flare removal, and of top of that this cluster have a point source very close to the cluster center, thus decreasing the photon statistics with the resulting effect being larger error bars for the temperature, translating into large error bars on the mass since $M \sim T$.

\begin{figure}
    \centering
    \includegraphics[width=0.8\textwidth]{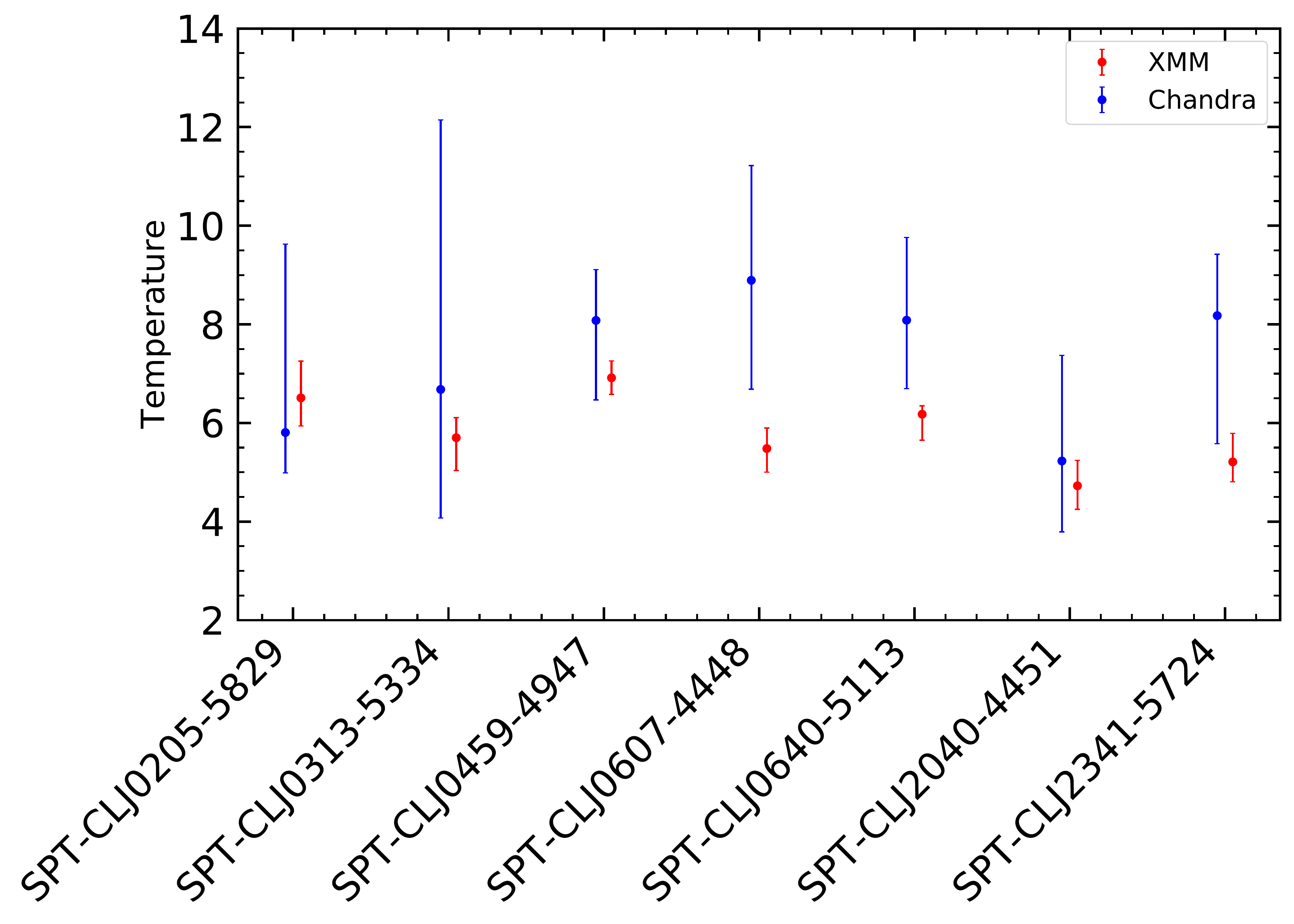}
    \caption{Comparison of a single temperature recovered from both \chandra\ and \xmm\ from a circular region of width equal in radius to $R_{500}$}
    \label{fig:T_cxo_xmm}
\end{figure}

\begin{figure}
    \centering
    \includegraphics[width=0.8\textwidth]{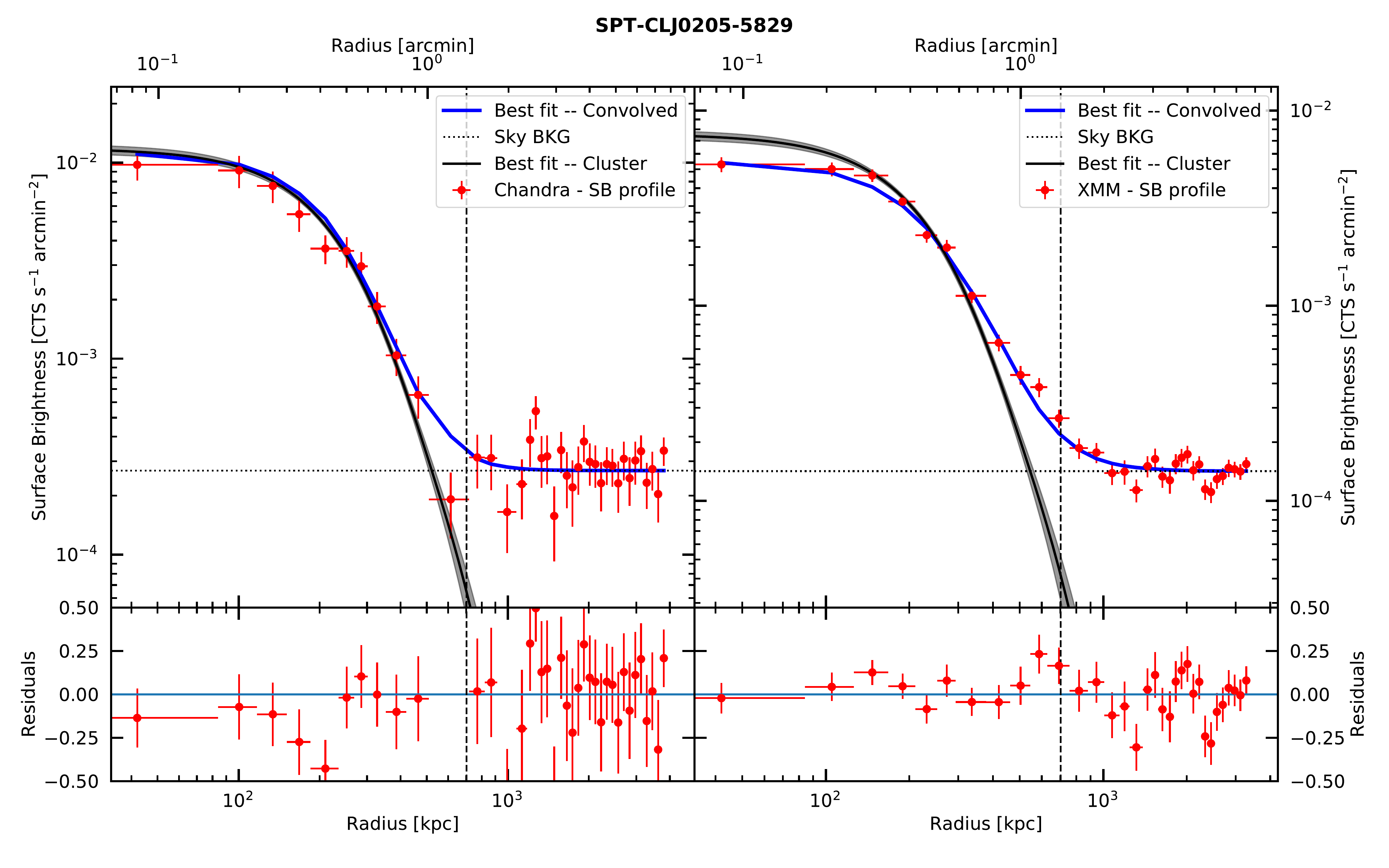}
    
    \includegraphics[width=0.8\textwidth]{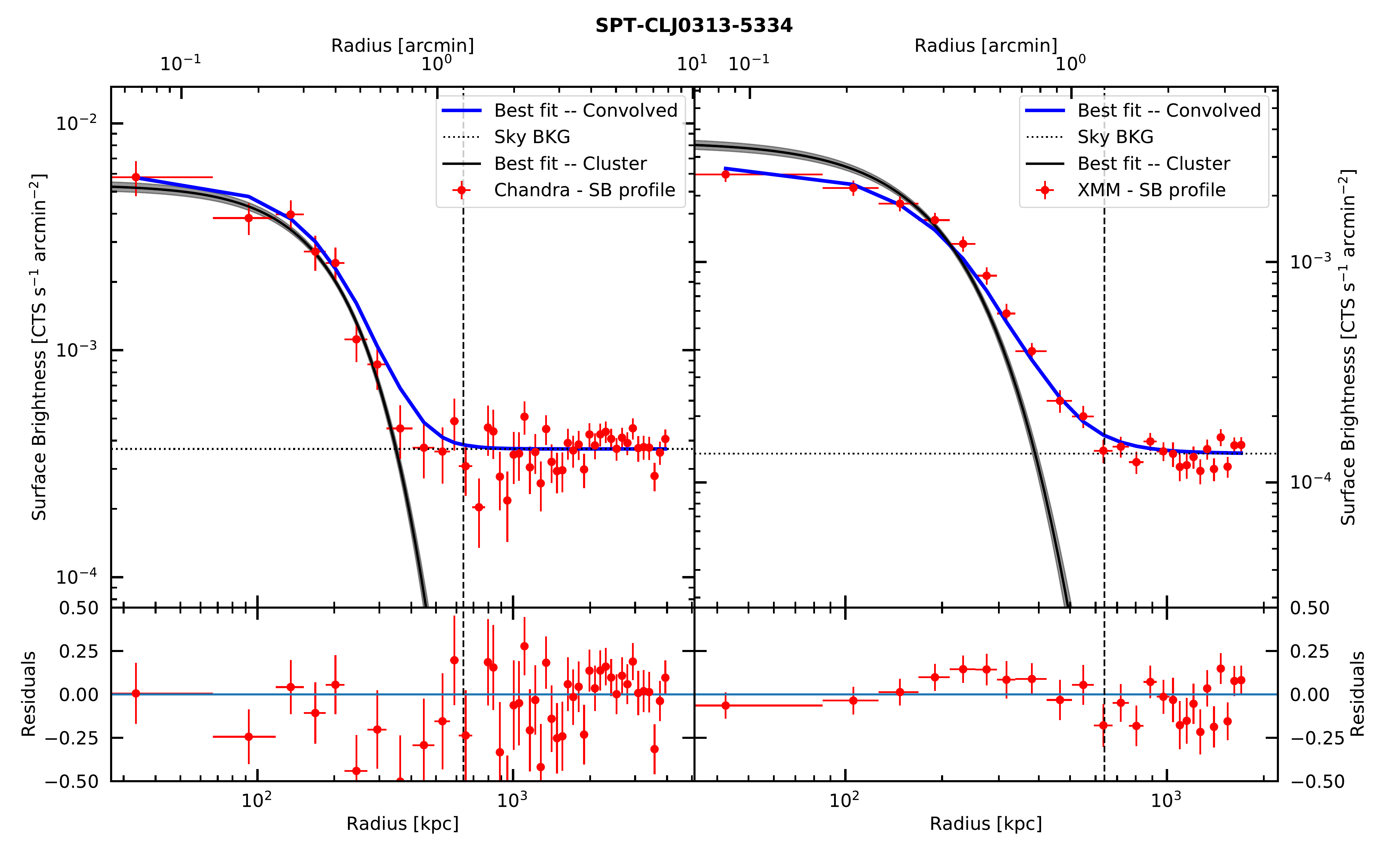}
\caption{\chandra\ (\emph{left}) and \xmm\ (\emph{right}) measured NXB subtracted surface brightness (red points).  The best-fitting model is the \cite{vikhlini+06} functional  (solid black line) form plus a constant sky background (horizontal dotted line); it is convolved with the intrumental PSF and it is shown with a blue line. In case of \chandra\ the PSF is simply a diagonal matrix with ones on the diagonal, while for \xmm\ it is calculated as in Sect.~\ref{sec:psf}. In the bottom panels we show the residuals ($\frac{\mu_i - N_{c,i}}{N_{c,i}}$). The dashed vertical line represents the location of $R_{500}$, as measured by solving HE equation (Equation~\ref{eq:HEE}) using the ``forward T'' method (see Sect.~\ref{sec:mass}). }
    \label{fig:gallery_sb}
 
\end{figure}
\begin{figure}
    \ContinuedFloat
    \centering

    \includegraphics[width=0.8\textwidth]{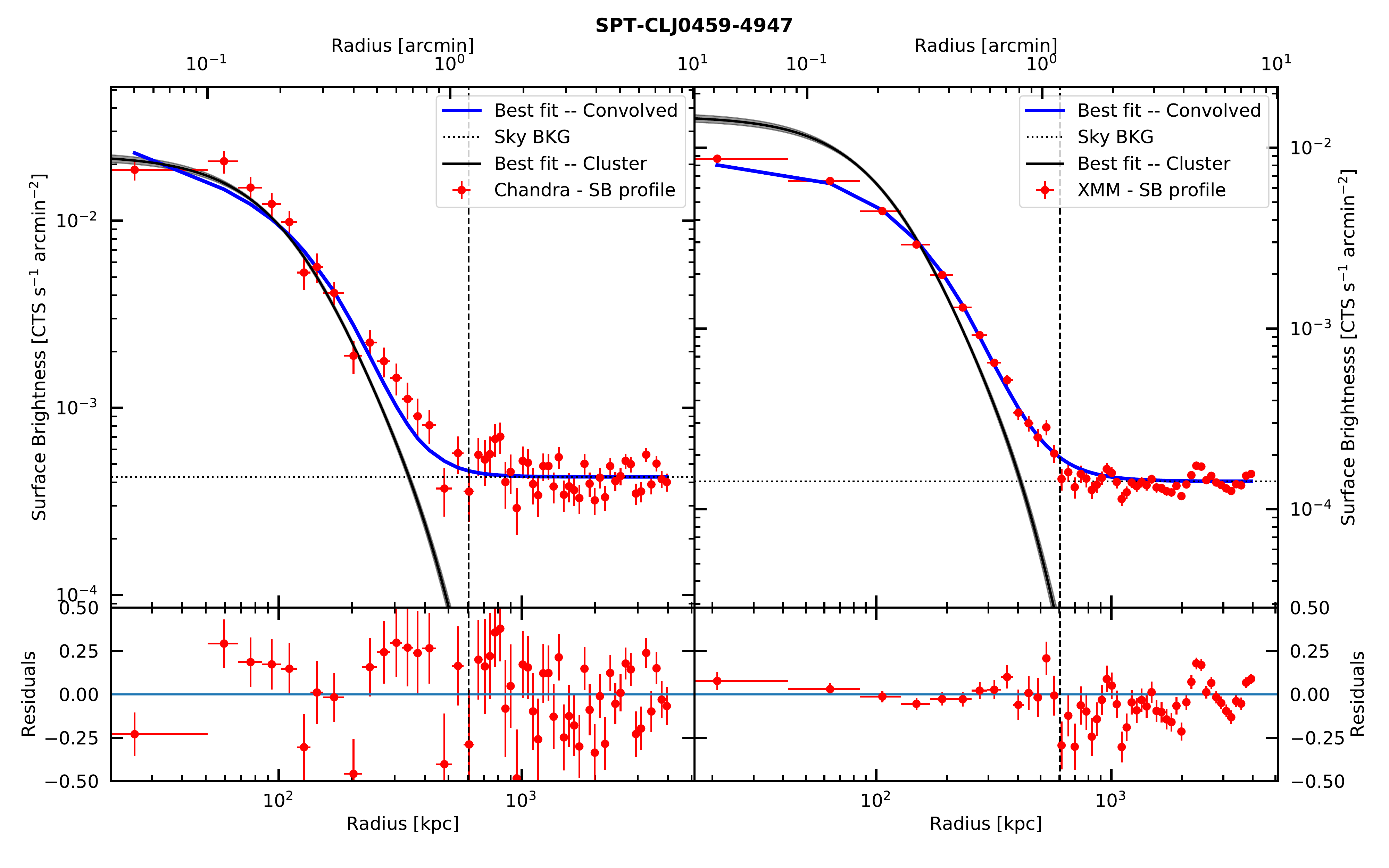}
    
    \includegraphics[width=0.8\textwidth]{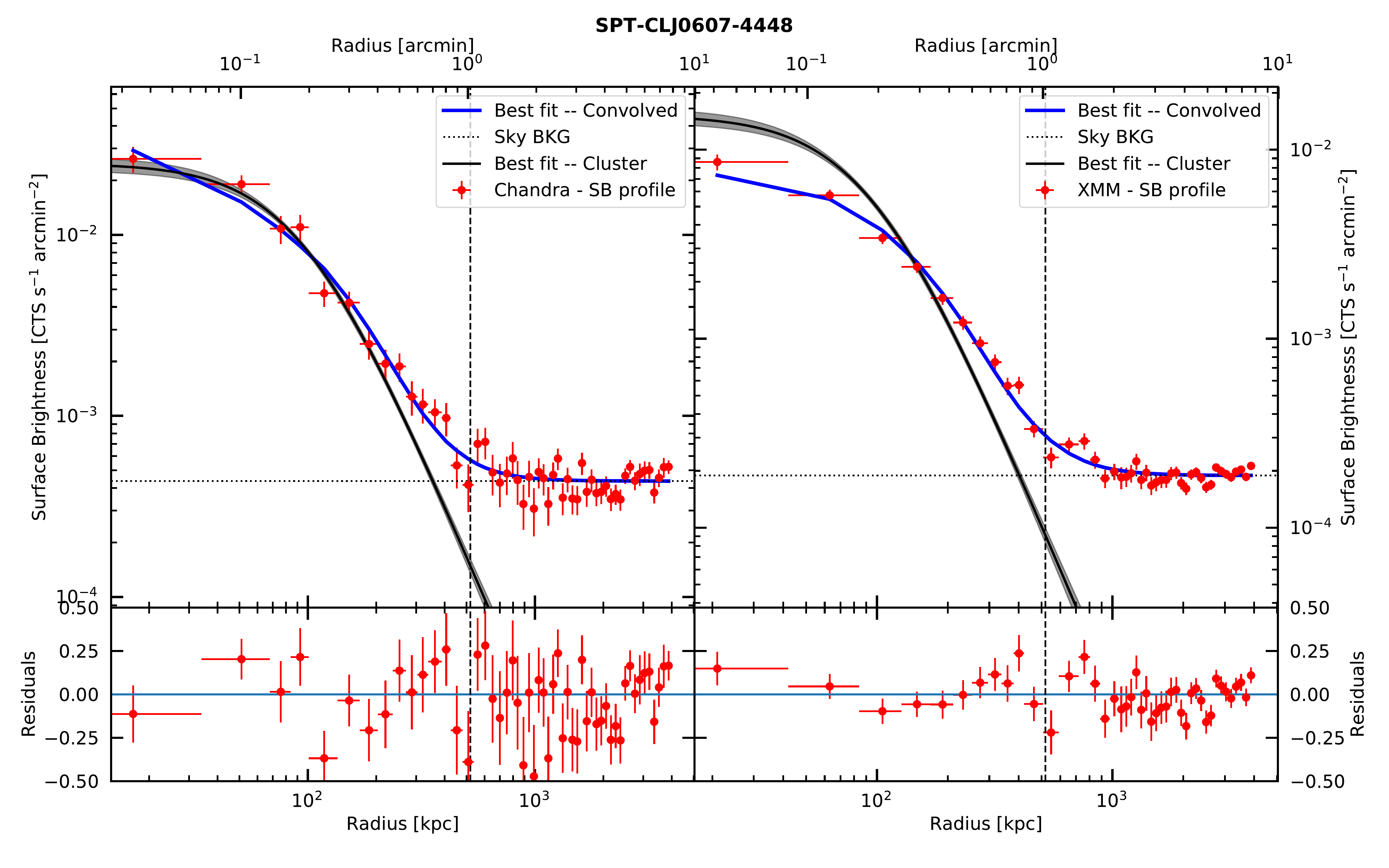}
    \caption{Continued}
\end{figure}
\begin{figure}
    \ContinuedFloat
    \centering
    
    \includegraphics[width=0.8\textwidth]{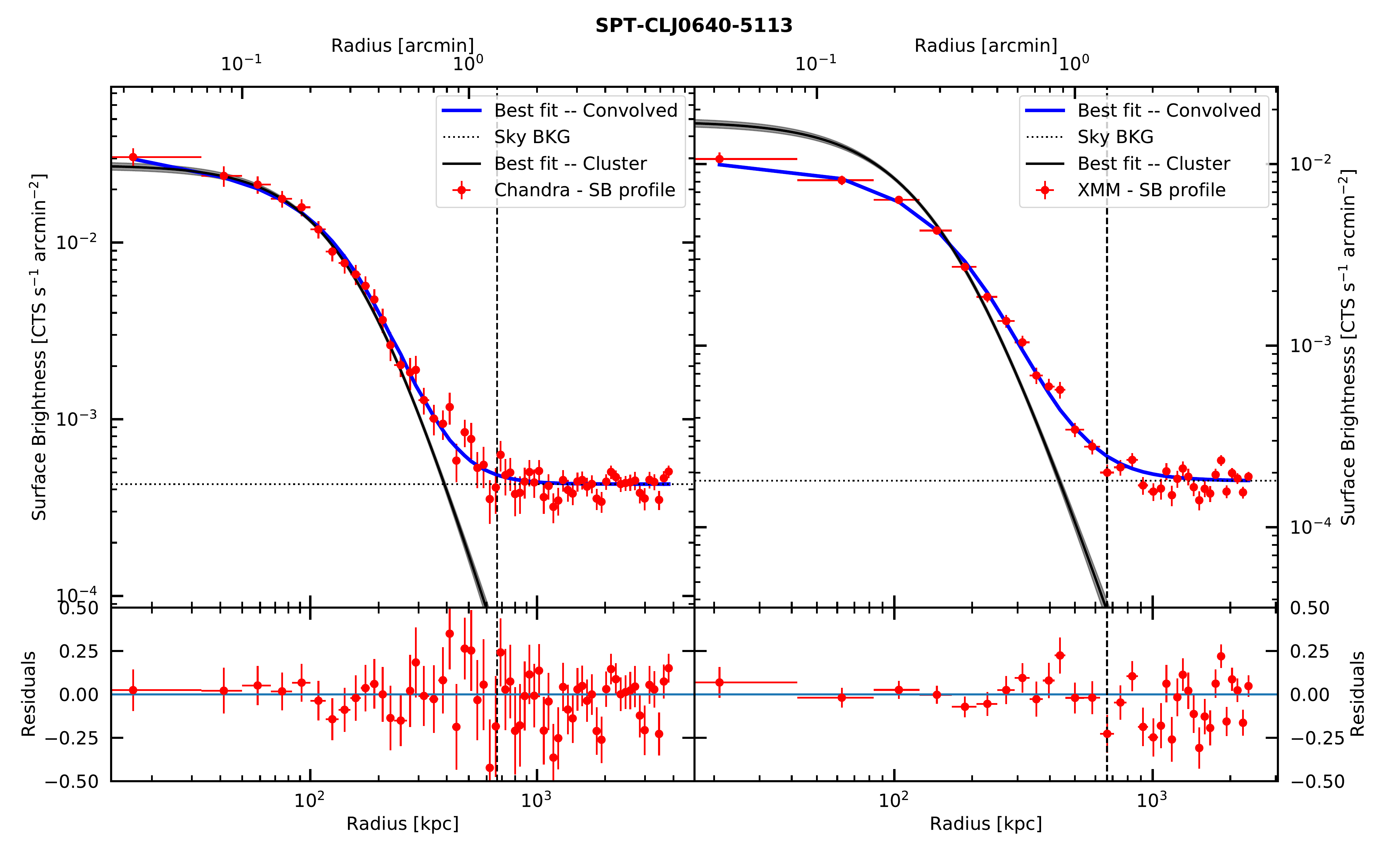}
    
    \includegraphics[width=0.8\textwidth]{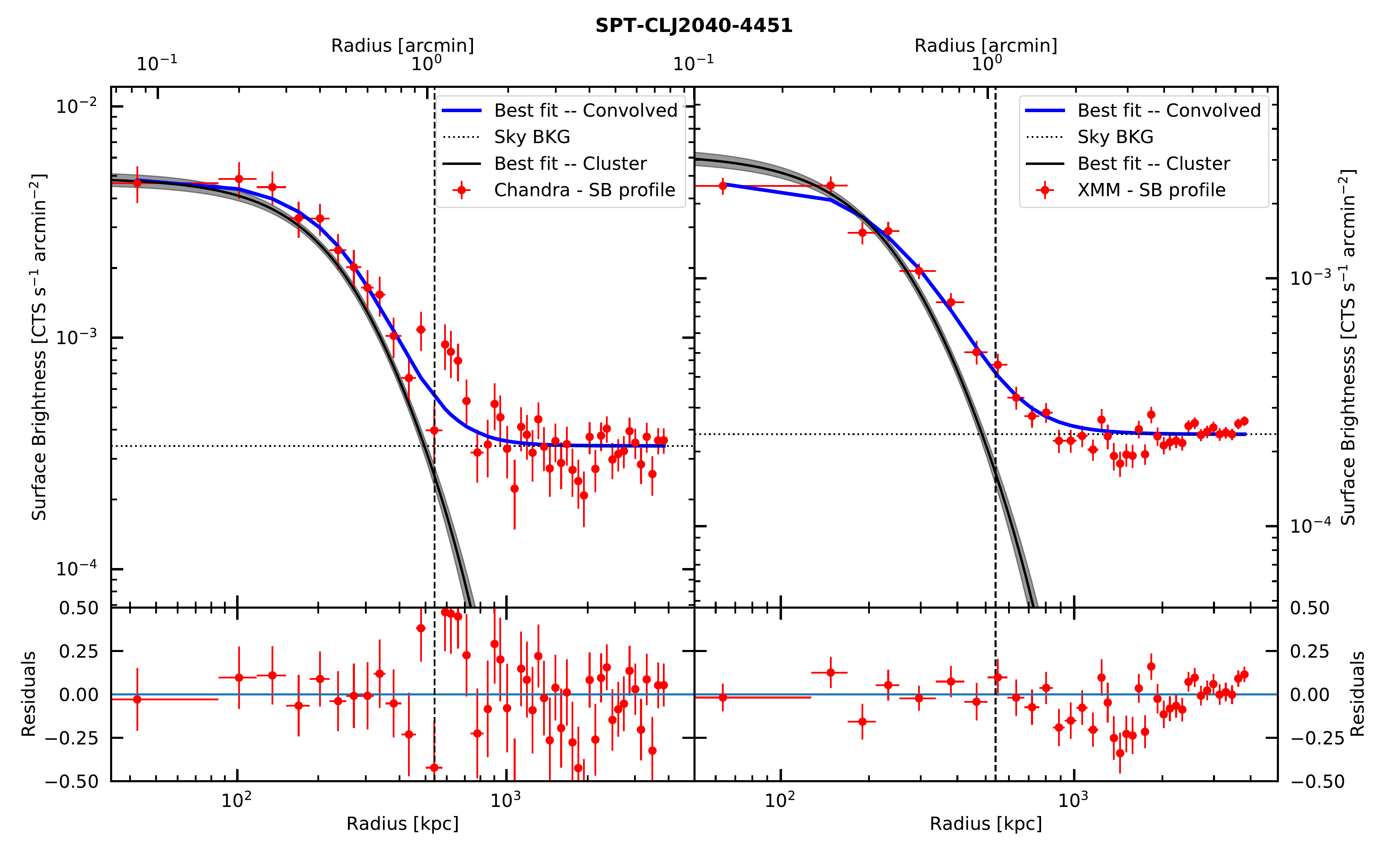}
    \caption{Continued}

\end{figure}
\begin{figure}
    \ContinuedFloat
    \centering
    
    \includegraphics[width=0.8\textwidth]{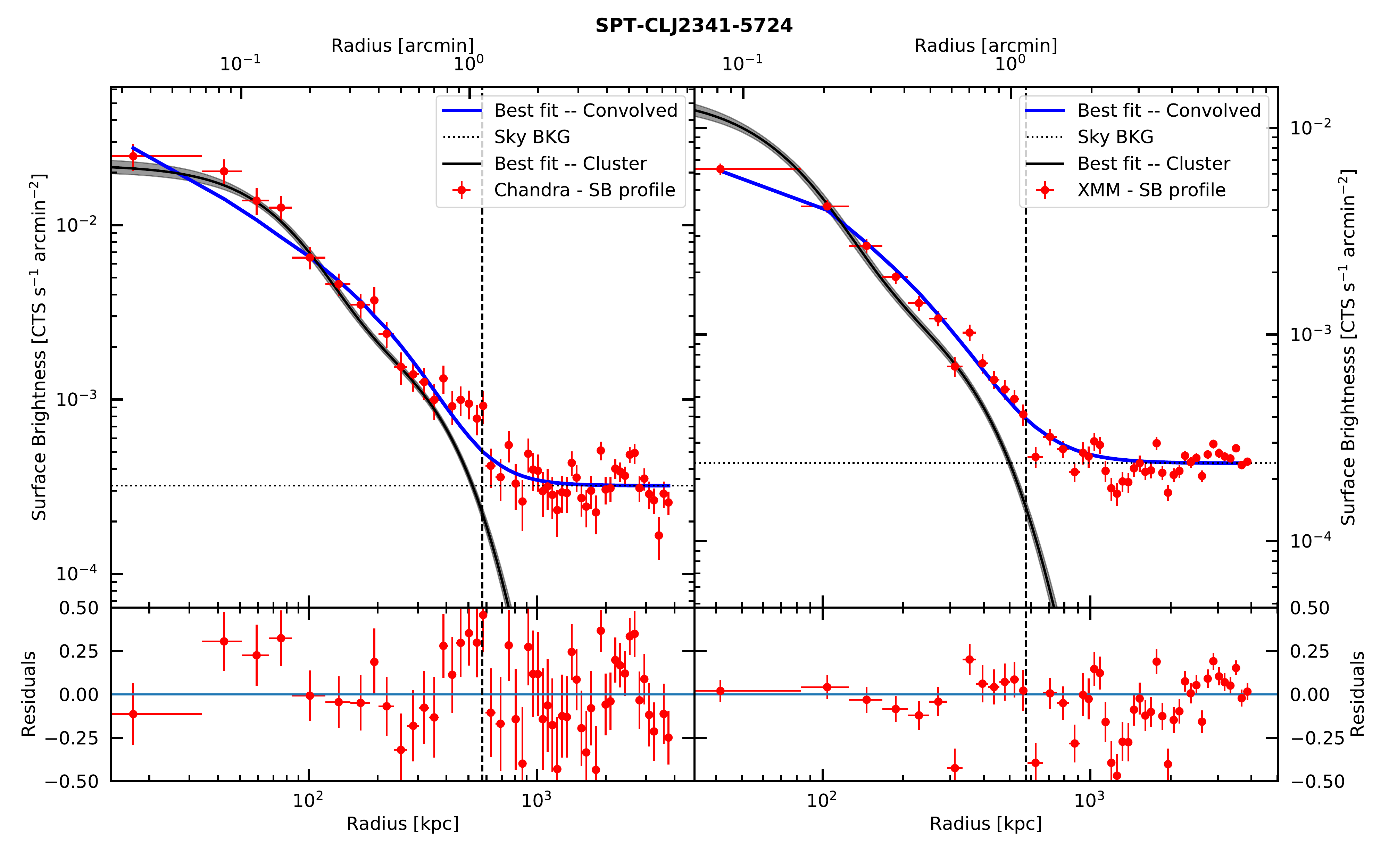}
    
    \caption{Continued}

\end{figure}

\begin{figure*}
\begin{center}
   \includegraphics[width=0.8\textwidth]{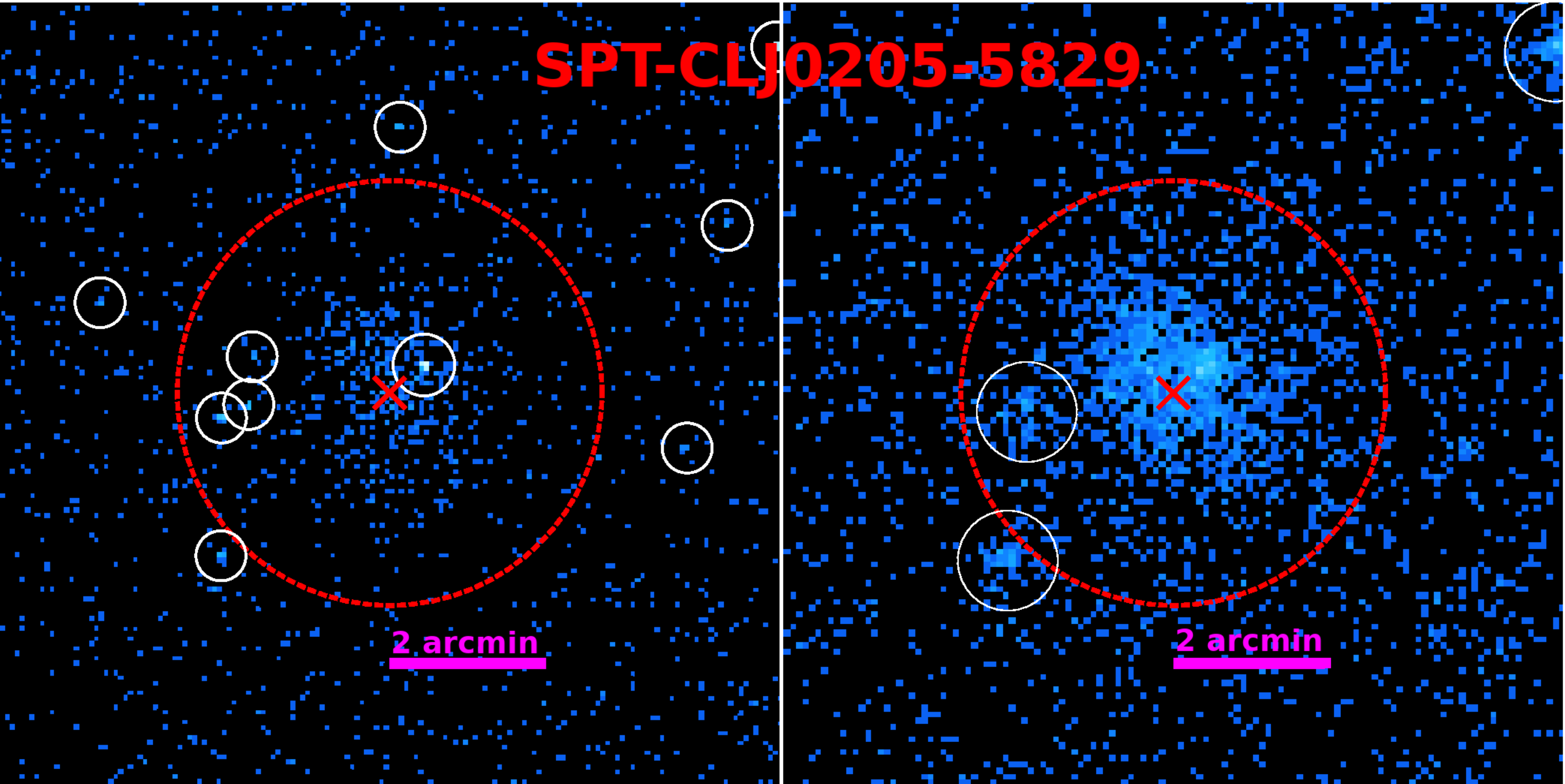}
  
   \includegraphics[width=0.8\textwidth]{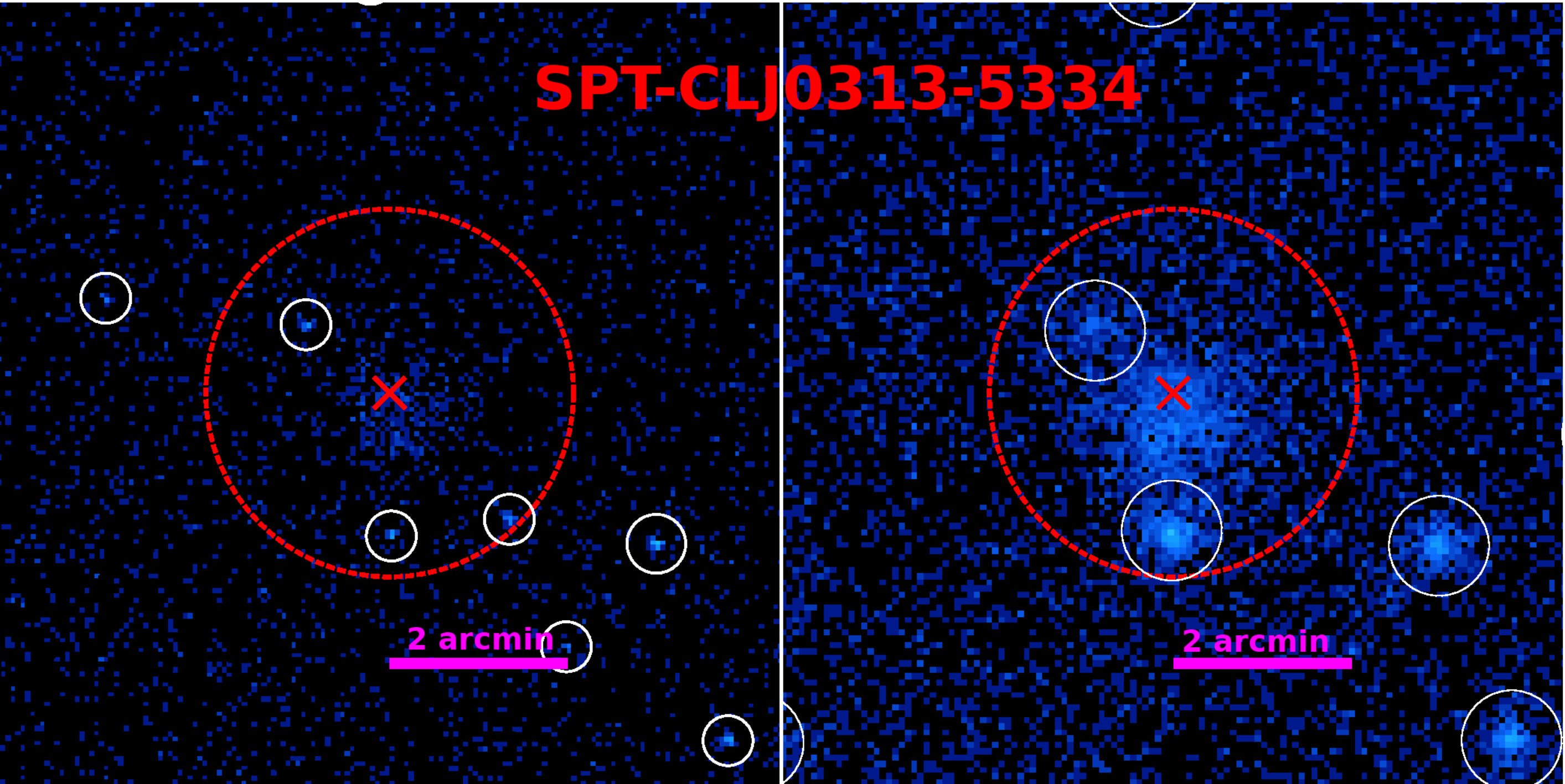}
   
   \includegraphics[width=0.8\textwidth]{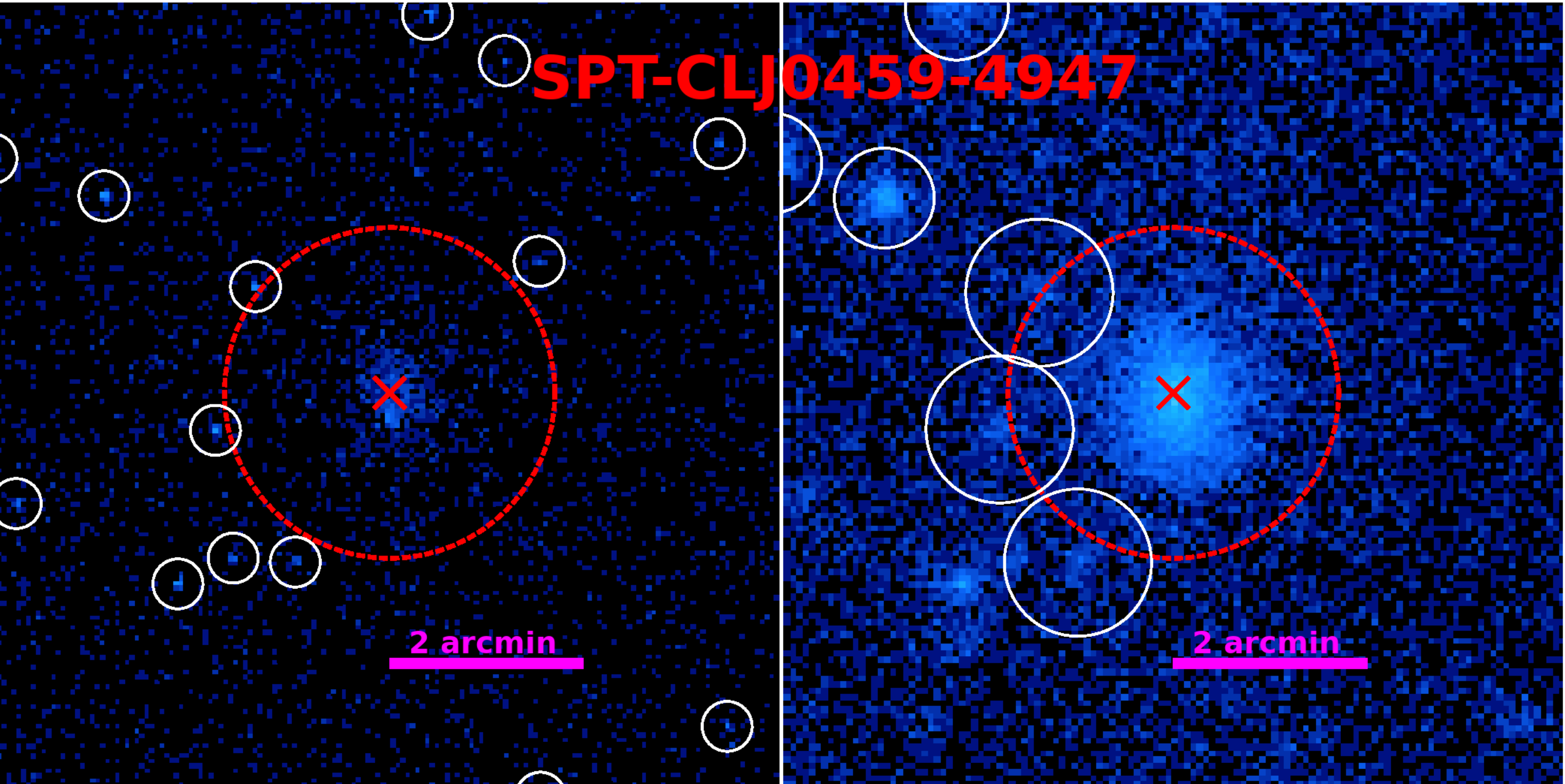}
   
    \caption{Raw count images of the clusters, \chandra\ counts are shown on the left, while the  \xmm\ image is shown on the right. The blue cross indicates the location of the center used, and the white circles indicate the point sources masked.}
    \label{fig:raw_images}
\end{center} 
\end{figure*}

\begin{figure*}
    \ContinuedFloat
\begin{center}
   \includegraphics[width=0.8\textwidth]{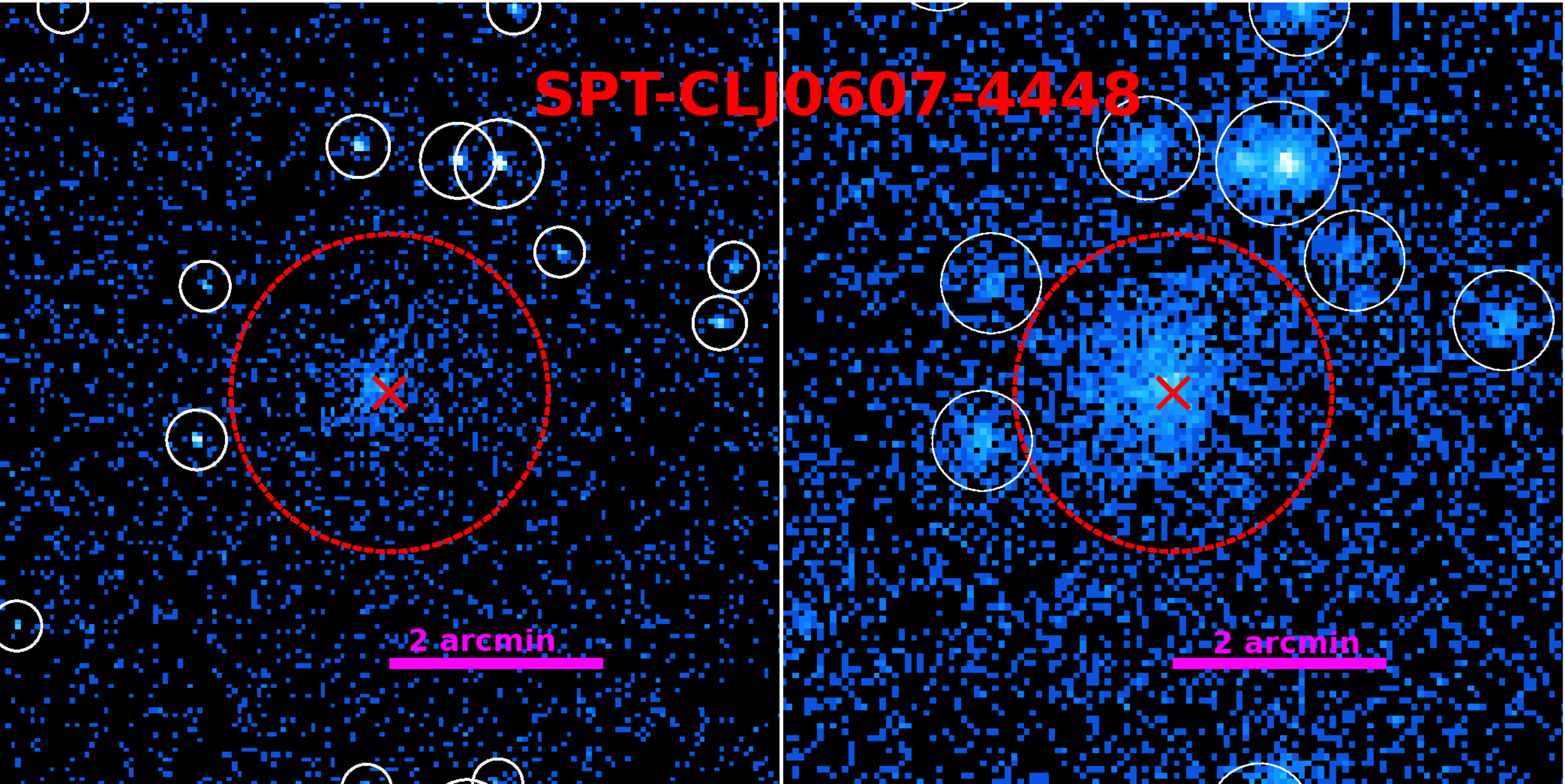}

   \includegraphics[width=0.8\textwidth]{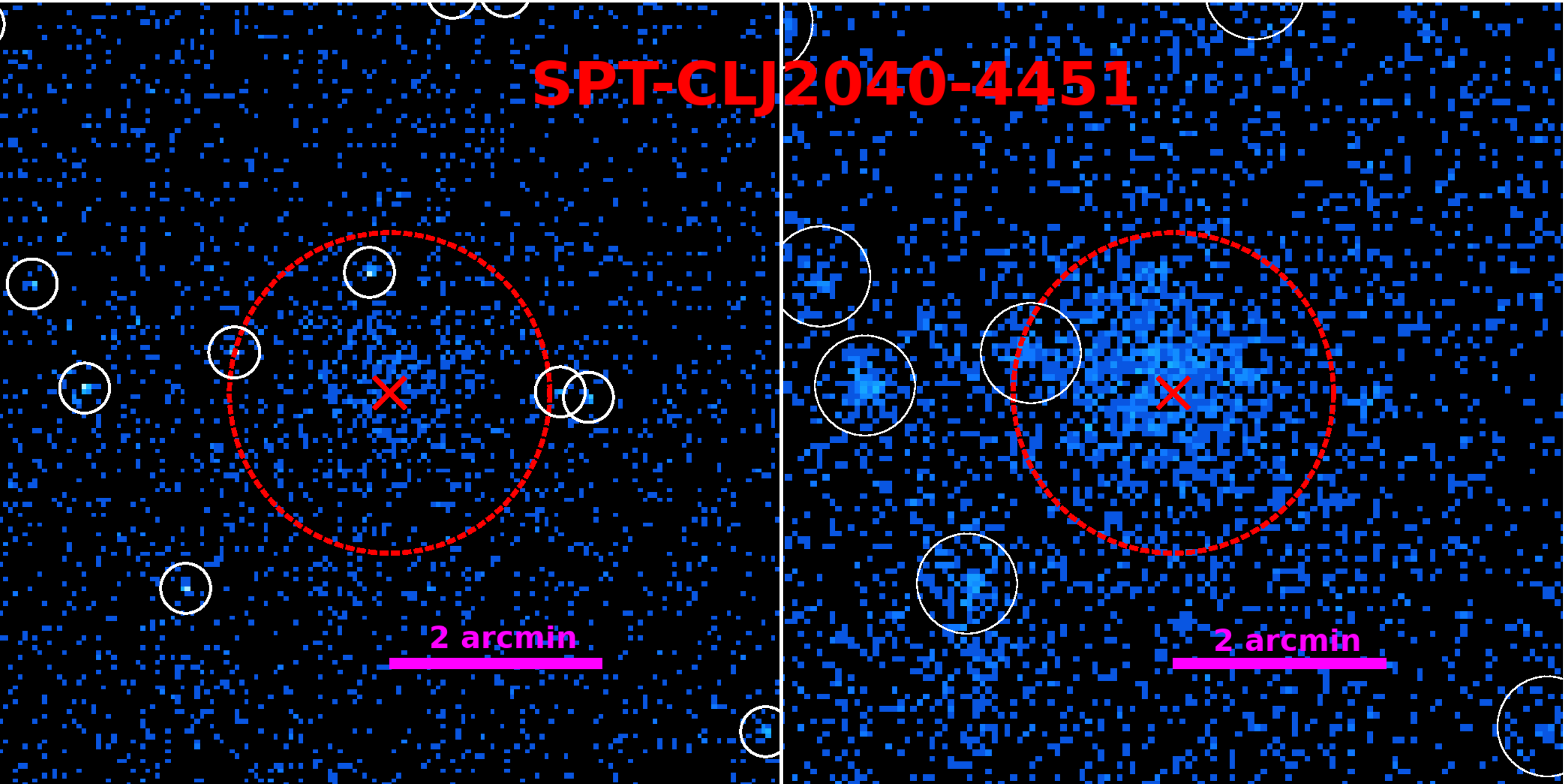}

   \includegraphics[width=0.8\textwidth]{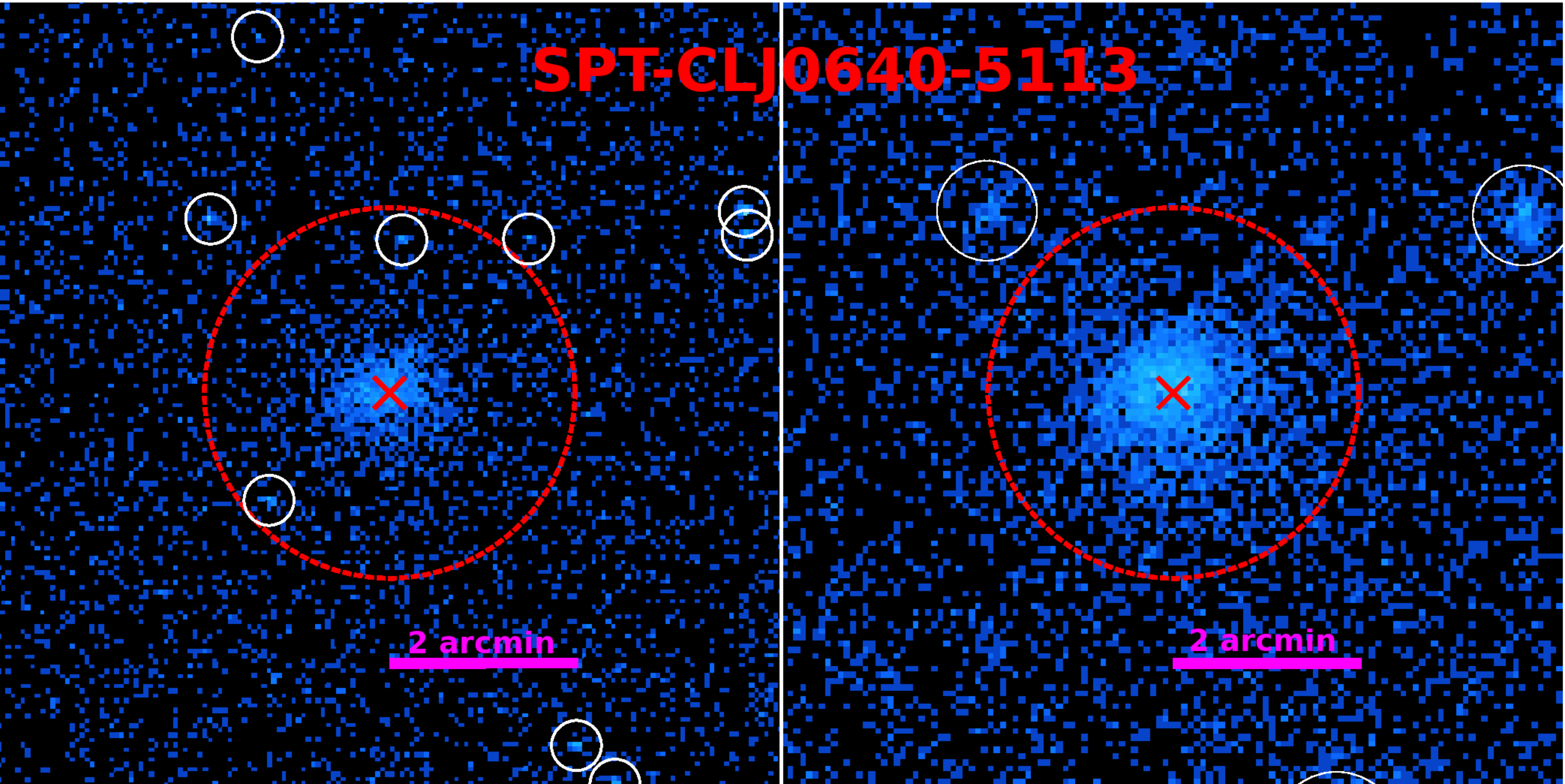} 
      \caption{Continued}

\end{center} 
\end{figure*}

\begin{figure*}
    \ContinuedFloat
\begin{center}      
   \includegraphics[width=0.8\textwidth]{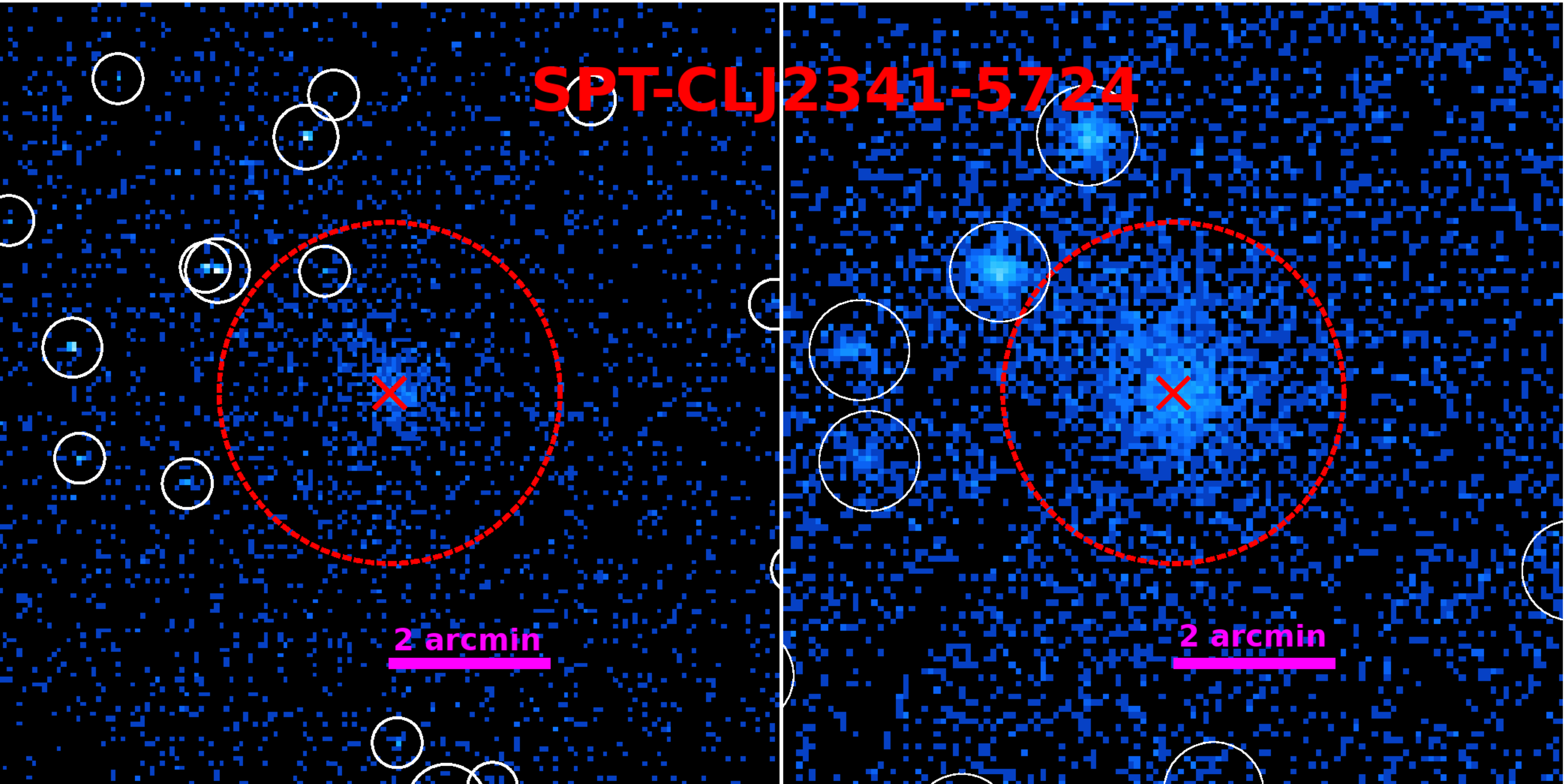}
   \caption{Continued}
\end{center} 

\end{figure*}

\begin{table*}[]
    \centering   
\begin{tabular}{ c c c c c c c }
\hline
Cluster & $T_{\rm XMM}$ & $T_{\rm CXO}$ & $M_{\rm forward \ T}$ & $M_{\rm forward \ P}$ & $M_{\rm NFW, \ no \ c-M}$ & $M_{\rm NFW, \ with \ c-M}$\\
 & [keV] & [keV] & [$10^{14} M_{\odot}$] & [$10^{14} M_{\odot}$] & [$10^{14} M_{\odot}$] & [$10^{14} M_{\odot}$] \\
\hline
SPT-CLJ0205-5829 & $6.51_{-0.57}^{+0.75}$ & $5.81_{-0.82}^{+3.82}$ & $4.59 \pm 1.04$ & $6.30 \pm 3.31$ & $4.88 \pm 3.50$  & $4.80 \pm 1.58$ \\
SPT-CLJ0313-5334 & $5.70_{-0.67}^{+0.41}$ & $6.68_{-2.61}^{+5.47}$ & $3.18 \pm 0.92$ & $4.31 \pm 1.20$ & $4.45 \pm 1.88$  & $3.07 \pm 1.08$ \\
SPT-CLJ0459-4947 & $6.92_{-0.33}^{+0.34}$ & $8.08_{-1.61}^{+1.03}$ & $3.29 \pm 0.37$ & $3.35 \pm 0.49$ & $4.83 \pm 2.44$  & $4.28 \pm 0.96$ \\
SPT-CLJ0607-4448 & $5.48_{-0.48}^{+0.42}$ & $8.89_{-2.21}^{+2.33}$ & $2.19 \pm 0.36$ & $2.71 \pm 0.73$ & $2.69 \pm 1.89$  & $1.94 \pm 0.71$ \\
SPT-CLJ0640-5113 & $6.18_{-0.52}^{+0.17}$ & $8.08_{-1.39}^{+1.68}$ & $2.77 \pm 0.39$ & $2.95 \pm 0.53$ & $3.14 \pm 0.96$  & $3.59 \pm 0.59$ \\
SPT-CLJ2040-4451 & $4.73_{-0.48}^{+0.52}$ & $5.23_{-1.44}^{+2.14}$ & $2.39 \pm 0.30$ & $1.84 \pm 0.32$ & $2.44 \pm 0.45$  & $2.72 \pm 0.51$ \\
SPT-CLJ2341-5724 & $5.21_{-0.40}^{+0.58}$ & $8.18_{-2.60}^{+1.25}$ & $2.55 \pm 0.36$ & $2.25 \pm 0.77$ & $1.54 \pm 1.61$  & $1.66 \pm 0.78$ \\
\hline
\end{tabular}

\caption{Information on the cluster recoverd temperatures within $R_{500}$, see Figure~\ref{fig:T_cxo_xmm}, and the recovered masses using different techniques, see Appendix~\ref{app:mass} and Figure~\ref{fig:mass}.}
  \label{tab:props}
\end{table*}

\begin{table*}[]
    \centering   
\begin{tabular}{ c c c c }
\multicolumn{4}{c}{(a) \bf Density} \\
\hline
$\left( R/R_{500} \right)_{in}$ & $\left( R/R_{500} \right)_{out}$ & 2+C$_\rho$ & Sign.  \\
\hline
0.01 & 0.02 & $-0.00 \pm 0.08 \pm 0.16$ & 11.4  \\
0.02 & 0.05 & $0.47 \pm 0.06 \pm 0.15$ & 9.4  \\
0.05 & 0.08 & $0.95 \pm 0.06 \pm 0.15$ & 6.6  \\
0.08 & 0.12 & $1.32 \pm 0.06 \pm 0.14$ & 4.5  \\
0.12 & 0.20 & $1.71 \pm 0.04 \pm 0.13$ & 2.2  \\
0.20 & 0.30 & $1.99 \pm 0.03 \pm 0.12$ & 0.1  \\
0.30 & 0.45 & $2.08 \pm 0.03 \pm 0.10$ & 0.7  \\
0.45 & 0.60 & $2.06 \pm 0.04 \pm 0.10$ & 0.6  \\
0.60 & 0.80 & $2.01 \pm 0.04 \pm 0.10$ & 0.1  \\
0.80 & 1.00 & $2.02 \pm 0.06 \pm 0.11$ & 0.2  \\
1.00 & 1.20 & $1.94 \pm 0.07 \pm 0.12$ & 0.4  \\
\hline
\end{tabular}

\begin{tabular}{ c c c c }
\multicolumn{4}{c}{(b) \bf Temperature} \\
\hline
$\left( R/R_{500} \right)_{in}$ & $\left( R/R_{500} \right)_{out}$ & 2/3+C$_T$ & Sign.  \\
\hline
0.05 & 0.20 & $0.52 \pm 0.12 \pm 0.17$ & 0.7\\
0.20 & 0.40 & $0.76 \pm 0.09 \pm 0.14$ & 0.5\\
0.40 & 0.75 & $0.60 \pm 0.09 \pm 0.13$ & 0.4\\
0.75 & 1.40 & $0.72 \pm 0.11 \pm 0.12$ & 0.4\\
\hline
\end{tabular}

\vspace{1cm}
\begin{tabular}{ c c c c }
\multicolumn{4}{c}{(c) \bf Pressure} \\
\hline
$\left( R/R_{500} \right)_{in}$ & $\left( R/R_{500} \right)_{out}$ & 8/3+C$_P$ & Sign. \\
\hline
0.01 & 0.02 & $1.31 \pm 0.08 \pm 0.12$ & 9.2  \\
0.02 & 0.05 & $1.66 \pm 0.07 \pm 0.12$ & 7.2  \\
0.05 & 0.12 & $2.07 \pm 0.06 \pm 0.13$ & 4.3  \\
0.12 & 0.20 & $2.39 \pm 0.07 \pm 0.13$ & 1.9  \\
0.20 & 0.30 & $2.56 \pm 0.07 \pm 0.15$ & 0.7  \\
0.30 & 0.50 & $2.65 \pm 0.05 \pm 0.17$ & 0.1  \\
0.50 & 0.80 & $2.65 \pm 0.06 \pm 0.21$ & 0.1  \\
0.80 & 1.20 & $2.69 \pm 0.05 \pm 0.24$ & 0.1  \\
\hline
\end{tabular}

\begin{tabular}{ c c c c }
\multicolumn{4}{c}{(d) \bf Entropy} \\
\hline
$\left( R/R_{500} \right)_{in}$ & $\left( R/R_{500} \right)_{out}$ & -2/3+C$_K$ & Sign. \\
\hline
0.01 & 0.02 & $0.42 \pm 0.16 \pm 0.27$ & 3.4  \\
0.02 & 0.05 & $0.30 \pm 0.10 \pm 0.26$ & 3.4  \\
0.05 & 0.12 & $0.03 \pm 0.06 \pm 0.24$ & 2.8  \\
0.12 & 0.20 & $-0.28 \pm 0.05 \pm 0.22$ & 1.7  \\
0.20 & 0.30 & $-0.50 \pm 0.05 \pm 0.19$ & 0.9  \\
0.30 & 0.50 & $-0.64 \pm 0.03 \pm 0.16$ & 0.2  \\
0.50 & 0.80 & $-0.63 \pm 0.04 \pm 0.13$ & 0.3  \\
0.80 & 1.20 & $-0.57 \pm 0.04 \pm 0.12$ & 0.8  \\
\hline
\end{tabular}

\caption{Evolution of the thermodynamic quantities with cosmic time, density (a) top left, temperature (b) top right, pressure (c) lower left, and entropy (d) lower right. In each single table the first two columns represent the inner and outer radial ranges in which we have look for the evolution. The third column represents the measured evolution with redshifts, its statistical and systematic uncertainty. The fourth column represents the significance measured in number of sigmas of the difference between the measured evolution and the evolution predicted by the self-similar expectation.}
  \label{tab:evo}

\end{table*}

\bibliography{SPT_high_z_eb}

\begin{thebibliography}{}
\expandafter\ifx\csname natexlab\endcsname\relax\def\natexlab#1{#1}\fi

\bibitem[{{Ameglio} {et~al.}(2007){Ameglio}, {Borgani}, {Pierpaoli}, \&
  {Dolag}}]{ameglio+07}
{Ameglio}, S., {Borgani}, S., {Pierpaoli}, E., \& {Dolag}, K. 2007, \mnras,
  382, 397

\bibitem[{{Amodeo} {et~al.}(2016){Amodeo}, {Ettori}, {Capasso}, \&
  {Sereno}}]{amodeo+16}
{Amodeo}, S., {Ettori}, S., {Capasso}, R., \& {Sereno}, M. 2016, \aap, 590,
  A126

\bibitem[{{Andreon}(2012)}]{Andreon+12}
{Andreon}, S. 2012, \aap, 546, A6

\bibitem[{{Arnaud}(1996)}]{xspec}
{Arnaud}, K.~A. 1996, in Astronomical Society of the Pacific Conference Series,
  Vol. 101, Astronomical Data Analysis Software and Systems V, ed. G.~H.
  {Jacoby} \& J.~{Barnes}, 17

\bibitem[{{Arnaud} {et~al.}(2010){Arnaud}, {Pratt}, {Piffaretti},
  {B{\"o}hringer}, {Croston}, \& {Pointecouteau}}]{arnaud+10}
{Arnaud}, M., {Pratt}, G.~W., {Piffaretti}, R., {et~al.} 2010, \aap, 517, A92

\bibitem[{{Ascasibar} {et~al.}(2006){Ascasibar}, {Sevilla}, {Yepes},
  {M{\"u}ller}, \& {Gottl{\"o}ber}}]{Ascasibar+06}
{Ascasibar}, Y., {Sevilla}, R., {Yepes}, G., {M{\"u}ller}, V., \&
  {Gottl{\"o}ber}, S. 2006, \mnras, 371, 193

\bibitem[{{Asplund} {et~al.}(2009){Asplund}, {Grevesse}, {Sauval}, \&
  {Scott}}]{aspl}
{Asplund}, M., {Grevesse}, N., {Sauval}, A.~J., \& {Scott}, P. 2009, \araa, 47,
  481

\bibitem[{{Avestruz} {et~al.}(2016){Avestruz}, {Nagai}, \& {Lau}}]{avestruz16}
{Avestruz}, C., {Nagai}, D., \& {Lau}, E.~T. 2016, \apj, 833, 227

\bibitem[{{Bartalucci} {et~al.}(2017{\natexlab{a}}){Bartalucci}, {Arnaud},
  {Pratt}, {D{\'e}mocl{\`e}s}, {van der Burg}, \&
  {Mazzotta}}]{bartalucci+17_general}
{Bartalucci}, I., {Arnaud}, M., {Pratt}, G.~W., {et~al.} 2017{\natexlab{a}},
  \aap, 598, A61

\bibitem[{{Bartalucci} {et~al.}(2017{\natexlab{b}}){Bartalucci}, {Arnaud},
  {Pratt}, {Vikhlinin}, {Pointecouteau}, {Forman}, {Jones}, {Mazzotta}, \&
  {Andrade-Santos}}]{bartalucci17}
---. 2017{\natexlab{b}}, \aap, 608, A88

\bibitem[{{Bayliss} {et~al.}(2014){Bayliss}, {Ashby}, {Ruel}, {Brodwin},
  {Aird}, {Bautz}, {Benson}, {Bleem}, {Bocquet}, {Carlstrom}, {Chang}, {Cho},
  {Clocchiatti}, {Crawford}, {Crites}, {Desai}, {Dobbs}, {Dudley}, {Foley},
  {Forman}, {George}, {Gettings}, {Gladders}, {Gonzalez}, {de Haan},
  {Halverson}, {High}, {Holder}, {Holzapfel}, {Hoover}, {Hrubes}, {Jones},
  {Joy}, {Keisler}, {Knox}, {Lee}, {Leitch}, {Liu}, {Lueker}, {Luong-Van},
  {Mantz}, {Marrone}, {Mawatari}, {McDonald}, {McMahon}, {Mehl}, {Meyer},
  {Miller}, {Mocanu}, {Mohr}, {Montroy}, {Murray}, {Padin}, {Plagge}, {Pryke},
  {Reichardt}, {Rest}, {Ruhl}, {Saliwanchik}, {Saro}, {Sayre}, {Schaffer},
  {Shirokoff}, {Song}, {Stalder}, {{\v S}uhada}, {Spieler}, {Stanford},
  {Staniszewski}, {Stark}, {Story}, {Stubbs}, {van Engelen}, {Vanderlinde},
  {Vieira}, {Vikhlinin}, {Williamson}, {Zahn}, \& {Zenteno}}]{Bayliss2040}
{Bayliss}, M.~B., {Ashby}, M.~L.~N., {Ruel}, J., {et~al.} 2014, \apj, 794, 12

\bibitem[{{B{\^\i}rzan} {et~al.}(2017){B{\^\i}rzan}, {Rafferty}, {Br{\"u}ggen},
  \& {Intema}}]{birzan17}
{B{\^\i}rzan}, L., {Rafferty}, D.~A., {Br{\"u}ggen}, M., \& {Intema}, H.~T.
  2017, \mnras, 471, 1766

\bibitem[{{Bleem} {et~al.}(2015){Bleem}, {Stalder}, {de Haan}, {Aird}, {Allen},
  {Applegate}, {Ashby}, {Bautz}, {Bayliss}, {Benson}, {Bocquet}, {Brodwin},
  {Carlstrom}, {Chang}, {Chiu}, {Cho}, {Clocchiatti}, {Crawford}, {Crites},
  {Desai}, {Dietrich}, {Dobbs}, {Foley}, {Forman}, {George}, {Gladders},
  {Gonzalez}, {Halverson}, {Hennig}, {Hoekstra}, {Holder}, {Holzapfel},
  {Hrubes}, {Jones}, {Keisler}, {Knox}, {Lee}, {Leitch}, {Liu}, {Lueker},
  {Luong-Van}, {Mantz}, {Marrone}, {McDonald}, {McMahon}, {Meyer}, {Mocanu},
  {Mohr}, {Murray}, {Padin}, {Pryke}, {Reichardt}, {Rest}, {Ruel}, {Ruhl},
  {Saliwanchik}, {Saro}, {Sayre}, {Schaffer}, {Schrabback}, {Shirokoff},
  {Song}, {Spieler}, {Stanford}, {Staniszewski}, {Stark}, {Story}, {Stubbs},
  {Vanderlinde}, {Vieira}, {Vikhlinin}, {Williamson}, {Zahn}, \&
  {Zenteno}}]{bleem15}
{Bleem}, L.~E., {Stalder}, B., {de Haan}, T., {et~al.} 2015, \apjs, 216, 27

\bibitem[{{Bocquet} {et~al.}(2019){Bocquet}, {Dietrich}, {Schrabback}, {Bleem},
  {Klein}, {Allen}, {Applegate}, {Ashby}, {Bautz}, {Bayliss}, {Benson},
  {Brodwin}, {Bulbul}, {Canning}, {Capasso}, {Carlstrom}, {Chang}, {Chiu},
  {Cho}, {Clocchiatti}, {Crawford}, {Crites}, {de Haan}, {Desai}, {Dobbs},
  {Foley}, {Forman}, {Garmire}, {George}, {Gladders}, {Gonzalez}, {Grandis},
  {Gupta}, {Halverson}, {Hlavacek-Larrondo}, {Hoekstra}, {Holder}, {Holzapfel},
  {Hou}, {Hrubes}, {Huang}, {Jones}, {Khullar}, {Knox}, {Kraft}, {Lee}, {von
  der Linden}, {Luong-Van}, {Mantz}, {Marrone}, {McDonald}, {McMahon}, {Meyer},
  {Mocanu}, {Mohr}, {Morris}, {Padin}, {Patil}, {Pryke}, {Rapetti},
  {Reichardt}, {Rest}, {Ruhl}, {Saliwanchik}, {Saro}, {Sayre}, {Schaffer},
  {Shirokoff}, {Stalder}, {Stanford}, {Staniszewski}, {Stark}, {Story},
  {Strazzullo}, {Stubbs}, {Vanderlinde}, {Vieira}, {Vikhlinin}, {Williamson},
  \& {Zenteno}}]{Bocquet+19}
{Bocquet}, S., {Dietrich}, J.~P., {Schrabback}, T., {et~al.} 2019, \apj, 878,
  55

\bibitem[{{Bonamente} {et~al.}(2012){Bonamente}, {Hasler}, {Bulbul},
  {Carlstrom}, {Culverhouse}, {Gralla}, {Greer}, {Hawkins}, {Hennessy}, {Joy},
  {Kolodziejczak}, {Lamb}, {Landry}, {Leitch}, {Marrone}, {Miller},
  {Mroczkowski}, {Muchovej}, {Plagge}, {Pryke}, {Sharp}, \&
  {Woody}}]{bonamente2012}
{Bonamente}, M., {Hasler}, N., {Bulbul}, E., {et~al.} 2012, New Journal of
  Physics, 14, 025010

\bibitem[{{Brodwin} {et~al.}(2016){Brodwin}, {McDonald}, {Gonzalez},
  {Stanford}, {Eisenhardt}, {Stern}, \& {Zeimann}}]{brodwin+16}
{Brodwin}, M., {McDonald}, M., {Gonzalez}, A.~H., {et~al.} 2016, \apj, 817, 122

\bibitem[{{Bulbul} {et~al.}(2016){Bulbul}, {Randall}, {Bayliss}, {Miller},
  {Andrade-Santos}, {Johnson}, {Bautz}, {Blanton}, {Forman}, {Jones},
  {Paterno-Mahler}, {Murray}, {Sarazin}, {Smith}, \& {Ezer}}]{bulbul2016}
{Bulbul}, E., {Randall}, S.~W., {Bayliss}, M., {et~al.} 2016, \apj, 818, 131

\bibitem[{{Bulbul} {et~al.}(2019){Bulbul}, {Chiu}, {Mohr}, {McDonald},
  {Benson}, {Bautz}, {Bayliss}, {Bleem}, {Brodwin}, \& {Bocquet}}]{bulbul19}
{Bulbul}, E., {Chiu}, I.~N., {Mohr}, J.~J., {et~al.} 2019, \apj, 871, 50

\bibitem[{{Bulbul} {et~al.}(2010){Bulbul}, {Hasler}, {Bonamente}, \&
  {Joy}}]{bulbul+10}
{Bulbul}, G.~E., {Hasler}, N., {Bonamente}, M., \& {Joy}, M. 2010, \apj, 720,
  1038

\bibitem[{{Bulbul} {et~al.}(2012){Bulbul}, {Smith}, {Foster}, {Cottam},
  {Loewenstein}, {Mushotzky}, \& {Shafer}}]{bulbul12}
{Bulbul}, G.~E., {Smith}, R.~K., {Foster}, A., {et~al.} 2012, \apj, 747, 32

\bibitem[{{Capelo} {et~al.}(2012){Capelo}, {Coppi}, \& {Natarajan}}]{capelo+12}
{Capelo}, P.~R., {Coppi}, P.~S., \& {Natarajan}, P. 2012, \mnras, 422, 686

\bibitem[{{Carlstrom} {et~al.}(2011){Carlstrom}, {Ade}, {Aird}, {Benson},
  {Bleem}, {Busetti}, {Chang}, {Chauvin}, {Cho}, {Crawford}, {Crites}, {Dobbs},
  {Halverson}, {Heimsath}, {Holzapfel}, {Hrubes}, {Joy}, {Keisler}, {Lanting},
  {Lee}, {Leitch}, {Leong}, {Lu}, {Lueker}, {Luong-Van}, {McMahon}, {Mehl},
  {Meyer}, {Mohr}, {Montroy}, {Padin}, {Plagge}, {Pryke}, {Ruhl}, {Schaffer},
  {Schwan}, {Shirokoff}, {Spieler}, {Staniszewski}, {Stark}, {Tucker},
  {Vanderlinde}, {Vieira}, \& {Williamson}}]{Carlstrom11}
{Carlstrom}, J.~E., {Ade}, P.~A.~R., {Aird}, K.~A., {et~al.} 2011, \pasp, 123,
  568

\bibitem[{{Cash}(1979)}]{cash79}
{Cash}, W. 1979, \apj, 228, 939

\bibitem[{{Cavagnolo} {et~al.}(2009){Cavagnolo}, {Donahue}, {Voit}, \&
  {Sun}}]{cavagnolo+09}
{Cavagnolo}, K.~W., {Donahue}, M., {Voit}, G.~M., \& {Sun}, M. 2009, \apjs,
  182, 12

\bibitem[{{Croston} {et~al.}(2006){Croston}, {Arnaud}, {Pointecouteau}, \&
  {Pratt}}]{croston+06}
{Croston}, J.~H., {Arnaud}, M., {Pointecouteau}, E., \& {Pratt}, G.~W. 2006,
  \aap, 459, 1007

\bibitem[{{De Grandi} \& {Molendi}(2002)}]{degrandi2002}
{De Grandi}, S., \& {Molendi}, S. 2002, \apj, 567, 163

\bibitem[{{Diehl} \& {Statler}(2006)}]{voronoi}
{Diehl}, S., \& {Statler}, T.~S. 2006, \mnras, 368, 497

\bibitem[{{Diemer}(2017)}]{Diemer+17}
{Diemer}, B. 2017, ArXiv e-prints, arXiv:1712.04512

\bibitem[{{Diemer} \& {Joyce}(2018)}]{diemer18}
{Diemer}, B., \& {Joyce}, M. 2018, arXiv e-prints, arXiv:1809.07326

\bibitem[{{Eckert} {et~al.}(2013{\natexlab{a}}){Eckert}, {Ettori}, {Molendi},
  {Vazza}, \& {Paltani}}]{eckert13n}
{Eckert}, D., {Ettori}, S., {Molendi}, S., {Vazza}, F., \& {Paltani}, S.
  2013{\natexlab{a}}, \aap, 551, A23

\bibitem[{{Eckert} {et~al.}(2017){Eckert}, {Ettori}, {Pointecouteau},
  {Molendi}, {Paltani}, \& {Tchernin}}]{xcop}
{Eckert}, D., {Ettori}, S., {Pointecouteau}, E., {et~al.} 2017, Astronomische
  Nachrichten, 338, 293

\bibitem[{{Eckert} {et~al.}(2011){Eckert}, {Molendi}, \& {Paltani}}]{Eckert+11}
{Eckert}, D., {Molendi}, S., \& {Paltani}, S. 2011, \aap, 526, A79

\bibitem[{{Eckert} {et~al.}(2013{\natexlab{b}}){Eckert}, {Molendi}, {Vazza},
  {Ettori}, \& {Paltani}}]{eckert13a}
{Eckert}, D., {Molendi}, S., {Vazza}, F., {Ettori}, S., \& {Paltani}, S.
  2013{\natexlab{b}}, \aap, 551, A22

\bibitem[{{Eckert} {et~al.}(2015){Eckert}, {Roncarelli}, {Ettori}, {Molendi},
  {Vazza}, {Gastaldello}, \& {Rossetti}}]{eckert15}
{Eckert}, D., {Roncarelli}, M., {Ettori}, S., {et~al.} 2015, \mnras, 447, 2198

\bibitem[{{Eckert} {et~al.}(2016){Eckert}, {Ettori}, {Coupon}, {Gastaldello},
  {Pierre}, {Melin}, {Le Brun}, {McCarthy}, {Adami}, {Chiappetti}, {Faccioli},
  {Giles}, {Lavoie}, {Lef{\`e}vre}, {Lieu}, {Mantz}, {Maughan}, {McGee},
  {Pacaud}, {Paltani}, {Sadibekova}, {Smith}, \& {Ziparo}}]{eckert16}
{Eckert}, D., {Ettori}, S., {Coupon}, J., {et~al.} 2016, \aap, 592, A12

\bibitem[{{Ettori} {et~al.}(2013){Ettori}, {Donnarumma}, {Pointecouteau},
  {Reiprich}, {Giodini}, {Lovisari}, \& {Schmidt}}]{ettori+13}
{Ettori}, S., {Donnarumma}, A., {Pointecouteau}, E., {et~al.} 2013, \ssr, 177,
  119

\bibitem[{{Ettori} {et~al.}(2010){Ettori}, {Gastaldello}, {Leccardi},
  {Molendi}, {Rossetti}, {Buote}, \& {Meneghetti}}]{ettori+10}
{Ettori}, S., {Gastaldello}, F., {Leccardi}, A., {et~al.} 2010, \aap, 524, A68

\bibitem[{{Ettori} \& {Molendi}(2011)}]{ettori+11}
{Ettori}, S., \& {Molendi}, S. 2011, Memorie della Societa Astronomica Italiana
  Supplementi, 17, 47

\bibitem[{{Ettori} {et~al.}(2018){Ettori}, {Ghirardini}, {Eckert},
  {Pointecouteau}, {Gastaldello}, {Sereno}, {Gaspari}, {Ghizzardi},
  {Roncarelli}, \& {Rossetti}}]{ettori+18}
{Ettori}, S., {Ghirardini}, V., {Eckert}, D., {et~al.} 2018, ArXiv e-prints,
  arXiv:1805.00035

\bibitem[{{Fabian} {et~al.}(2003){Fabian}, {Sanders}, {Crawford}, \&
  {Ettori}}]{fabian+03}
{Fabian}, A.~C., {Sanders}, J.~S., {Crawford}, C.~S., \& {Ettori}, S. 2003,
  \mnras, 341, 729

\bibitem[{{Fakhouri} \& {Ma}(2009)}]{Fakhouri09}
{Fakhouri}, O., \& {Ma}, C.-P. 2009, \mnras, 394, 1825

\bibitem[{{Fakhouri} {et~al.}(2010){Fakhouri}, {Ma}, \&
  {Boylan-Kolchin}}]{Fakhouri10}
{Fakhouri}, O., {Ma}, C.-P., \& {Boylan-Kolchin}, M. 2010, \mnras, 406, 2267

\bibitem[{{Feroz} {et~al.}(2009){Feroz}, {Hobson}, \& {Bridges}}]{multinest}
{Feroz}, F., {Hobson}, M.~P., \& {Bridges}, M. 2009, \mnras, 398, 1601

\bibitem[{{Foreman-Mackey} {et~al.}(2013){Foreman-Mackey}, {Hogg}, {Lang}, \&
  {Goodman}}]{emcee}
{Foreman-Mackey}, D., {Hogg}, D.~W., {Lang}, D., \& {Goodman}, J. 2013, \pasp,
  125, 306

\bibitem[{{Foster} {et~al.}(2012){Foster}, {Ji}, {Smith}, \&
  {Brickhouse}}]{foster+2012}
{Foster}, A.~R., {Ji}, L., {Smith}, R.~K., \& {Brickhouse}, N.~S. 2012, \apj,
  756, 128

\bibitem[{{Fowler} {et~al.}(2007){Fowler}, {Niemack}, {Dicker}, {Aboobaker},
  {Ade}, {Battistelli}, {Devlin}, {Fisher}, {Halpern}, {Hargrave}, {Hincks},
  {Kaul}, {Klein}, {Lau}, {Limon}, {Marriage}, {Mauskopf}, {Page}, {Staggs},
  {Swetz}, {Switzer}, {Thornton}, \& {Tucker}}]{Fowler+07}
{Fowler}, J.~W., {Niemack}, M.~D., {Dicker}, S.~R., {et~al.} 2007, Applied
  Optics, 46, 3444

\bibitem[{{Fruscione} {et~al.}(2006){Fruscione}, {McDowell}, {Allen},
  {Brickhouse}, {Burke}, {Davis}, {Durham}, {Elvis}, {Galle}, {Harris},
  {Huenemoerder}, {Houck}, {Ishibashi}, {Karovska}, {Nicastro}, {Noble},
  {Nowak}, {Primini}, {Siemiginowska}, {Smith}, \& {Wise}}]{ciao}
{Fruscione}, A., {McDowell}, J.~C., {Allen}, G.~E., {et~al.} 2006, in
  \procspie, Vol. 6270, Society of Photo-Optical Instrumentation Engineers
  (SPIE) Conference Series, 62701V

\bibitem[{Gao \& Han(2012)}]{Gao2012}
Gao, F., \& Han, L. 2012, Computational Optimization and Applications, 51, 259

\bibitem[{{Gaskin} {et~al.}(2019){Gaskin}, {Swartz}, {Vikhlinin}, {{\"O}zel},
  {Gelmis}, {Arenberg}, {Bandler}, {Bautz}, {Civitani}, {Dominguez}, {Eckart},
  {Falcone}, {Figueroa-Feliciano}, {Freeman}, {G{\"u}nther}, {Havey},
  {Heilmann}, {Kilaru}, {Kraft}, {McCarley}, {McEntaffer}, {Pareschi},
  {Purcell}, {Reid}, {Schattenburg}, {Schwartz}, {Schwartz}, {Tananbaum},
  {Tremblay}, {Zhang}, \& {Zuhone}}]{gaskin2019}
{Gaskin}, J.~A., {Swartz}, D.~A., {Vikhlinin}, A., {et~al.} 2019, Journal of
  Astronomical Telescopes, Instruments, and Systems, 5, 021001

\bibitem[{{Ghirardini} {et~al.}(2019){Ghirardini}, {Ettori}, {Eckert}, \&
  {Molendi}}]{Ghirardini+19}
{Ghirardini}, V., {Ettori}, S., {Eckert}, D., \& {Molendi}, S. 2019, \aap, 627,
  A19

\bibitem[{{Ghirardini} {et~al.}(2018{\natexlab{a}}){Ghirardini}, {Ettori},
  {Eckert}, {Molendi}, {Gastaldello}, {Pointecouteau}, {Hurier}, \&
  {Bourdin}}]{ghirardini18}
{Ghirardini}, V., {Ettori}, S., {Eckert}, D., {et~al.} 2018{\natexlab{a}},
  \aap, 614, A7

\bibitem[{{Ghirardini} {et~al.}(2018{\natexlab{b}}){Ghirardini}, {Eckert},
  {Ettori}, {Pointecouteau}, {Molendi}, {Gaspari}, {Rossetti}, {De Grandi},
  {Roncarelli}, {Bourdin}, {Mazzotta}, {Rasia}, \& {Vazza}}]{ghirardini18_b}
{Ghirardini}, V., {Eckert}, D., {Ettori}, S., {et~al.} 2018{\natexlab{b}},
  ArXiv e-prints, arXiv:1805.00042

\bibitem[{{Gonzalez} {et~al.}(2013){Gonzalez}, {Sivanandam}, {Zabludoff}, \&
  {Zaritsky}}]{gonzalez+13}
{Gonzalez}, A.~H., {Sivanandam}, S., {Zabludoff}, A.~I., \& {Zaritsky}, D.
  2013, \apj, 778, 14

\bibitem[{{Goodman} \& {Weare}(2010)}]{Goodman+10}
{Goodman}, J., \& {Weare}, J. 2010, Communications in Applied Mathematics and
  Computational Science, Vol.~5, No.~1, p.~65-80, 2010, 5, 65

\bibitem[{{Grant} {et~al.}(2005){Grant}, {Bautz}, {Kissel}, {LaMarr}, \&
  {Prigozhin}}]{grant05}
{Grant}, C.~E., {Bautz}, M.~W., {Kissel}, S.~M., {LaMarr}, B., \& {Prigozhin},
  G.~Y. 2005, in \procspie, Vol. 5898, UV, X-Ray, and Gamma-Ray Space
  Instrumentation for Astronomy XIV, ed. O.~H.~W. {Siegmund}, 201--211

\bibitem[{{Hasler} {et~al.}(2012){Hasler}, {Bulbul}, {Bonamente}, {Carlstrom},
  {Culverhouse}, {Gralla}, {Greer}, {Hawkins}, {Hennessy}, {Joy},
  {Kolodziejczak}, {Lamb}, {Landry}, {Leitch}, {Mantz}, {Marrone}, {Miller},
  {Mroczkowski}, {Muchovej}, {Plagge}, {Pryke}, \& {Woody}}]{hasler2012}
{Hasler}, N., {Bulbul}, E., {Bonamente}, M., {et~al.} 2012, \apj, 748, 113

\bibitem[{{Hlavacek-Larrondo} {et~al.}(2012){Hlavacek-Larrondo}, {Fabian},
  {Edge}, {Ebeling}, {Sanders}, {Hogan}, \& {Taylor}}]{Hlavacek-Larrondo+12}
{Hlavacek-Larrondo}, J., {Fabian}, A.~C., {Edge}, A.~C., {et~al.} 2012, \mnras,
  421, 1360

\bibitem[{{Kaiser}(1986)}]{Kaiser+86}
{Kaiser}, N. 1986, \mnras, 222, 323

\bibitem[{{Kalberla} {et~al.}(2005){Kalberla}, {Burton}, {Hartmann}, {Arnal},
  {Bajaja}, {Morras}, \& {P{\"o}ppel}}]{kalberla05}
{Kalberla}, P.~M.~W., {Burton}, W.~B., {Hartmann}, D., {et~al.} 2005, \aap,
  440, 775

\bibitem[{{Khedekar} {et~al.}(2013){Khedekar}, {Churazov}, {Kravtsov},
  {Zhuravleva}, {Lau}, {Nagai}, \& {Sunyaev}}]{khedekar13}
{Khedekar}, S., {Churazov}, E., {Kravtsov}, A., {et~al.} 2013, \mnras, 431, 954

\bibitem[{{Khullar} {et~al.}(2019){Khullar}, {Bleem}, {Bayliss}, {Gladders},
  {Benson}, {McDonald}, {Allen}, {Applegate}, {Ashby}, {Bocquet}, {Brodwin},
  {Bulbul}, {Canning}, {Capasso}, {Chiu}, {Crawford}, {de Haan}, {Dietrich},
  {Gonzalez}, {Hlavacek-Larrondo}, {Hoekstra}, {Holzapfel}, {von der Linden},
  {Mantz}, {Patil}, {Reichardt}, {Saro}, {Sharon}, {Stalder}, {Stanford},
  {Stark}, \& {Strazzullo}}]{khullar19}
{Khullar}, G., {Bleem}, L.~E., {Bayliss}, M.~B., {et~al.} 2019, \apj, 870, 7

\bibitem[{{Komatsu} \& {Seljak}(2001)}]{komatsu+01}
{Komatsu}, E., \& {Seljak}, U. 2001, \mnras, 327, 1353

\bibitem[{{Kravtsov} \& {Borgani}(2012)}]{kravtsov12}
{Kravtsov}, A.~V., \& {Borgani}, S. 2012, \araa, 50, 353

\bibitem[{{Leccardi} \& {Molendi}(2008)}]{lm08}
{Leccardi}, A., \& {Molendi}, S. 2008, \aap, 486, 359

\bibitem[{{Markevitch} {et~al.}(1998){Markevitch}, {Forman}, {Sarazin}, \&
  {Vikhlinin}}]{Markevitch+98}
{Markevitch}, M., {Forman}, W.~R., {Sarazin}, C.~L., \& {Vikhlinin}, A. 1998,
  \apj, 503, 77

\bibitem[{{Mazzotta} {et~al.}(2004){Mazzotta}, {Rasia}, {Moscardini}, \&
  {Tormen}}]{mazzotta+04}
{Mazzotta}, P., {Rasia}, E., {Moscardini}, L., \& {Tormen}, G. 2004, \mnras,
  354, 10

\bibitem[{{McDonald} {et~al.}(2013){McDonald}, {Benson}, {Vikhlinin},
  {Stalder}, {Bleem}, {de Haan}, {Lin}, {Aird}, {Ashby}, {Bautz}, {Bayliss},
  {Bocquet}, {Brodwin}, {Carlstrom}, {Chang}, {Cho}, {Clocchiatti}, {Crawford},
  {Crites}, {Desai}, {Dobbs}, {Dudley}, {Foley}, {Forman}, {George},
  {Gettings}, {Gladders}, {Gonzalez}, {Halverson}, {High}, {Holder},
  {Holzapfel}, {Hoover}, {Hrubes}, {Jones}, {Joy}, {Keisler}, {Knox}, {Lee},
  {Leitch}, {Liu}, {Lueker}, {Luong-Van}, {Mantz}, {Marrone}, {McMahon},
  {Mehl}, {Meyer}, {Miller}, {Mocanu}, {Mohr}, {Montroy}, {Murray},
  {Nurgaliev}, {Padin}, {Plagge}, {Pryke}, {Reichardt}, {Rest}, {Ruel}, {Ruhl},
  {Saliwanchik}, {Saro}, {Sayre}, {Schaffer}, {Shirokoff}, {Song}, {{\v
  S}uhada}, {Spieler}, {Stanford}, {Staniszewski}, {Stark}, {Story}, {van
  Engelen}, {Vanderlinde}, {Vieira}, {Williamson}, {Zahn}, \&
  {Zenteno}}]{mcdonald+13}
{McDonald}, M., {Benson}, B.~A., {Vikhlinin}, A., {et~al.} 2013, \apj, 774, 23

\bibitem[{{McDonald} {et~al.}(2014){McDonald}, {Benson}, {Vikhlinin}, {Aird},
  {Allen}, {Bautz}, {Bayliss}, {Bleem}, {Bocquet}, {Brodwin}, {Carlstrom},
  {Chang}, {Cho}, {Clocchiatti}, {Crawford}, {Crites}, {de Haan}, {Dobbs},
  {Foley}, {Forman}, {George}, {Gladders}, {Gonzalez}, {Halverson},
  {Hlavacek-Larrondo}, {Holder}, {Holzapfel}, {Hrubes}, {Jones}, {Keisler},
  {Knox}, {Lee}, {Leitch}, {Liu}, {Lueker}, {Luong-Van}, {Mantz}, {Marrone},
  {McMahon}, {Meyer}, {Miller}, {Mocanu}, {Mohr}, {Murray}, {Padin}, {Pryke},
  {Reichardt}, {Rest}, {Ruhl}, {Saliwanchik}, {Saro}, {Sayre}, {Schaffer},
  {Shirokoff}, {Spieler}, {Stalder}, {Stanford}, {Staniszewski}, {Stark},
  {Story}, {Stubbs}, {Vanderlinde}, {Vieira}, {Williamson}, {Zahn}, \&
  {Zenteno}}]{mcdonald+14}
---. 2014, \apj, 794, 67

\bibitem[{{McDonald} {et~al.}(2016){McDonald}, {Stalder}, {Bayliss}, {Allen},
  {Applegate}, {Ashby}, {Bautz}, {Benson}, {Bleem}, {Brodwin}, {Carlstrom},
  {Chiu}, {Desai}, {Gonzalez}, {Hlavacek-Larrondo}, {Holzapfel}, {Marrone},
  {Miller}, {Reichardt}, {Saliwanchik}, {Saro}, {Schrabback}, {Stanford},
  {Stark}, {Vieira}, \& {Zenteno}}]{mcdonald16}
{McDonald}, M., {Stalder}, B., {Bayliss}, M., {et~al.} 2016, \apj, 817, 86

\bibitem[{{McDonald} {et~al.}(2017){McDonald}, {Allen}, {Bayliss}, {Benson},
  {Bleem}, {Brodwin}, {Bulbul}, {Carlstrom}, {Forman}, {Hlavacek-Larrondo},
  {Garmire}, {Gaspari}, {Gladders}, {Mantz}, \& {Murray}}]{mcdonald17}
{McDonald}, M., {Allen}, S.~W., {Bayliss}, M., {et~al.} 2017, \apj, 843, 28

\bibitem[{{Nagai} {et~al.}(2007){Nagai}, {Kravtsov}, \& {Vikhlinin}}]{nagai+07}
{Nagai}, D., {Kravtsov}, A.~V., \& {Vikhlinin}, A. 2007, \apj, 668, 1

\bibitem[{{Nandra} {et~al.}(2013){Nandra}, {Barret}, {Barcons}, {Fabian}, {den
  Herder}, {Piro}, {Watson}, {Adami}, {Aird}, {Afonso}, \& et~al.}]{nandra+13}
{Nandra}, K., {Barret}, D., {Barcons}, X., {et~al.} 2013, ArXiv e-prints,
  arXiv:1306.2307

\bibitem[{{Navarro} {et~al.}(1997){Navarro}, {Frenk}, \& {White}}]{nfw+97}
{Navarro}, J.~F., {Frenk}, C.~S., \& {White}, S.~D.~M. 1997, \apj, 490, 493

\bibitem[{{Neto} {et~al.}(2007){Neto}, {Gao}, {Bett}, {Cole}, {Navarro},
  {Frenk}, {White}, {Springel}, \& {Jenkins}}]{Neto+07}
{Neto}, A.~F., {Gao}, L., {Bett}, P., {et~al.} 2007, \mnras, 381, 1450

\bibitem[{{Ostriker} {et~al.}(2005){Ostriker}, {Bode}, \&
  {Babul}}]{Ostriker+05}
{Ostriker}, J.~P., {Bode}, P., \& {Babul}, A. 2005, \apj, 634, 964

\bibitem[{{Planck Collaboration} {et~al.}(2014){Planck Collaboration}, {Ade},
  {Aghanim}, {Armitage-Caplan}, {Arnaud}, {Ashdown}, {Atrio-Barandela},
  {Aumont}, {Baccigalupi}, {Banday}, \& et~al.}]{planck_14_scaling}
{Planck Collaboration}, {Ade}, P.~A.~R., {Aghanim}, N., {et~al.} 2014, \aap,
  571, A20

\bibitem[{{Planck Collaboration} {et~al.}(2016){Planck Collaboration}, {Ade},
  {Aghanim}, {Arnaud}, {Ashdown}, {Aumont}, {Baccigalupi}, {Banday},
  {Barreiro}, {Bartlett}, \& et~al.}]{planck+16}
---. 2016, \aap, 594, A13

\bibitem[{{Pratt} {et~al.}(2009){Pratt}, {Croston}, {Arnaud}, \&
  {B{\"o}hringer}}]{pratt+09}
{Pratt}, G.~W., {Croston}, J.~H., {Arnaud}, M., \& {B{\"o}hringer}, H. 2009,
  \aap, 498, 361

\bibitem[{{Pratt} {et~al.}(2010){Pratt}, {Arnaud}, {Piffaretti},
  {B{\"o}hringer}, {Ponman}, {Croston}, {Voit}, {Borgani}, \&
  {Bower}}]{pratt+10}
{Pratt}, G.~W., {Arnaud}, M., {Piffaretti}, R., {et~al.} 2010, \aap, 511, A85

\bibitem[{{Read} {et~al.}(2011){Read}, {Rosen}, {Saxton}, \&
  {Ramirez}}]{read+11}
{Read}, A.~M., {Rosen}, S.~R., {Saxton}, R.~D., \& {Ramirez}, J. 2011, \aap,
  534, A34

\bibitem[{{Roncarelli} {et~al.}(2013){Roncarelli}, {Ettori}, {Borgani},
  {Dolag}, {Fabjan}, \& {Moscardini}}]{roncarelli+13}
{Roncarelli}, M., {Ettori}, S., {Borgani}, S., {et~al.} 2013, \mnras, 432, 3030

\bibitem[{{Salvetti} {et~al.}(2017){Salvetti}, {Marelli}, {Gastaldello},
  {Ghizzardi}, {Molendi}, {De Luca}, {Moretti}, {Rossetti}, \&
  {Tiengo}}]{salvetti17}
{Salvetti}, D., {Marelli}, M., {Gastaldello}, F., {et~al.} 2017, ArXiv
  e-prints, arXiv:1705.04172

\bibitem[{{Sanders} {et~al.}(2018){Sanders}, {Fabian}, {Russell}, \&
  {Walker}}]{Sanders+18}
{Sanders}, J.~S., {Fabian}, A.~C., {Russell}, H.~R., \& {Walker}, S.~A. 2018,
  \mnras, 474, 1065

\bibitem[{{Sanderson} {et~al.}(2003){Sanderson}, {Ponman}, {Finoguenov},
  {Lloyd-Davies}, \& {Markevitch}}]{Sanderson+03}
{Sanderson}, A.~J.~R., {Ponman}, T.~J., {Finoguenov}, A., {Lloyd-Davies},
  E.~J., \& {Markevitch}, M. 2003, \mnras, 340, 989

\bibitem[{{Schellenberger} {et~al.}(2015){Schellenberger}, {Reiprich},
  {Lovisari}, {Nevalainen}, \& {David}}]{schellenberger+15}
{Schellenberger}, G., {Reiprich}, T.~H., {Lovisari}, L., {Nevalainen}, J., \&
  {David}, L. 2015, \aap, 575, A30

\bibitem[{{Shaw} {et~al.}(2010){Shaw}, {Nagai}, {Bhattacharya}, \&
  {Lau}}]{shaw+10}
{Shaw}, L.~D., {Nagai}, D., {Bhattacharya}, S., \& {Lau}, E.~T. 2010, \apj,
  725, 1452

\bibitem[{{Shitanishi} {et~al.}(2018){Shitanishi}, {Pierpaoli}, {Sayers},
  {Golwala}, {Ameglio}, {Mantz}, {Mroczkowski}, {Rasia}, \&
  {Siegel}}]{Shitanishi+18}
{Shitanishi}, J.~A., {Pierpaoli}, E., {Sayers}, J., {et~al.} 2018, \mnras, 481,
  749

\bibitem[{{Snowden} {et~al.}(2008){Snowden}, {Mushotzky}, {Kuntz}, \&
  {Davis}}]{snowden08}
{Snowden}, S.~L., {Mushotzky}, R.~F., {Kuntz}, K.~D., \& {Davis}, D.~S. 2008,
  \aap, 478, 615

\bibitem[{{Stalder} {et~al.}(2013){Stalder}, {Ruel}, {{\v S}uhada}, {Brodwin},
  {Aird}, {Andersson}, {Armstrong}, {Ashby}, {Bautz}, {Bayliss}, {Bazin},
  {Benson}, {Bleem}, {Carlstrom}, {Chang}, {Cho}, {Clocchiatti}, {Crawford},
  {Crites}, {de Haan}, {Desai}, {Dobbs}, {Dudley}, {Foley}, {Forman}, {George},
  {Gettings}, {Gladders}, {Gonzalez}, {Halverson}, {Harrington}, {High},
  {Holder}, {Holzapfel}, {Hoover}, {Hrubes}, {Jones}, {Joy}, {Keisler}, {Knox},
  {Lee}, {Leitch}, {Liu}, {Lueker}, {Luong-Van}, {Mantz}, {Marrone},
  {McDonald}, {McMahon}, {Mehl}, {Meyer}, {Mocanu}, {Mohr}, {Montroy},
  {Murray}, {Natoli}, {Nurgaliev}, {Padin}, {Plagge}, {Pryke}, {Reichardt},
  {Rest}, {Ruhl}, {Saliwanchik}, {Saro}, {Sayre}, {Schaffer}, {Shaw},
  {Shirokoff}, {Song}, {Spieler}, {Stanford}, {Staniszewski}, {Stark}, {Story},
  {Stubbs}, {van Engelen}, {Vanderlinde}, {Vieira}, {Vikhlinin}, {Williamson},
  {Zahn}, \& {Zenteno}}]{Stalder0205}
{Stalder}, B., {Ruel}, J., {{\v S}uhada}, R., {et~al.} 2013, \apj, 763, 93

\bibitem[{{Tillson} {et~al.}(2011){Tillson}, {Miller}, \&
  {Devriendt}}]{Tillson11}
{Tillson}, H., {Miller}, L., \& {Devriendt}, J. 2011, \mnras, 417, 666

\bibitem[{{Tozzi} {et~al.}(2015){Tozzi}, {Santos}, {Jee}, {Fassbender},
  {Rosati}, {Nastasi}, {Forman}, {Sartoris}, {Borgani}, {Boehringer},
  {Altieri}, {Pratt}, {Nonino}, \& {Jones}}]{tozzi+15}
{Tozzi}, P., {Santos}, J.~S., {Jee}, M.~J., {et~al.} 2015, \apj, 799, 93

\bibitem[{{Urban} {et~al.}(2011){Urban}, {Werner}, {Simionescu}, {Allen}, \&
  {B{\"o}hringer}}]{Urban+11}
{Urban}, O., {Werner}, N., {Simionescu}, A., {Allen}, S.~W., \&
  {B{\"o}hringer}, H. 2011, \mnras, 414, 2101

\bibitem[{{Urban} {et~al.}(2014){Urban}, {Simionescu}, {Werner}, {Allen},
  {Ehlert}, {Zhuravleva}, {Morris}, {Fabian}, {Mantz}, {Nulsen}, {Sanders}, \&
  {Takei}}]{urban14}
{Urban}, O., {Simionescu}, A., {Werner}, N., {et~al.} 2014, \mnras, 437, 3939

\bibitem[{{Vikhlinin} {et~al.}(2006){Vikhlinin}, {Kravtsov}, {Forman}, {Jones},
  {Markevitch}, {Murray}, \& {Van Speybroeck}}]{vikhlini+06}
{Vikhlinin}, A., {Kravtsov}, A., {Forman}, W., {et~al.} 2006, \apj, 640, 691

\bibitem[{{Vikhlinin} {et~al.}(2009){Vikhlinin}, {Kravtsov}, {Burenin},
  {Ebeling}, {Forman}, {Hornstrup}, {Jones}, {Murray}, {Nagai}, {Quintana}, \&
  {Voevodkin}}]{vikhlinin+09}
{Vikhlinin}, A., {Kravtsov}, A.~V., {Burenin}, R.~A., {et~al.} 2009, \apj, 692,
  1060

\bibitem[{{Voit} {et~al.}(2005){Voit}, {Kay}, \& {Bryan}}]{voit+05}
{Voit}, G.~M., {Kay}, S.~T., \& {Bryan}, G.~L. 2005, \mnras, 364, 909

\bibitem[{{Walker} {et~al.}(2019){Walker}, {Simionescu}, {Nagai}, {Okabe},
  {Eckert}, {Mroczkowski}, {Akamatsu}, {Ettori}, \& {Ghirardini}}]{walker2019}
{Walker}, S., {Simionescu}, A., {Nagai}, D., {et~al.} 2019, \ssr, 215, 7

\bibitem[{{Walker} {et~al.}(2012){Walker}, {Fabian}, {Sanders}, \&
  {George}}]{walker12a}
{Walker}, S.~A., {Fabian}, A.~C., {Sanders}, J.~S., \& {George}, M.~R. 2012,
  \mnras, 424, 1826

\bibitem[{{Wechsler} {et~al.}(2002){Wechsler}, {Bullock}, {Primack},
  {Kravtsov}, \& {Dekel}}]{Wechsler02}
{Wechsler}, R.~H., {Bullock}, J.~S., {Primack}, J.~R., {Kravtsov}, A.~V., \&
  {Dekel}, A. 2002, \apj, 568, 52

\bibitem[{{Zhuravleva} {et~al.}(2013){Zhuravleva}, {Churazov}, {Kravtsov},
  {Lau}, {Nagai}, \& {Sunyaev}}]{zhuravleva13}
{Zhuravleva}, I., {Churazov}, E., {Kravtsov}, A., {et~al.} 2013, \mnras, 428,
  3274

\end{thebibliography}



\end{document}